\documentclass[12pt]{article}
\pdfoutput=1 
\usepackage{amsmath}
\usepackage{graphicx}
\usepackage{enumerate}
\usepackage{url} 
\usepackage{epsfig,amssymb,amsfonts,amsthm,epsfig,verbatim,
  enumitem}
\usepackage{mathabx}
\usepackage{esint}
\usepackage{color,graphicx,lscape,longtable,xr}
\usepackage{multirow,soul}
\usepackage{booktabs}
\usepackage{algorithm}
\usepackage{algorithmic}
\usepackage[authoryear, sort]{natbib}

\newcommand{\mystrut}{\vphantom{\int_0^1}}

\newcommand{\sumi}{\ensuremath{\sum_{i=1}^{n}}}

\newtheorem{Th}{\underline{\bf Theorem}}

\newtheorem{Rem}{\underline{\bf Remark}}

\newtheorem{Lem}{\underline{\bf Lemma}}

\def\boxit#1{\vbox{\hrule\hbox{\vrule\kern6pt  \vbox{\kern6pt#1\kern6pt}\kern6pt\vrule}\hrule}}
\def\bse{\begin{eqnarray*}}
\def\ese{\end{eqnarray*}}
\def\be{\begin{eqnarray}}
\def\ee{\end{eqnarray}}
\def\bsq{\begin{equation*}}
\def\esq{\end{equation*}}
\def\bq{\begin{equation}}
\def\eq{\end{equation}}

\def\supp{\hbox{supp}}
\def\vec{\mbox{vec}}
\def\argmin{\mbox{argmin}}

\def\sumi{\sum_{i=1}^n}

\def\trans{^{\rm T}}

\def\ba{{\boldsymbol\alpha}}

\def\wc{\widecheck}

\def\bb{{\boldsymbol\beta}}
\def\bPsi{{\boldsymbol\Psi}}
\def\brho{{\boldsymbol\rho}}
\def\bomega{{\boldsymbol\omega}}

\def\bG{{\boldsymbol\Gamma}}

\def\bg{{\boldsymbol\gamma}}

\def\A{{\bf A}}

\def\a{{\bf a}}
\def\B{{\bf B}}
\def\c{{\bf c}}
\def\C{{\bf C}}
\def\D{{\bf D}}
\def\Q{{\bf Q}}
\def\V{{\bf V}}

\def\h{{\bf h}}

\def\I{{\bf I}}

\def\M{{\bf M}}

\def\mL{\mbox{$\mathcal{L}$}}

\def\t{{\bf t}}
\def\e{{\bf e}}

\def\P{{\bf P}}

\def\E{{\bf E}}

\def\u{{\bf u}}
\def\v{{\bf v}}

\def\X{{\bf X}}
\def\H{{\bf H}}

\def\x{{\bf x}}

\def\Y{{\bf Y}}

\def\Z{{\bf Z}}

\def\bSig{{\bf \Sigma}}

\def\log{\hbox{log}}

\def\squarebox#1{\hbox to #1{\hfill\vbox to #1{\vfill}}}
\def\btheta{{\boldsymbol \theta}}

\def\0{{\bf 0}}
\def\vec{\mathrm{vec}}

\def\mU{\mathcal{U}}
\def\mF{\mathcal{F}}
\def\mM{\mathcal{M}}

\def\mS{\mathcal{S}}

\def\wh{\widehat}

\def\log{\hbox{log}}

\newcommand{\blind}{1}

\begin{document}

\def\spacingset#1{\renewcommand{\baselinestretch}%
{#1}\small\normalsize} \spacingset{1}


\if1\blind
{
  \title{\bf High dimensional test for  functional covariates}
  \author{Huaqing Jin
    and 
    Fei Jiang \thanks{Email: fei.jiang@ucsf.edu}\hspace{.2cm}\\
    Department of Epidemiology and Biostatistics,  \\
    University of California, San Francisco}
  \maketitle
} \fi

\if0\blind
{
  \bigskip
  \bigskip
  \bigskip
  \begin{center}
    {\LARGE\bf High dimensional test for  functional covariates}
\end{center}
  \medskip
} \fi

\bigskip
\begin{abstract}
As medical devices become more complex, they routinely collect
extensive and complicated data.  While classical regressions typically examine the relationship between an outcome and a vector of predictors,  
it becomes imperative to identify the relationship with predictors possessing
functional structures. In this article, we introduce a novel inference
procedure for examining the relationship between outcomes and large-scale functional predictors.
We target testing the linear hypothesis on the functional parameters 
under the generalized functional linear regression framework, where the
number of the functional parameters grows with the sample size. We develop
the estimation procedure for the high dimensional generalized
functional linear model incorporating B-spline functional
approximation and amenable regularization. Furthermore, 
we construct a procedure that is able to test the
local alternative hypothesis on the linear combinations of the
functional parameters. We establish the statistical guarantees in
terms of non-asymptotic convergence of the parameter estimation and
the oracle property and asymptotic normality of the
estimators. Moreover, we derive the asymptotic distribution of the test
statistic. We carry out intensive simulations and illustrate with a
new dataset from an Alzheimer’s disease magnetoencephalography study.
\end{abstract}

\noindent%
\vfill

\newpage
\spacingset{1.9} 

\section{Introduction}\label{sec:intro}

Functional  regression is widely used to
model the relationship between a scalar response and
functional predictors.   Traditional approaches 
typically only involve low-dimensional functional predictors.  
Among the studies, \cite{yuan2010reproducing} and
\cite{cai2012minimax} investigate parameter estimation in functional
linear regression using a single functional
predictor. \cite{lian2013shrinkage} introduces a shrinkage functional estimator for functional linear regression in a finite-dimensional setting. Furthermore, \cite{kong2016partially}
explore a partial functional linear regression with finite-dimensional
functional predictors and  high-dimensional baseline covariates.  In
addition, \cite{lei2014adaptive}, \cite{hilgert2013minimax} and
\cite{shang2015nonparametric} have formulated testing procedures to
assess hypotheses concerning the finite-dimensional functional predictors in the generalized functional linear models. 
However, in numerous applications, the number
of functional predictors can grow with the sample size. Examples can
be found in neuroimaging analyses that focus on the relationship
between a disease marker and temporal signals from serveral brain regions of
interest (ROI) recorded by magnetoencephalography (MEG), electroencephalogram (EEG) or 
functional magnetic resonance imaging (fMRI).  Under this context, 
\cite{gertheiss2013variable} 
adopt a penalized likelihood method for variable selection under the generalized functional linear model framework with 
an application to a neuroimaging dataset.
\cite{fan2015functional} propose a functional additive regression model 
to handle a large number of functional predictors.
\cite{zhang2022subgroup} consider the subgroup analysis problem 
under the high-dimensional functional regression framework.
\cite{zhang2022multivariate} propose to estimate the high
dimensional functional parameters in the generalized additive model through a
penalized regression method utilizing group minimax concave penalty. 
\cite{xue2023optimal} introduce a procedure for estimating high-dimensional functional parameters, 
designed specifically to analyze EEG data. 
Despite the existence of various estimation techniques, options for
hypothesis testing on  high-dimensional functional parameters in the generalized functional
linear model framework remain scarce.   To the best of our knowledge, the bootstrap method proposed by
\cite{xue2021hypothesis} stands out as the sole known testing
approach in this area specifically for the functional linear
regression. However, this method does not extend to the nonlinear
regression models. Moreover, it lacks the capability to test for local
alternative hypotheses, incorporate baseline covariates and does not
aim to test  linear hypotheses.  
To address these limitations, we introduce a novel hypothesis testing
procedure that applies to high-dimensional functional predictors and
goes beyond the functional linear regression. This innovative approach is
built on the  generalized functional linear regression and extended to
test local alternatives and linear hypotheses of the functional parameters of interest.

\begin{figure}[!t]
\centering
\includegraphics[scale=0.11]{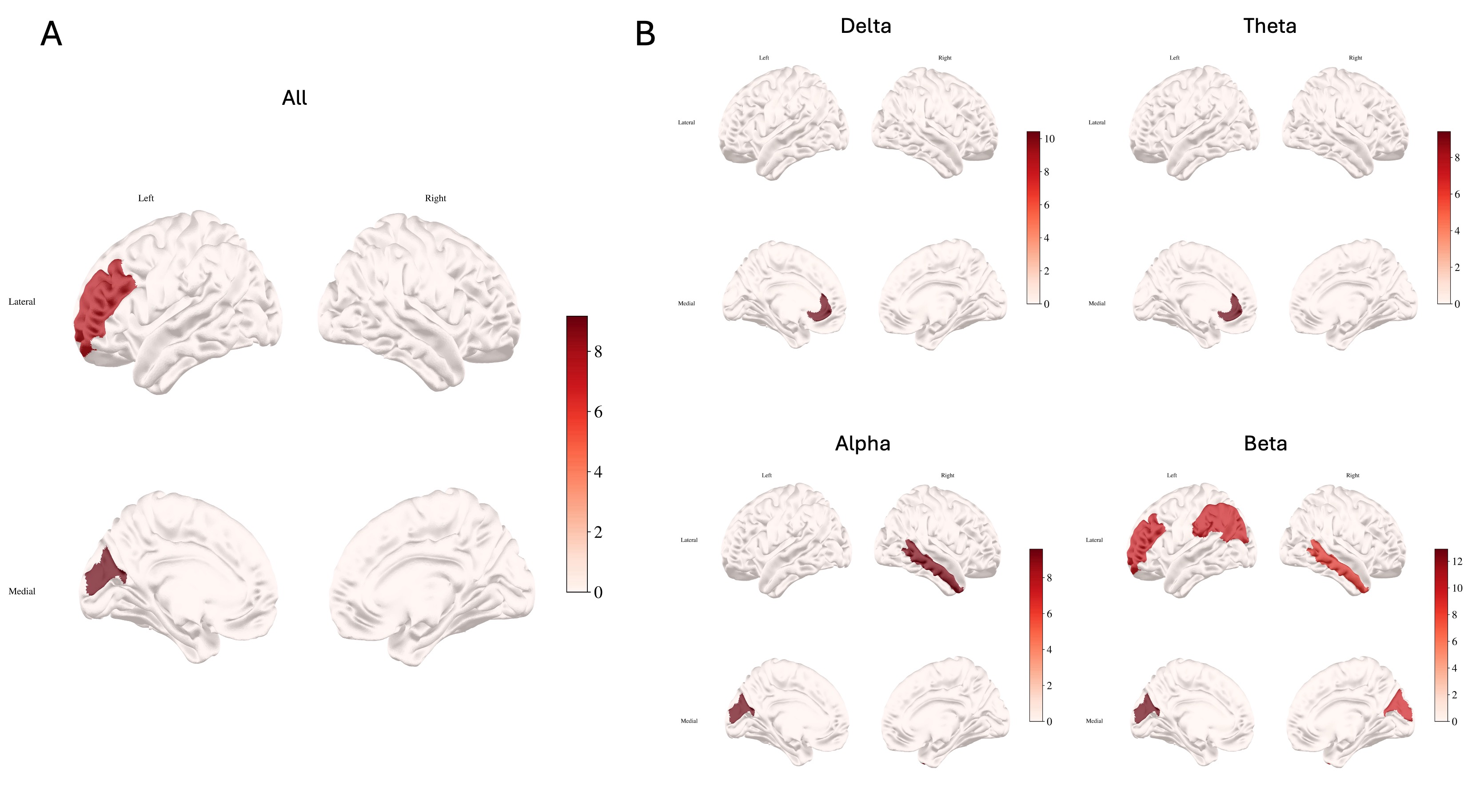}
\caption{
{\bf A:}
Negative log p-values from the significant ROIs selected via fitting 
MMSE score on each ROI's average PSD, adjusted for age, gender, and education level.
The significant ROIs are identified by the p-values less than $0.05/68$.
{\bf B:}
Negative log p-values from the significant ROIs selected via fitting MMSE score 
on each ROI's average PSD at delta, theta, alpha, and beta bands, 
adjusted for age, gender, and education level.
The significant ROIs are identified by the p-values less than $0.05/68/4$.
}
\label{fig:naive}
\end{figure}

We formulate a high-dimensional hypothesis testing procedure for
functional predictors (HDFP) under the generalized functional 
linear regression framework.
Our work is motivated by the Alzheimer's disease (AD) study \citep{jin2023} 
that aims to investigate the
association between the features from MEG recordings and the cognitive
function in 82 AD patients and 61 healthy controls. 
A critical question to explore is the relationship between
the 
power spectrum densities (PSD) of the MEG data and cognitive
function, which can be measured by cognitive scores such as the mini-mental state examination
(MMSE) score \citep{kurlowicz1999mini}.  Here, a PSD from one brain region of interest (ROI) is characterized as the square of the Fourier
transform of the raw time series across  a sequence of
frequencies. Understanding the association is crucial as it can lead
us to create new treatments aiming at suppressing abnormal waveforms,
thereby potentially slowing down the neurodegenerative process
\citep{bosco2023antiseizure}.  To improve interpretability, rather
than using raw time series data as predictors, we map the signals
recorded from the scalp onto the brain surface. We then consolidate
the voxel-level time series into region-level aggregates, which leads
to a 68-dimensional functional predictor based on the
Desikan-Killiany (DK) atlas \citep{desikan2006automated}. Finally,
Fourier transformations are applied independently to each brain signal
to generate region-wise PSDs over the whole brain. Conventionally, the average PSDs
over all frequencies or a subset of frequencies are used as a predictor
of the cognitive function \citep{wang2015power}. This approach ignores the
functional pattern of PSD, resulting in the predictors that do not
sufficiently explain the variations in the outcome. We apply this
approach to study the  relationship between the MMSE score and the 68
dimensional average
PSDs (over all frequencies) through a linear regression model adjusted for age, gender, and education level. 
Figure \ref{fig:naive}A shows that  the PSDs from only two ROIs 
significantly affect the MMSE score with the p-values less than
$0.05/68$. 
This finding deviates from the results in the existing literature \citep{jin2023,jagust2018imaging}.
We further calculate the average PSD within the specific
frequency bands, namely delta (2HZ--4HZ), theta (4HZ--8HZ), alpha
(8HZ--12HZ), and beta (12HZ--35HZ) bands
and conduct similar linear regression analysis.
We then identify the significant ROIs corresponding to
each frequency band.
Figure \ref{fig:naive}B shows that the
significant ROI varies across the frequency bands,  suggesting
that the effect of the PSDs on the MMSE score varies across the
frequencies.  
This preliminary study indicates that incorporating 
functional parameters that vary with frequency is essential for explaining the relationship
between PSD and cognitive function.

To study the functional relationship between the PSDs over a large number
of ROIs and the cognitive function or the disease indicator, 
we pursue a high dimensional
generalized functional linear regression.  We develop a hypothesis
testing procedure to identify the important linear
combinations of the function predictors. Specifically, we target  the
association between a response variable $Y$ and a 
high-dimensional 
covariate of interest $\X(s)$ adjusted by
the confounder $\Z$. The association between $Y$ and $\X(s)$  varies with
$s$. 
In our motivating example, $Y$ is the MMSE score, $\X(s)$ is the
PSD at the frequency $s$, and $\Z$ contains information about a
subject's age, gender, 
and education level. We model the conditional distribution of $Y$
given $\X$ and $\Z$ through $f\left\{Y,
  \ba\trans\Z+ \int_{0}^1 \bb\trans(s) \X(s)
    ds \right\} $, where $f(\cdot)$ is a conditional density function in the
  exponential family. When $\X(s)$ is a one dimensional function, we can
  approximate  $\bb(s) $ through a basis expansion
  $\B(s)\trans\bg$, where $\bg$ is an unknown parameter. In this case,
  the regression problem simplifies to a generalized linear model, and
  the hypothesis on  $\bb(s) $  can be transformed to a linear hypothesis on
  $\bg$. However when
  $\X(s)$ is a high-dimensional function, it is unclear whether any of the
  existing proposed tests are applicable. The first obstacle in
  constructing valid hypothesis testing procedures is that 
  the asymptotic distribution of  the resulting estimator has not been
  established. We utilize the B-splines to approximate the functional
  parameters and derive the asymptotic normality of the functional
  estimator when the amenable penalties \citep{loh2015, loh2017} are used
  for the parameter estimation.  Even after establishing the
  asymptotic normal properties, it is still challenging to construct
  the testing under the high dimensional setting for the 
  generalized functional linear regression. This is because under the
  amenable penalties \citep{loh2015, loh2017}, the asymptotic
  normality of the estimators is built on a minimal signal condition,
  which requires the norm of the nonzero elements in $\bb(s)$ to be at least of order
  $\lambda$. Here $\lambda$  is the penalty parameter which goes to
  zero when the sample
  size increases. 
  Now consider testing the null hypothesis $\beta_1(s)
  =0$
  versus the alternative $\beta_1(s)
  =h_n(s)$, where $\beta_1(s)$ is the first element of
  $\bb(s)$. The minimal signal condition implies that the test will have no
  power in testing the local alternative when $(\int h_n(s)^2 ds)^{1/2}  \leq \lambda$. 
To address this issue, we eliminate the penalties on the subset of
parameters involved in the test. However, it remains uncertain how
quickly the dimension of this parameter subset can increase while
still maintaining adequate test power and the number of basis
functions required to approximate the function. To clarify this, we
have derived the convergence properties of the estimators. These
properties specify the rate at which the dimension of the parameter
can grow relative to the sample size, while considering an increasing number of B-spline bases. 
This analysis ensures consistency, asymptotic normality, and adequate testing power.

We briefly summarize our contributions as follows. First, we develop a
new estimation procedure for the high-dimensional generalized functional linear model  with amenable regularization. Second, we show
the variable selection consistency and the consistency of the
resulting estimator. We provide the explicit convergence rate of the
estimator. Third, we derive the asymptotic normality of the estimator
for the nonzero parameters and the parameters to test. Fourth, we
construct a  test statistic for linear hypothesis in the form of
$\C\bb(s) = \t(s)$, where $\C$ is a known matrix and $\t$ is a known
function. Fifth, we derive the asymptotic distribution of the test
statistics. These five essential components collectively allow us to perform hypothesis testing 
for a generalized functional linear model with high-dimensional functional parameters.

Our proposal is related to but also clearly different from
existing high-dimensional test procedure. First, our model captures
varying effects through functional parameters.  In this sense, it is similar in spirit to
\cite{xue2021hypothesis}'s bootstrap-based test. Nevertheless, our
approach is extended to the generalized functional linear model, 
accommodating models with baseline covariates. Additionally, our
approach is capable of testing linear and local alternatives hypotheses, which is
similar to \cite{shi2019linear}'s solution. However, we consider
functional parameters and primarily focus on a group sparsity type of
penalties  \citep{yuan2006model}.

The rest of the paper is organized as follows.
In Section \ref{sec:model} , we present the model, discuss the estimation procedure and introduce the test statistic.
In Section \ref{sec:theory}
we present the theoretic properties of the estimators and the test
statistic.  In Section \ref{sec:numerical}, we validate the theoretical properties
of the proposed method, and perform the benchmark comparison through
simulation studies. Finally, 
we apply the proposed method to analyze the data from the AD 
MEG study in Section \ref{sec:realdata}.

{\bf Notations:} We first introduce some general notations that will be used throughout the text.
For a matrix $\M$, let $\|\M\|_{\max}$ be the matrix maximum norm,
$\|\M\|_\infty$ be the $L_\infty$ norm and
$\|\M\|_p$ be the $L_p$ norm. For a general vector $\a$, let
$\|\a\|_\infty$ be the vector sup-norm, $\|\a\|_p$ be the vector
$l_p$-norm. Let $\e_j$ be the unit vector with 1 on its $j$th
entry. For a vector ${\bf v} = (v_1,
\ldots, v_p)\trans$, let
$\supp({\bf v})$ be the set of indices with $v_i \neq 0$ and $\|{\bf v}\|_0 =
|\supp({\bf v})|$, where $|\mU|$ stands for the cardinality of the set $\mU$.
For a vector ${\bf v}\in \mathbb{R}^p$ and a subset $S \subseteq (1,
\ldots, p)$, we use ${\bf v}_S \in \mathbb{R}^S$ to  denote the vector
obtained by restricting
${\bf v}$  on the set $S$. For a matrix $\M$, denote $\M_{ k\cdot}$
and $\M_{\cdot j}$ be
the $k$th row and $j$th column of matrix $\M$. Further, we define 
$
\|X\|_{\psi_1} \equiv \sup_{k \geq 1} k^{-1} E(|X|^k)^{1/k}, 
$
and 
$
\|X\|_{\psi_2} \equiv\sup_{k \geq 1} k^{-1/2} E(|X|^k)^{1/k}. 
$
Following \cite{fletcher1980},  for an arbitrary norm
$\|\cdot\|_A$ and its dual normal $\|\cdot\|_D$, we define $\partial
\|\x\|_A$ as the set $\{{\bf v}: \|\x\|_A = {\bf v}\trans \x, \|{\bf v}\|_D \leq
1\}$. Thus, for an arbitrary  vector
$\x = (x_1, \ldots, x_p)\trans$,
$\partial \|\x\|_1 = \{{\bf v} =
(v_1, \ldots, v_p)\trans: v_j =
{\rm sign}(x_j) \text{ if } x_j\ne0,  \text{ and } |v_j| \leq 1 \text{ if } x_j=
0  \}$, and
$\partial \|\x\|_2 = \{{\bf v} =
(v_1, \ldots, v_p)\trans: v_j = x_j/\|\x\|_2\}$. Following the
definition in \cite{loh2012}, a random vector $\V$
is sub-Gaussian or sub-exponential if for any given unit vector $\u$,
$\u\trans \V$ is sub-Gaussian or sub-exponential, respectively. 

\section{Model, Estimation, and Test Statistic}\label{sec:model}
\subsection{Problem formulation}

Consider a high-dimensional functional covariate 
$\X(s)  = \{X_1(s), \ldots, X_d(s)\}\trans\in \mathbb{R}^{d}, s
\in [0, \tau] $,
where $d$ is the number of the functional predictors. Without loss of generality, we assume $\tau = 1$, because we
can always project any compact intervals to $[0, 1]$. 
In our application, we consider $X_j(s)$ to be the PSD (in $\log_{10}$ scale) of MEG/EEG
recording from $j$th region of interest.
We consider the PSD rather than the original time series because  
the PSD analysis of the MEG/EEG data is a common practice in the neuroimaging community, 
due to its ability to provide a more stable and less noisy representation of the brain signals 
\citep{ranasinghe2022altered,wang2015power}.
Note that while we focus on the PSD in this paper,
our method can also be applied to the original time series data.

 Let $\Z = (Z_1,
\ldots, Z_q)\trans$ be a vector of baseline covariates with $q<\infty$. Let $p = d
+ q$ and $Y$ be the outcome of interest,
 such as cognitive scores or disease indicator.
We consider
the conditional distribution of $Y_i$ given $\X_i, \Z_i$  for the
independent identically distributed $\{(Y_i, \X_i, \Z_i), i = 1,
\ldots, n\}$ as 
\be
&&f\left\{Y_i,
  \ba_0\trans\Z_i + \int_{0}^1 \bb\trans(s) \X_i(s)
    ds \right\} \nonumber \\
  &=& \exp \left[\frac{Y \left\{\ba_0\trans\Z_i+ \int_{0}^1\bb\trans
  (s)\X(s) ds  \right\}- \psi\left\{\ba_0\trans\Z_i+ \int_{0}^1\bb\trans
  (s)\X(s) ds\right\} }{c(\sigma)}\right],  \label{eq:model}
\ee
where $\bb(s) = \{\beta_{1}(s),
\ldots, \beta_{d}(s)\}\trans$ represents the  functional effects, $\ba_0$ is an unknown true effect of the baseline covariates, 
$c(\sigma)$ is a scale 
parameter, and 
$f$ is a known conditional density function of $Y$ given
$\{\Z, \X(s)\}$ in the exponential family. 
Since we are interested in $\ba_0, \bb(s)$, we
rescale the loss function and assume $c(\sigma) = 1$. 

We aim to test the hypothesis that
\be \label{eq:hypotest}
\mathbf{H}_0: \C\bb_{\mM}(s)=\t(s) 
\mathrm{\ vs. \ } 
\mathbf{H}_1: \C\bb_{\mM}(s) = \t(s)+ \h_n(s), 
\ee
where $\mM \subset (1, \ldots,
d)$ with $|\mM|=m$ is the subset of indices for the functional
coefficients that we are interested in to test; $\C$ is an $r\times m$
matrix and $\t(s)$ and $\h_n(s)$ are the known functions. Here $\h_n(s)$ quantifies the effect size
under the alternative hypothesis. 
For example, if we are interested in testing the difference of the
effects from the first two  functional predictors, 
we can set $\C=[1, -1]$, $\t(s)\equiv \0$ and $\h_n(s)$ to be some
nonzero function. In summary, by varying $\C, \t, \h_n,$ and $\mM$, we can generate different linear hypotheses to test.

\subsection{Parameter estimation}

Assume $\beta_j(s) \in C^{\omega}([0, 1]), j = 1, \ldots, d$, 
is a smooth function with $\omega$th order derivatives,
we approximate $\beta_j(s)$ by a B-spline function
$\B(s)\trans \bg_{0j}$, where $\B(s)$ is the $N$ dimensional B-spline basis
function and $\bg_{0j}$ is the true spline coefficients. 
Let $\mS=\{j: \beta_j(s) \neq 0, j \in \mM^c\}$ and $ k_0 = |\mS|$,
and $\bG_0 \equiv (\bg_{0j}, j = 1, \ldots, d)$.

 Furthermore, we consider the loss function as the 
negative of the logarithm of the likelihood function 
\bse
\mL (\ba, \bG) = - n^{-1}\sumi \log\left[ f\left\{Y_i,
    \ba\trans\Z_i+ \int_{0}^1 \B(s)\trans \bG \X_i(s)
  ds \right\}\right], 
\ese
where $\bG = (\bg_{j}, j = 1, \ldots, d)$. 
Let $\psi'_i(\ba, \bG),  \psi''_i(\ba, \bG)$ be the first and
second derivatives of $\psi(t)$ evaluated at $t =  \ba\trans\Z_i +\int_{0}^1
\B(s)\trans \bG \X_i(s)ds$. Slightly abusing the notations, we define
$\psi_i\{\ba, \bb(\cdot)\}$,  $\psi'_i\{\ba, \bb(\cdot)\},
\psi''_i\{\ba, \bb(\cdot)\}$ be the  cumulant function and its first and
second derivatives of $\psi_i(t)$ evaluated at $t =  \ba\trans\Z_i +\int_{0}^1
\bb(s)\trans \X_i(s)ds$.   
In a generalized
linear model $\psi''_i(\ba, \bG) >0$ and we assume that  there is a
constant $C_{\psi}$ such that $|\psi_i''(\cdot, \cdot)| \leq C_{\psi}$
uniformly.  These conditions are often used in the generalized linear model
to ensure the estimation consistency \citep{loh2015}.

Suppose we are interested in testing whether $\C 
\bb_\mM(s) =\t(s)$ or not, 
when estimating $\ba, \bG$, we minimize the regularized loss
function
\be\label{eq:lapHT}
\wh{\ba}, \wh{\bG}=\argmin_{\ba, \bG}\left\{\mL
(\ba,  \bG) + \sum_{j \in \mM^c} \rho_{\lambda}(\|N^{-1/2}\bg_j\|_2)\right\},
 \text{ such that } \|\ba\|_1 + \sum_{j=1}^d \|N^{-1/2}\bg_j\|_2 \leq R, 
 \ee
 where $\rho_{\lambda}$ is some regularized function and $\lambda$ is
the tuning parameter, 
and  $R$ is a constant that greater than $2\|\ba_0\|_1 +
2\sum_{j=1}^d \|N^{-1/2}\bg_{0j}\|_2 $, 
$\wh{\ba}$ and $\wh{\bG}=(\wh{\bg}_j, j = 1, \ldots, d)$  are the estimators of $\ba$ and $\bG$, 
respectively. The additional constraints $\|\ba_0\|_1 +
2\sum_{j=1}^d \|N^{-1/2}\bg_{0j}\|_2 $ are only used to facilitate the
theoretical derivation to control the magnitude of the estimator and
have little empirical effect.  Note that in (\ref{eq:lapHT}), we do
not impose the penalization on $\beta_j(s)$ for $j \in \mM$, and hence
do not shrink the estimators corresponding to 
the functional coefficients involved in the hypothesis.
Since we do not shrink the estimators, the procedure can effectively test the local hypothesis.

\subsection{Computational algorithm}
For notation simplicity,
we define a generic $q + d N$ dimensional vector 
$\btheta \equiv (\btheta_1\trans, \btheta_{2j}\trans, j =
1, \ldots, d)\trans \equiv \{\ba\trans,
N^{-1/2}\vec(\bG)\trans\}\trans$,  $\btheta_0 \equiv (\btheta_{01}\trans, \btheta_{02j}\trans, j =
1, \ldots, d)\trans \equiv \{\ba_0\trans,
N^{-1/2}\vec(\bG_0)\trans\}\trans$ and $\wh\btheta =(\wh\btheta_1\trans, \wh\btheta_{2j}\trans, j =
1, \ldots, d)\trans \equiv \{\wh\ba\trans,
N^{-1/2}\vec(\wh\bG)\trans\}\trans$. Without loss
of generality, we assume the last $(p - q- m - k_0) N$ element of
$\btheta_0$ are zeros.
We define $\btheta_{\mF} = \{\btheta_1\trans,
\btheta_{2j}\trans, j \in \mF \}$ where $\mF$ is a subset of $\{1, \ldots, d\}$
and define 
$\btheta_{0\mF}$ and $\wh{\btheta}_{\mF}$ in the similar way 
and denote $\mL(\btheta) \equiv \mL(\ba, \bG)$.  

We utilize the
contractive Peaceman-Rachford splitting method (CPRSM) \citep{he2014strictly} to solve for $\ba$ and $\bG$ in (\ref{eq:lapHT}).
More specifically, let $\D = [\0_{dN, q}, \I_{dN\times dN}
]$ and  let $f_1(\btheta) = \mL
(\btheta)$, where $\btheta = (\btheta_1\trans, \btheta_{2j}\trans, j =
1, \ldots, d)$,  and $f_2(\bG) =
\sum_{j = 1}^{T-1}\rho_{\lambda}(\|N^{-1/2}\bg_{j}\|_2)$. 
We estimate $\ba$ and $\bG$ by minimizing
\be\label{eq:PRSM}
f_1(\btheta)   + f_2(\bG) 
\text{ subject to } \D \btheta -
N^{-1/2}\vec(\bG)= \0 \text{ and
$\|\btheta_1\|_1  +\sum_{j = 1}^d \|\btheta_{2j}\|_2 \leq
R$ }.
\ee
At the $k$th iteration,  CPRSM updates $\btheta^{(k)}$  as
\be
\btheta^{k+1/2} 
&=& \argmin_{\btheta} f_1(\btheta) - \brho^{(k)\rm T} \{\D\btheta-N^{-1/2} \vec(\bG ^{(k)})\}  \nonumber \\
&& + \frac{\beta}{2} \|\D\btheta- N^{-1/2} \vec(\bG ^{(k)})\|_2^2,   \label{eq:updatetheta}\\
\wc\btheta_1, \|\wc\btheta_{2j}\|_2
&=& \mathrm{Proj}_{\left\{\btheta: \|\btheta_1\|_1  + \sum_{j =1}^d\|\btheta_{2j}\|_2 \leq R \right\}}([\btheta_1^{(k+1/2)\rm T},  \|\btheta_{2j}^{(k+1/2)}\|_2
    , j = 1, \ldots, d]\trans), \label{eq:proj}\\
\btheta^{(k+1)} 
&=& (\wc\btheta_{1}\trans,
    \btheta_{2j}^{ (k+1/2)\rm T}\|\wc{\btheta}_{2j}\|_2
    /\|\btheta_{2j}^{(k+1/2)}\|_2, j = 1, \ldots, d)\trans, \label{eq:updatetheta1}\\\
\brho^{k+1/2} 
&=& \brho^{(k)} - \alpha \beta
    \{\D\btheta^{(k + 1)}-N^{-1/2} \vec(\bG ^{(k)})\}, \label{eq:updaterhohalf}\\
\bG^{(k+1)} 
&=& \argmin_{\bG}  f_2(\bG)
                         - \brho^{(k + 1/2)\rm T}\{\D\btheta^{(k +
                           1)}-N^{-1/2} \vec(\bG)\}  \nonumber\\
        &&+ \frac{\beta}{2}
                         \|\D\btheta^{(k + 1)}-
                       N^{-1/2} \vec(\bG)\|_2^2, \label{eq:updategam}\\
\brho^{(k+1)} 
&=& \brho^{(k+1/2)} -\alpha\beta \{\D\btheta^{(k + 1)}-N^{-1/2} \vec(\bG^{(k + 1)})\}, \label{eq:updaterho}
\ee
where 
$(\alpha,\beta)$ are the relaxation and penalty parameters in CPRSM and  
$\mathrm{Proj}_C(\cdot)$ projects a vector to the feasible set $C$.
When $\mL(\btheta)$ is the logarithm of a Gaussian likelihood,
(\ref{eq:updatetheta}) reduces to a quadratic minimization 
which can be solved analytically.
When $\mL(\btheta)$ is the log-likelihood from  the other generalized
linear models, such as the logistic model,
(\ref{eq:updatetheta})  can be solved using Newton-Raphson method iteratively.
The projection step (\ref{eq:proj}) 
can be achieved by adopting the linear programming
method \citep{duchi2008}.
(\ref{eq:updategam}) is a standard penalized quadratic minimization step.
Throughout the paper, we choose $\rho_{\lambda}(\cdot)$ as the  smoothly clipped absolute deviation (SCAD) penalty 
\citep{fan2001variable}, which is an amenable penalty satisfying
Conditions (A1)--(A6) in \ref{sec:regcon} of the supplementary material
\citep{loh2015, loh2017}.
When using the SCAD penalty, the minimization in (\ref{eq:updategam})
can be addressed by first calculating the analytical solution of the
minimization without the penalty term, and then applying SCAD shrinkage to
the resulting solution. 

We summarize the CPRSM algorithm in Algorithm \ref{alg:PRSM} 
in \ref{sec:cprsm} of the supplementary material
and 
more details of the computational algorithm are also provided there.

\subsection{Test statistic}\label{sec:teststat}
We propose a test statistic to test the hypothesis (\ref{eq:hypotest}).
Define
\bse
\Q\{\ba, \bb(\cdot)\} \equiv E\left\{\psi''_i(\ba, \bb(\cdot)) \left(\left[\Z_i\trans, N^{1/2}\left\{\int_{0}^1\X_i (s) \otimes
  \B(s)  ds\right\}\trans\right]\trans\right)^{\otimes2}\right\}.
\ese
The covariance matrix of the residuals is 
\bse
\bSig\{\ba, \bb(\cdot)\} \equiv E
\left\{\left( [\psi'_i\{\ba, \bb(\cdot)\} - Y_i]
\left[\Z_i\trans, N^{1/2}\left\{\int_{0}^1\X_i (s) \otimes
  \B(s)  ds\right\}\trans\right]\trans\right)^{\otimes2}\right\}.
\ese
Further, define 
\bse
\A &=& (\C\otimes \I_{N\times N} )[\0_{mN\times q}, \I_{mN\times mN}, \0_{mN\times k_0N}],   \\
\bPsi(\bSig, \Q,\bb) &=& \Q\{\ba_0, \bb(\cdot)\}_{\mM\cup S, \mM\cup S}^{-1} \bSig_{\mM\cup
  S, \mM\cup S} \{\ba, \bb(\cdot)\}  \Q\{\ba_0, \bb(\cdot)\}_{\mM\cup
  S, \mM\cup S}^{-1}, 
\ese
where $\M_{\mM\cup S, \mM\cup S}$ is the top left  $\{q+ (m+k_0)N\}
\times \{q+ (m+k_0)N\}$  of an arbitrary matrix $\M$.

The sample versions of $\Q\{\ba, \bb(\cdot)\}$, $\bSig\{\ba, \bb(\cdot)\}$ 
and $\bPsi(\bSig, \Q,\bb)$ 
based on the estimator $\wh\btheta$
 are denoted by $\wh\Q(\wh\btheta)$, $\wh\bSig(\wh\btheta)$
and $\bPsi(\wh\bSig, \wh\Q,\wh\btheta)$, respectively.
Then the test statistic is defined as 
\be
T &=& n \left(\A\wh{\btheta}_{\mM\cup S} -N^{-1/2}\vec\left[  \left\{\int_0^1 \B(s)\B(s)\trans ds
    \right\}^{-1} \int_0^1 \B(s) \t\trans(s)ds
  \right]\right)\trans \A\bPsi(\wh{\bSig}, \wh{\Q},\wh{\btheta})^{-1}\A\trans \nonumber \\
&&\left(\A\wh{\btheta}_{\mM\cup S}-N^{-1/2}\vec\left[  \left\{\int_0^1 \B(s)\B(s)\trans ds
    \right\}^{-1} \int_0^1 \B(s) \t\trans(s)ds
   \right]\right). \label{eq:teststat}
 \ee

As we will show in the next section, 
the test statistic $T$ asymptotically follows central chi-square distribution
with degree of freedom $rN$ under the null hypothesis. 
Therefore, to conduct the hypothesis test (\ref{eq:hypotest})
with significance level $\alpha$, we reject the null hypothesis if
$T> \chi^2_{1-\alpha, rN}$, 
where $\chi^2_{1-\alpha, rN}$ is the $(1-\alpha)$th quantile of 
the central chi-square distribution with degree of freedom $rN$.

\section{Theoretical properties}\label{sec:theory}
We demonstrate the statistical consistency of $\wh{\ba}$ and $\wh{\bG}$. This is followed by establishing the asymptotic normality of the resulting estimators. 
Finally, we prove that the test statistic constructed in Section \ref{sec:teststat} follows a chi-squared distribution.

\subsection{Conditions} 
Before presenting the main results, 
we firstly list a set of necessary conditions. 

\begin{enumerate}[label=(C\arabic*)]
\item \label{con:boundpsi} Assume  $c_{\psi} <\psi''(\cdot ) <C_{\psi}$ and $0 <|\psi'''(\cdot)| <\infty$.

  \item \label{con:subexp} Assume for an arbitrary unit vector $\u =
    (u_1, \ldots, u_d)\trans$,
    $\sum_{j=1}^d u_{j} \int_0^1 X^2_{ij}(s) ds $ is a sub-exponential
    distributed random variable with $\|\sum_{j=1}^d u_{j} \int_0^1
    X^2_{ij}(s) ds\|_{\psi_1} \leq A_1$. Furthermore,  $\int_0^1\X_i(s) ds$ is a
    sub-Gaussian random vector, with $\|\int_0^1 \u\trans \X_i(s)
    ds\|_{\psi_2} \leq A_2$, and   $\Z_i$ is a sub-Gaussian random
    vector with $\|\v\trans \Z_i\|_{\psi_2} \leq A_3$ for any unit
    vector $\v$. And $Y_i - \psi_i'\{\ba, \bb(\cdot)\} $ is a mean zero sub-Gaussian
    random variable. 
    \item \label{con:bspline} Define the knots $\tau_{-b + 1}=\dots= \tau_0 = 0<\tau_1<\dots<\tau_N<1=\tau_{N+1}=\dots=
\tau_{N+ b}$, where $K$ is the 
  number of interior knots and $[0, 1]$ is divided into $K+1 $
  subintervals.   Assume $\beta_j(s) \in C^{\omega}([0, 1])$, and there is $N = K + b$ dimensional
  $\bg_{0j}$,  and a $b$th order Bspline with $b >\omega$ such that
  $\sup_{s \in [0,
    1]}|\B\trans (s) \bg_{0j} - \beta_j(s)| = O_p(N^{-\omega})$, $j = 1,
  \ldots, d$. The
  smallest and  largest
  distances between two knots are $h_{\min}, h_{\max}$,
  respectively. Assume $h_{\max}/h_{\min} \leq H
  <\infty$. Furthermore, assume $Nn^{-1/(2\omega+1)}\to \infty$. 

  \item\label{con:npN} Assume $n \geq c_0\log(p)$ 
   ,  $k_0\leq c_0' \log(p)$, $N \leq c_0'\log(p)$ and $q \leq
   c_0'\log(p)$ for constants
   $c_0, c_0' >0$. 

  \item \label{con:secondorder} Define $\alpha_{\min}(\M)$ and
    $\alpha_{\max}(\M)$ to be the minimal and maximal eigenvalue of
    $\M$, respectively. Assume
    \bse
    0 <\alpha_{\min }&\leq& \alpha_{\min}\left[E \left\{\left(\left[\Z_i\trans, N^{1/2}\left\{\int_{0}^1\X_i (s) \otimes
  \B(s)
  ds\right\}\trans\right]\trans\right)^{\otimes2}\right\}\right]\\
&\leq& \alpha_{\max}\left[E \left\{\left(\left[\Z_i\trans, N^{1/2}\left\{\int_{0}^1\X_i (s) \otimes
  \B(s)
  ds\right\}\trans\right]\trans\right)^{\otimes2}\right\}\right] \leq
\alpha_{\max} <\infty. 
\ese
  \end{enumerate}

Condition \ref{con:boundpsi} and \ref{con:secondorder} guarantee the boundedness and the
invertibility of the second derivative of the noise free negative log
likelihood. Condition (C2) implies that the integration of $\X_j$
should exhibit a sub-Gaussian decay in its tail, and it also suggests
that each element in $\Z_i$ is a sub-Gaussian random variable. Similar
requirement is also assumed in\cite{loh2012, loh2015, loh2017}. Condition
\ref{con:bspline} is the basic assumption to ensure the accuracy of
B-spline approximation. 
Condition \ref{con:npN} establishes the relationship between sample
size, the number of unknown parameters, and the number of knots in the
B-spline necessary to ensure consistent estimation.  
\subsection{Consistency} \label{sec:consistency}
We first show  $\wh{\btheta}$ converges to $\btheta_0$ 
with high probability in Theorem \ref{th:1}.

\begin{Th}\label{th:1}
  Denote $\wh{\v} \equiv (\wh{\v}_1\trans,
\wh{\v}_{2j}\trans, j = 1, \ldots, d)\trans \equiv  \wh\btheta - \btheta_0 \equiv  \{\wh{\ba}\trans -\ba_0\trans,
N^{-1/2}\vec(\wh{\bG})\trans - N^{-1/2}\vec(\bG_0)\trans\}\trans, $
where $\wh{\ba}, \wh{\bG}$ are the solution for (\ref{eq:lapHT}). 
Assume Conditions (A1)--(A6) in the supplementary material and
\ref{con:boundpsi}--\ref{con:secondorder} hold and $c_{\psi}\alpha_{\min} >3\mu/2$. Furthermore,
assume $$\lambda \geq
  \max\left\{2 \tau_1 R \log\{\max(p, n)\}/n,  C_0 \sqrt{\log\{\max(p, n)\}/n} +
    C k_0^{1/2} N^{-\omega}\right\}/4, $$ where $C_0, C$ are positive
  constants defined in Lemma \ref{lem:firstinfinty},  we have there
  are constants $c_1 >0$ such that $
\|\wh{\v}\|_2  \leq  3\lambda \sqrt{q + m + k_0}/ (c_{\psi}\alpha_{\min}
-3\mu/2 ), 
$
with probability greater than $1 - 2 \exp\left\{-  c_1 n\min \left(
   c_{\psi}^2 \alpha_{\min}^2/54 ^2, 1\right)/2\right\} - 20\max(p,
n)^{-1}$. And
\bse
\|\wh{\v}_1\|_1 + \sum_{j = 1}^d \|\wh{\v}_{2j}\|_2
&\leq& 12 \lambda (q + m + k_0)/  (c_{\psi}\alpha_{\min}
-3\mu/2), 
\ese
with probability greater than $1 - 2 \exp\left\{-  c_1 n\min \left(
 c_{\psi}^2   \alpha_{\min}^2/54 ^2, 1\right)/2\right\} - 20\max(p, n)^{-1}$.
\end{Th}
When choosing $\lambda = O(\log\{\max(p, n)\}/n+ N^{-\omega})$ and $q,
m, k_0$ are finite constants,  Theorem~\ref{th:1} shows that the $L_1$
distance $\|\wh{\ba}-\ba_{0}\|_1$ and the $L_2$ distance
$\sum_{j=1}^d\|N^{-1/2}(\wh{\bg}_j)\trans - N^{-1/2}(\bg_{0j})\|_2$ is vanished in the rate of
$O(\log\{\max(p, n)\}/n+ N^{-\omega})$ with high probability.

In Theorem~\ref{th:3}, 
we show that that 
under the condition that $(q+m+k_0)\sqrt{\log\{\max(p, n)\}/n} \to 0$ 
and  $\lambda = O_p \{\left[\log\{\max(p, n)\}/n\right]^{1/4}\}$,
minimizing the objective function (\ref{eq:lapHT}) yeilds a unique solution 
which approaches true estimator $\btheta_0$ with high probability.

  \begin{Th}\label{th:3}
    Assume Conditions (A1)--(A6) in the supplementary material and
   and \ref{con:boundpsi} --
    \ref{con:secondorder} hold. Further assume that $\mu
    <c_{\psi}\alpha_{\min}/2$ and  $24 \lambda (q + m + k_0)/  (c_{\psi}\alpha_{\min}
  -3\mu/2 ) \leq R$, $$\lambda \geq
    \max\left\{2 \tau_1 R \log(p)/n,  C_0 \sqrt{\log\{\max(p, n)\}/n} +
      C k_0^{1/2} N^{-\omega}\right\}/4, $$ and  \bse
    n&\geq& \max\left[8\tau_1 (\log(p) + N + q)\{\sqrt{q} + \sqrt{m} + (2/\delta + 2)
   \sqrt{k_0}\}^2/ (c_{\psi}\alpha_{\min} /2-
   \mu),\right.\\
   &&\left.\tau_1 \{\log(p) + N +
   q\} (q + m + k_0 )   /(c_{\psi}\alpha_{\min} /2- \mu) \right].
  \ese
  Furthermore, assume $(q + m + k_0) \sqrt{\log\{\max(p, n)\}/n} \to
  0$, $k_0
    N^{-2\omega} = o[\log\{\max(p, n)\}/n]$,  and $\lambda =
    O_p \{\left[\log\{\max(p, n)\}/n\right]^{1/4}\}$. 
     Let $\E_j$ be
  a matrix so that $\E_j\trans(\wh{\btheta} - \btheta_0) = \wh{\bg}_{ j}
  - {\bG}_{0\cdot j}$, assume $$\sup_{j \in S} \sum_{k \in S} \|\E_j\trans\{{\Q}_{\mM \cup S,
                             \mM\cup S}(\btheta^*)\}^{-1} \E_k \|_2 \leq
                           c_{\infty}. $$  Then we have $\wh{\btheta}$
                           is an unique minimizer for (\ref{eq:lapHT})
                           and 
                          \bse
\wh{\btheta}_{\mM \cup S} -  \btheta_{ 0\mM \cup S}&=&- \{{\Q}_{\mM \cup S,
    \mM\cup S}\{\ba_0, \bb(\cdot)\}\}^{-1}\\
  &&\times \left\{\sumi
  \{\psi'_i\{\ba_0, \bb(s)\} - Y_i\}
  \left[\Z_i\trans, N^{1/2}\left\{\int_{0}^1\X_i (s) \otimes
    \B(s)  ds\right\}\trans\right]\trans\right\}_{\mM
  \cup S} \{1+ o_p(1)\}, 
                             \ese
                             and $\wh{\btheta}_{(\mM \cup S)^c}  = \0$
                             with probability greater than $1 -
                             \exp\{-O(n)\} - O\{\max(p, n)^{-1}\}$. 
  \end{Th}
The proof of the theorem follows the standard prime-dual construction
introduced in \cite{mj2009sharp} that is illustrated in Theorem \ref{th:2} in the supplementary material. We first show when $\mM\cup S$ is
given, 
\bse
&& \mL
(\ba,  \bG) + N^{-1/2}\sum_{j \in \mM^c} \rho_{\lambda}(\|\bg_{
  j}\|_2), \nonumber\\
&& \text{ such that } \|\ba\|_1 + N^{-1/2}\sum_{j=1}^d \|\bg_{
  j}\|_2 \leq R, \text{ and } \bg_{ j} = \0 \text{ for } j \in
(\mM\cup S)^c, 
\ese
has an unique local minimizer in the interior of the feasible set. Then we show that
all the functional stationary points of (\ref{eq:lapHT}) must have
support in $\mM\cup S$. Therefore, the local minimizers of
(\ref{eq:lapHT}) are actually the local minimizers of the above oracle
loss function, so are also unique.

\begin{Rem}\label{rem:1}
Let $\E_0$ be 
the matrix so that $\E_0\trans(\wh{\btheta} - \btheta_0) = \wh{\ba} -
\ba_0$  and $\E_j$ be 
the  matrix so that $\E_j\trans(\wh{\btheta} - \btheta_0) = N^{-1/2}\wh{\bg}_{ j}
-N^{-1/2} {\bG}_{0\cdot j}$. Theorem \ref{th:3} leads to  the asymptotic normality of the estimator
$\wh{\btheta}_{\mM\cup S} $ that 
\be\label{eq:asymnorm}
\sqrt{n} (\wh{\ba} -
\ba_0) \sim N(0, \E_0\trans \bPsi \E_0). 
\ee
Furthermore, for each $s$, 
\bse
N^{-1/2}n^{1/2}\{\B(s)\trans \wh{\bg}_j-\beta_j(s) \} =O_p[N^{-1/2}n^{1/2}\{\B(s)\trans \wh{\bg}_{j} -
\B(s)\trans {\bg}_{0j} \}].
\ese
This holds because $N^{-1/2}n^{1/2}\{\B(s)\trans \wh{\bg}_{j} -
\B(s)\trans {\bg}_{0j} \} = O_p(1)$ similar to the results in \cite{jiang2015fused} and $N^{-1/2}n^{1/2}\{\B(s)\trans {\bg}_{0j} -
\beta_{j}(s) \} = O(N^{-1/2-\omega} n^{1/2}) = o_p(1)$ by Condition \ref{con:bspline}. Therefore, 
\be\label{eq:asymnorm1}
N^{-1/2} n^{1/2}\{\B(s)\trans \wh{\bg}_j-\beta_j(s)\}\sim N\{0,
\B(s)\trans\E_j\trans \bPsi \E_j \B(s)\}. 
\ee
 We then can construct the confidence intervals (CIs) for $\ba_0$ and
 the point-wise CI for
$\beta_j(s)$ based on  (\ref{eq:asymnorm}) and (\ref{eq:asymnorm1})
\end{Rem}

We now show that the test statistic $T$ in (\ref{eq:teststat})  is 
asymptotically chi-square distributed under both null and alternative hypotheses.

Define
\bse \bomega_n &=& - \sqrt{n} \A \Q\{\ba_0, \bb(\cdot)\}_{\mM\cup S,
  \mM\cup S}^{-1}\\
&&\times \left\{n^{-1}\sumi
\{\psi'_i\{\ba_0, \bb(s)\} - Y_i\}
\left[\Z_i\trans, N^{1/2}\left\{\int_{0}^1\X_i (s) \otimes
 \B(s)  ds\right\}\trans\right]\trans\right\}_{\mM
\cup S},
\ese
and
define $\c_n = \vec\left[\sqrt{n} \bPsi^{-1/2}(\bSig,
 \Q, \bb) N^{-1/2} \left\{\int_0^1 \B(s)\B(s)\trans ds
   \right\}^{-1} \int_0^1 \B(s) \h_n\trans(s)ds\right]$. 
Let $T_0 = \bPsi^{-1/2}(\bSig, \Q, \bb) (\bomega_n  + \c_n)$.

We first construct the asymptotic distribution of $T_0$, 
Then we show that the test statistic $T$ is asymptotically equivalent to $T_0$. 

\begin{Lem}\label{lem:T0}
  Assume Conditions \ref{con:boundpsi} and \ref{con:secondorder}
  holds. Furthermore, assume $
  (rN)^{1/4}n^{-1/2}\to 0$ and 
  \bse
  && E\left[\mystrut \|\bPsi^{-1/2}(\bSig, \Q, \bb) \A \Q^{-1}_{\mM\cup S,
    \mM\cup S}\{\ba_0, \bb(\cdot)\}_{\mM\cup S,
    \mM\cup S}^{-1} \right.\\
  &&\left.\times \{\psi'_i\{\ba_0, \bb(s)\} - Y_i\}
  \left[\Z_i\trans, N^{1/2}\left\{\int_{0}^1\X_i (s) \otimes
    \B(s)  ds\right\}\trans\right]\trans_{\mM
  \cup S}\|_2^3\right] = O(1). 
  \ese
  We have $
  \lim_{n\to \infty } | \Pr(T_0\leq x) - \Pr\left\{\chi^2(rN, \c_n\trans \bPsi^{-1}(\bSig, \Q, \bb) \c_n) \leq
    x\right\}| = 0, 
$
  where $\chi^2(r, \gamma)$ is a $r$ degree of freedom non-central chi-square distribution,
  with non-centrality parameter $\gamma$. 
  \end{Lem}

 In Lemma~\ref{lem:T0}, we show that the test statistic $T_0$ is asymptotically chi-square distributed
 under both null and alternative hypotheses with different
 non-central parameters. Note that $\h_n(s) =0$ under the null
 hypothesis, and therefore the non-central
 parameter satisfies $\c_n\trans \bPsi^{-1}(\bSig, \Q, \bb) \c_n = 0$. 
  We then show that the test statistic $T$ is asymptotically equivalent to $T_0$ in Theorem~\ref{th:4}.

  \begin{Th} \label{th:4}
    Assume Conditions (A1)--(A6) in the supplementary material and \ref{con:boundpsi}
    --\ref{con:secondorder} hold. Furthermore, assume the conditions
    in Theorem \ref{th:3} hold,  we have $T- T_0 = o(rN)$. In addition, assume  $
  (rN)^{1/4}n^{-1/2}\to 0$ and 
  \bse
  && E\left[\mystrut \|\bPsi^{-1/2}(\bSig, \Q, \bb) \A \Q^{-1}_{\mM\cup S,
    \mM\cup S}\{\ba_0, \bb(\cdot)\}_{\mM\cup S,
    \mM\cup S}^{-1} \right.\\
  &&\left.\times \{\psi'_i\{\ba_0, \bb(s)\} - Y_i\}
  \left[\Z_i\trans, N^{1/2}\left\{\int_{0}^1\X_i (s) \otimes
    \B(s)  ds\right\}\trans\right]\trans_{\mM
  \cup S}\|_2^3\right] = O(1). 
  \ese
  Then $
  \lim_{n\to \infty } | \Pr(T\leq x) - \Pr\left\{\chi^2(rN, \c_n\trans \bPsi^{-1}(\bSig, \Q, \bb) \c_n) \leq
    x\right\}| = 0, $
  where $\chi^2(r, \gamma)$ is a $r$ degree of freedom non-central chi-square distribution,
  with non-centrality parameter $\gamma$.
  \end{Th}
Theorem~\ref{th:4} shows that the test statistic $T$ is asymptotically $\chi^2(rN)$ 
under the null hypothesis. Therefore, if we reject the null hypothesis
if $T > \chi^2_{1-\alpha, rN}$, we can perserve the type I error under
$\alpha$ nominal level asymptotically.

\section{Numerical studies}\label{sec:numerical}

\subsection{Simulation configuration} \label{sec:conf}

We carry out comprehensive numerical analyses to assess both the
theoretical attributes and the performance of the proposed HDFP
method. Data are generated using the model specified in
(\ref{eq:model}), with Gaussian and Bernoulli distributions
corresponding to the linear and logistic model for the continuous and
binary outcomes, respectively. For  the linear model, random errors are
drawn from either a standard normal distribution or a $t$-distribution
with three degrees of freedom, and these errors are subsequently
standardized to have a variance of one. For the logistic model, we model $Y$ given $\ba_0\trans\Z_i+ \int_{0}^1\bb\trans
  (s)\X(s) ds $ as following a Bernoulli distribution.

The $\bb(s)$ 
is constructed using  Fourier basis following \cite{xue2021hypothesis}. 
Specifically, we set 
\be \label{eq:betagen}
\beta_j(s) = c_{j}\sum_{k=1}^{50}  \eta_k \phi_k(s), 
\eta_k = I_{\{k\le 4\}}(1.2 - 0.2k) + 0.4I_{\{5\le k \le 50\}}(k-3)^{-4}, \  
\mathrm{for} \ k =1,\ldots, 50,\nonumber 
\ee
where $I_{\{\cdot\}}$ is an indicator function, 
$c_j = 0$ for $j=3, \ldots, d-1$
and $c_j$ is chosen differently for each $j$  with $j=1, 2, d$ under various settings
to control the signal-to-noise ratio 
and $\phi_k(s)$'s are the Fourier basis functions.
Specifically, $\phi_1(s) = 1$, 
$\phi_{2k}(s) = \sqrt{2}\cos\left\{ k \pi(2s-1) \right\}$
for $1 \le k \le 25$ and 
$\phi_{2k-1}(s) = \sqrt{2}\sin\left\{ (k-1) \pi(2s-1) \right\}$
for $2 \le k \le 25$.
We consider two types of functional predictors,
one is the uncorrelated functional predictors generated based on the B-spline basis 
and the other is highly-correlated functional predictors
generated with the spectral graph model (SGM) \citep{verma2022spectral}
to mimic the real PSD data in our motivating example.

The uncorrelated $\X(s)$ is generated as follows, 
\be\label{eq:uncorr}
X_j(s) = \sum_{k=1}^{10} \xi_{jk} B_k(s) + 0.5\epsilon_j(s),
\ee
where $\xi_{jk}$ is generated from $N(0, 5^2)$, 
$B_k(s)$ is the $k$th B-spline basis function and 
$\epsilon_j(s)$ is the standard Gaussian process.
Here  $\epsilon_j(s)$ and $\xi_{jk}$ are independent for different $j$.
Therefore, the correlation between different functional predictors is zero. 


The highly-correlated $\X(s)$ is generated from the SGM, 
which simulates the PSD of the human brain activity
at  the resting state through  
\cite{verma2022spectral}.
After creating the PSDs from the SGM, we centered the PSD at each ROI
and added Gaussian noise with a mean of 0 and a variance of 100 to the
PSD to mimic real data. 
The detail of generating the PSDs is presented in
\cite{verma2022spectral}. 
The resulting PSDs have strong spatially correlation
with average 
pair-wise correlation coefficient around $0.6$ empirically. 


Unless stated otherwise,  $\ba_0 = [\alpha_{00}, -1, 2]$ under both
the linear and logistic models. For the linear
model, the intercept $\alpha_{00}$ is $5$. For the logistic
model, $\alpha_{00}$ is adaptively adjusted to ensure the probability
of success remains at $0.5$. The two baseline covariates in $\Z$ are derived respectively from the standard normal distribution and a Bernoulli distribution with a probability of $0.5$.

In the implementation, 
the number of B-spline bases ($N$ and the penalty parameter $\lambda$
are optimized using a 10-fold cross-validation (CV) process aimed at
minimizing the prediction error. The number of B-spline bases is
chosen from the set $[4, 6, 8, 10, 12]$. We select  $\lambda$ from $[0, 1.5]$ with increments of $0.1$  under the linear
model and from $(0, 0.05, 0.1, 0.2, 0.3, 0.4)$ under the  logistic
model. The  parameter $a$ in the SCAD
penalty is set to be $3.7$ as recommended by \cite{fan2010sure}, and
$\alpha = 0.9$, $\beta \in [0.5, 10]$ to ensure rapid convergence
according to \cite{he2014strictly}. We set $R=2e5$ to guarantee it to be
larger than $\|\ba_0\|_1 +
2\sum_{j=1}^d \|N^{-1/2}\bg_{0j}\|_2 $. The CPRSM algorithm terminates when the norm of the difference between the parameters in two consecutive iterations is less than $5e-4$ or when the number of iterations reaches 2000. All results in this section are based on $1000$ repetitions.


\subsection{Evaluate the asymptotic property of $\ba$}

We use the asymptotic normality outlined in Remark \ref{rem:1} to
construct 95\% CIs for the coefficients of the
baseline covariates $\ba$. We validate the empirical
coverage of these theoretical CIs for the coefficients of the baseline
covariates $\ba$ and confirm their consistency. We choose $d=200$ and
vary the sample size $n$ from $100$ to $3200$. The data are derived from a linear model with standard normal random
errors and a logistic model. The uncorrelated functional predictors
are generated as specified in (\ref{eq:uncorr}). Additionally, we set
$c_j=2$ for $j=d$ and $c_j=0$ for all other values in
(\ref{eq:betagen}) to generate the functional coefficients. In this
scenario, only the last functional covariate is associated with the
outcome. Since only $\ba$ is the parameter of interest, we apply
penalties to all the functional coefficients in the estimation.

\begin{figure}[t]
\centering
\includegraphics[width=0.8\textwidth]{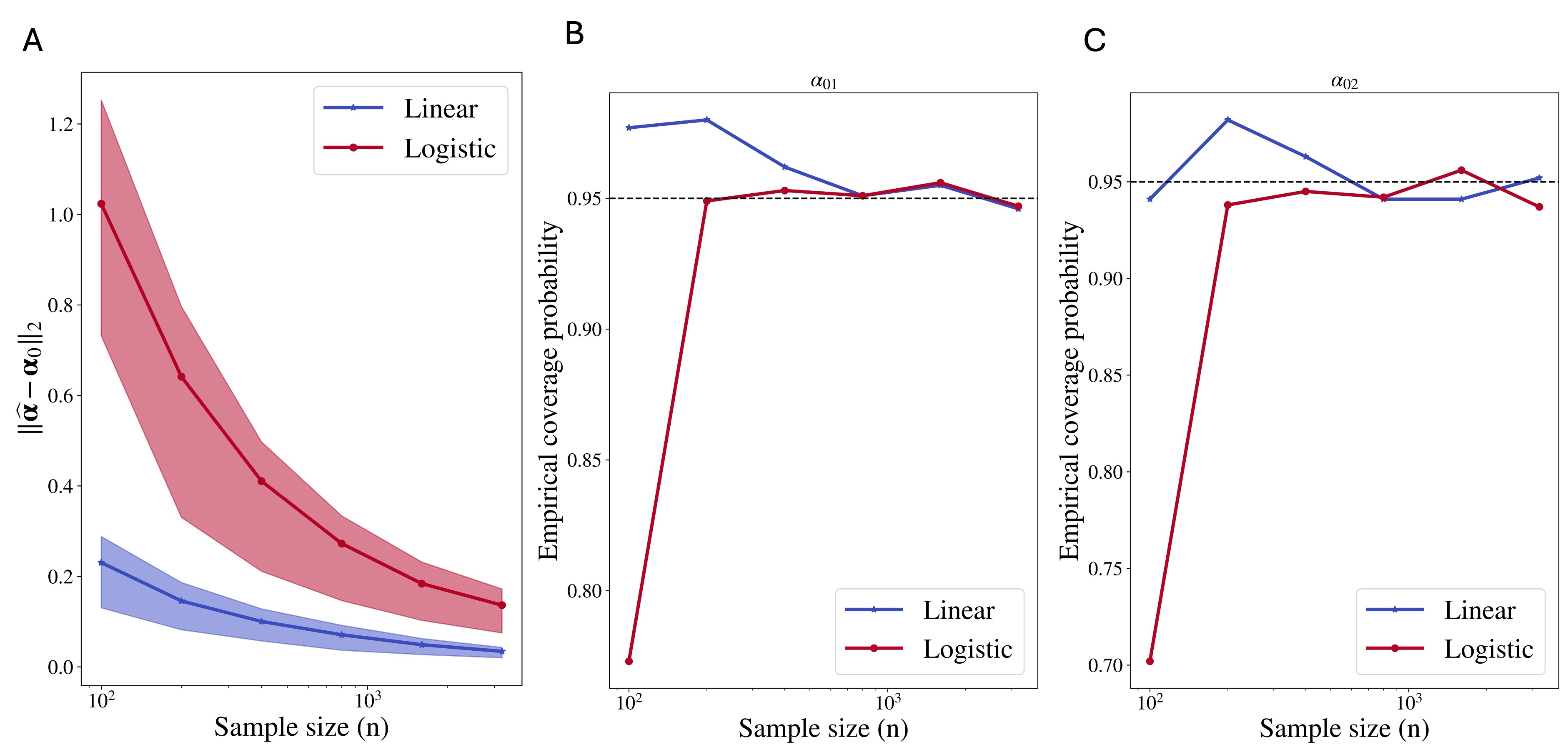}
\caption{
{\bf A:}
$\|\wh{\ba} -
  \ba_0\|_2$  for the coefficients of the baseline covariates 
 when the sample size $n$ varies under the linear and logistic
 scenarios. The shadow area represents the $25\%$ 
 and $75\%$ quantiles of the estimated error over 1000 simulations. 
{\bf B and C:}
Empirical coverage probabilities of the $95\%$ 
confidence interval of $\alpha_{01}$ and $\alpha_{02}$, respectively under linear and
logistic models. 
}
\label{fig:alp_cov}
\end{figure}

For each given sample size, we repeat the simulation $1000$ times and
the results are shown in Figure \ref{fig:alp_cov}, 
where we report both 
the average estimation error $\|\wh{\ba} -
  \ba_0\|_2$  and 
the empirical coverage of the $95\%$ CI for $\ba_0$
over  the 1000 simulations. Figure \ref{fig:alp_cov}A shows that 
the average estimation error decreases when the sample size
increases, which validates the consistency of the estimator. Figures \ref{fig:alp_cov}B and 2C illustrate that the 
CI coverage probabilities for $\alpha_{01}$ and $\alpha_{02}$ in the linear model are close to 95\% across all evaluated sample sizes. Additionally, the coverage probability approaches 95\% more closely as the sample size increases. A similar trend is observed under the logistic model, although the coverage probability only reaches 95\% when the sample size exceeds 300. This suggests that the estimation is more efficient under the linear model.


\subsection{Evaluate the hypothesis testing procedure}\label{sec:hypocomp}  
We then evaluate the performance of HDFP in
testing linear functional hypotheses.
We  compare our method 
with the bootstrap method proposed by  \cite{xue2021hypothesis}.
Note that the bootstrap method is developed under the linear model with no
baseline covariate, and therefore, our comparative study is conducted
without baseline covariates.
All the tuning paramters in the bootstap method are chosen through the
10-fold CV. 

\begin{table}[!t]
      \centering
      \caption{
        The empirical size and power  of hypothesis testing for the comparison 
        between HDFP and bootstrap methods
      under the linear model with normal noise and $t$ noise 
      based on $1000$ repetitions.
      We choose number of functional covariates $d=200$, sample size $n=100$
      and significance level $\alpha=0.05$.
      $c=0$ indicates the null hypothesis is true. Err. dist.
      standards for the distribution of the random error. 
      Type I errors are in \textbf{bold}.
      }

      \renewcommand{\arraystretch}{0.6}
      \begin{tabular}{lllcccc}
      \toprule[1.5pt]
    Type of $\X(s)$ &Err. dist. & $c$ & HDFP & Bootstrap 
       & HDFP & Bootstrap \\
      &&& \multicolumn{2}{c}{$\beta_1(s)=0$}  
       & \multicolumn{2}{c}{$\beta_1(s)=\beta_2(s)=0$}  \\
       \midrule[1.5pt]

    Uncorrelated&& \textbf{0.0} &  \textbf{0.045} & \textbf{0.047} & {\bf 0.050} & {\bf 0.049} \\
    &$N(0, 1)$
    & 0.1 &  0.115 & 0.115 & 0.167 & 0.099 \\
    && 0.2 &  0.377 & 0.361 & 0.637 & 0.388 \\
    && 0.4 &  0.862 & 0.930 & 0.975 & 0.854 
    \\ \cmidrule{2-7}
    && \textbf{0.0} &  \textbf{0.038} & \textbf{0.048} & \textbf{0.044} & \textbf{0.035} \\
    &$t(3)$
    & 0.1 &  0.134 & 0.118 & 0.154& 0.099 \\
    && 0.2 &  0.403 & 0.370 & 0.615& 0.379 \\
    && 0.4 &  0.909 & 0.933 & 0.941& 0.915 \\

      \midrule[1.5pt]

   Highly-correlated (PSD)&& \textbf{0.0} &  \textbf{0.038} & \textbf{0.036} & \textbf{0.040} & \textbf{0.024} \\
   &$N(0, 1)$
   & 0.1 &  0.114 & 0.037 & 0.150 & 0.035 \\
   && 0.2 &  0.187 & 0.048 & 0.496 & 0.063 \\
   && 0.4 &  0.624 & 0.051 & 0.935 & 0.055  \\ \cmidrule{2-7}
   && \textbf{0.0} &  \textbf{0.026} & \textbf{0.027} & \textbf{0.038}& \textbf{0.019} \\
   &$t(3)$
   & 0.1 &  0.066 & 0.030 & 0.130& 0.027 \\
   && 0.2 &  0.176 & 0.051 & 0.304& 0.020 \\
   && 0.4 &  0.628 & 0.144 & 0.944& 0.052 \\
  \bottomrule[1.5pt]
      \end{tabular}
      \label{tab:compXY}
  \end{table}

We choose $d=200$ and sample size $n=100$. We generate the
uncorrelated functional
predictor from (\ref{eq:uncorr}) and correlated predictor from the SGM as
described in Section \ref{sec:conf}. 
We choose $c_d=1$ and varying $c_1$ and $c_2$ in (\ref{eq:betagen}) under different simulation settings.
Specifically, we aim to test the null hypotheses that $\H_{0, 1}:
\beta_1(s) = 0$ and 
$\H_{0, 2}: \beta_1(s) = \beta_2(s) = 0$.
To assess the performance for testing $\H_{0, 1}$, we generate
functional parameters according to (\ref{eq:betagen}) with $c_1=c$ and
$c_2=0$, varying $c$ from $0$ to $0.4$.  To evaluate the
  performance  for testing $\H_{0, 2}$, we generate functional
  parameters from (\ref{eq:betagen}) with both $c_1=c_2=c$, varying $c$ from $0$ to $0.4$.

We show the size and power of testing $\H_{0, 1}$ and $\H_{0, 2}$ in Table
\ref{tab:compXY} under various settings. Table \ref{tab:compXY} shows that 
when the functional predictors are uncorrelated,
both HDFP and the bootstrap  method control the type I error
($c=0$) when testing $\H_{0, 1}$, while 
HDFP and the bootstrap yield similar powers under the  alternative hypotheses when $c\neq 0$.
Furthermore, when the hypothese involve  two functional parameters as those
in $\H_{0, 2}$, 
HDFP  yields greater power than the bootstrap  method does.
When the functional predictors are highly-correlated, 
both methods control the type I error under the nominal level. 
However, the bootstrap method only achieves a power of less than 0.15 in all scenarios.
The results indicate that the HDFP method outperforms the bootstrap method,
particularly when the functional predictors are highly correlated.

\subsection{Simulation evaluates the testing procedure  under the  real data setting}\label{sec:realsimu}

\begin{table}[!t]
      \centering
      \caption{
      The empirical size and power of hypothesis testing for the HDFP 
      under the linear  and logistic models based on $1000$ repetitions.
      The set of interested functional coefficients is $\beta_1(s)$.
      We choose number of functional covariates $d=68$, sample size $n=200$
      and significance level $\alpha=0.05$.
      The type I errors are in \textbf{bold}.
      }
      \renewcommand{\arraystretch}{0.6}
      \begin{tabular}{ll|ccc|ccc}
      \toprule[1.5pt]
    $\H_0$ & $c$ & \multicolumn{3}{c}{$n=200$} & \multicolumn{3}{|c}{$n=500$} \\
    & & $N(0, 1)$ & $t(3)$ & Logistic & $N(0, 1)$ & $t(3)$ & Logistic \\
       \midrule[1.5pt]

    & \textbf{0.0} &  {\bf 0.054} & {\bf 0.051} & {\bf 0.046} & {\bf
                                                                0.044}
                  & {\bf 0.048} & {\bf 0.032}\\
    $\H^1_{0, 1}$
    & 0.1 &  0.649 & 0.732 & 0.149 & 0.998 & 0.999 & 0.180\\
    & 0.2 &  0.997 & 1.000 & 0.314 & 1.000 & 1.000 & 0.594 \\
    & 0.4 &  1.000 & 1.000 & 0.794 & 1.000 & 1.000 & 1.000 \\ \midrule

    & 0.0 &  0.555 & 0.645 & 0.116 & 0.997 & 0.996 & 0.197 \\
    $\H^1_{0, 2}$
    & \textbf{0.1} & { \bf 0.054} & {\bf 0.053} & {\bf 0.064} & {\bf
                                                                0.041}
                  & {\bf 0.049} & {\bf 0.038} \\
    & 0.2 &  0.640 & 0.730 & 0.103 & 0.997 & 0.999 & 0.138\\
    & 0.4 &  1.000 & 1.000 & 0.486 & 1.000 & 1.000 & 0.884\\ \midrule

    & 0.0 &  0.996 & 1.000 & 0.383 & 1.000 & 1.000 & 0.659 \\
    $\H^1_{0, 3}$
    & 0.1 &  0.573 & 0.638 & 0.155 & 0.997 & 0.996 & 0.169\\
    & \textbf{0.2} & {\bf 0.053} & {\bf 0.049} & {\bf 0.054} & {\bf
                                                               0.043}
                  & {\bf 0.043} & {\bf 0.059} \\
    & 0.4 &  0.998 & 1.000 & 0.209 & 1.000 & 1.000 & 0.449 \\
  \bottomrule[1.5pt]
      \end{tabular}
      \label{tab:realsimu1}
  \end{table}
We  assess the performance of the HDFP method in a more realistic
setting when the distributions of the functional predictors mimic those
in the real data.
Therefore, we choose $d=68$ and consider $n=200$ and $n=500$.
Furthermore, we generate a continuous and a binary baseline covariate
from the standard normal distribution 
and the Bernoulli distribution with probability $0.5$, respectively.

The baseline effect  is $\ba_0=[5, -1, 2]$ under the linear
model. Whereas, the baseline effect is 
$\ba_0=[\alpha_{00}, -1, 2]$ for logistic model where $\alpha_{00}$ is
adaptively chosen in each scenario to ensure that the postive and negative samples remain
balanced.  
The functional predictors are generated by the SGM 
\citep{verma2022spectral}, and 
the functional coefficients $\bb(s)$ are generated from the Fourier
basis as described in (\ref{eq:betagen})
with $c_d=2$, $c_j=0$ for $j\neq 1, 2, d$. 
We vary $(c_1, c_2)$ under  different simulation settings.

We consider two sets of hypothesis testings, where we  aim to test
\bse
&\H^1_{0, 1}: \beta_1(s) = 2c \sum_{k=1}^{50}  \eta_k \phi_k(s),
c=0; &\H^2_{0, 1}: \beta_1(s) - \beta_2(s) =  2c \sum_{k=1}^{50}  \eta_k \phi_k(s), c=0.   \\
&\H^1_{0, 2}: \beta_1(s) = 2c \sum_{k=1}^{50}  \eta_k \phi_k(s),
c=0.1; & \H^2_{0, 2}: \beta_1(s) - 0.8\beta_2(s) = (0.2c+0.2) \sum_{k=1}^{50}  \eta_k \phi_k(s), c=0. \\
&\H^1_{0, 3}: \beta_1(s) = 2c \sum_{k=1}^{50}  \eta_k \phi_k(s),
c=0.2; &\H^2_{0, 3}: 4\beta_1(s) - 3\beta_2(s) =  (8c + 1)\sum_{k=1}^{50}  \eta_k \phi_k(s), c=0.
\ese

\begin{table}[!t]
      \centering
      \caption{
        The empirical size and power of linear hypothesis testing for the HDFP 
      under the linear  and logistic models based on $1000$ repetitions.
     The functional parameters of interested are $\beta_1(s), \beta_2(s)$.
      We choose number of functional covariates $d=68$, sample size $n=200$
      and significance level $\alpha=0.05$.
      The value of $c$ corresponding to the true null hypothesis is in \textbf{bold}.
      }
      \renewcommand{\arraystretch}{0.6}
      \begin{tabular}{ll|ccc|ccc}
      \toprule[1.5pt]
    $\H_0$ & $c$ & \multicolumn{3}{c}{$n=200$} & \multicolumn{3}{|c}{$n=500$} \\
    & & $N(0, 1)$ & $t(3)$ & Logistic & $N(0, 1)$ & $t(3)$ & Logistic \\
       \midrule[1.5pt]

    & \textbf{0.0} & {\bf 0.054} & {\bf 0.063} & {\bf 0.041} & {\bf
                                                               0.046}
                  & {\bf 0.049} & {\bf 0.051}\\
    $\H^2_{0, 1}$
    & 0.1 &  0.248 & 0.251 & 0.066 & 0.833 & 0.811 & 0.093\\
    & 0.2 &  0.849 & 0.858 & 0.088 & 1.000 & 1.000 & 0.202 \\
    & 0.4 &  1.000 & 1.000 & 0.227 & 1.000 & 1.000 & 0.608 \\ \midrule

    & \textbf{0.0} &  {\bf 0.052} & {\bf 0.063} & {\bf 0.041} & {\bf
                                                                0.048}
                  & {\bf 0.047} & {\bf 0.051} \\
    $\H^2_{0, 2}$
    & 0.1 &  0.311 & 0.323 & 0.059 & 0.916 & 0.893 & 0.080 \\
    & 0.2 &  0.921 & 0.923 & 0.062 & 1.000 & 1.000 & 0.180\\
    & 0.4 &  1.000 & 1.000 & 0.161 & 1.000 & 1.000 & 0.607\\ \midrule

    & \textbf{0.0} &  {\bf 0.053} & {\bf 0.062} & {\bf 0.040} & {\bf
                                                                0.047}
                  & {\bf 0.049} & {\bf 0.050} \\
    $\H^2_{0, 3}$
    & 0.1 &  0.336 & 0.345 & 0.057 & 0.930 & 0.919 & 0.075\\
    & 0.2 &  0.929 & 0.934 & 0.063 & 1.000 & 1.000 & 0.178 \\
    & 0.4 &  1.000 & 1.000 & 0.149 & 1.000 & 1.000 & 0.600 \\
  \bottomrule[1.5pt]
      \end{tabular}
      \label{tab:realsimu2}
  \end{table}
We set the significance level $\alpha=0.05$ and repeat the simulation $1000$ times.
Table \ref{tab:realsimu1}
shows the results of testing hypotheses $\H^1_{0, 1}$ to $\H^1_{0,
  3}$, 
where only one parameter is involved in the test. 
HDFP effectively controls the type I errors under the linear
model. However,  it shows slightly
inflated type I error rates when testing $\H^1_{0,
  3}$ under the logistic model, which is likely a finite sample phenomenon. Furthermore, the power increases as the discrepancy between the
true value of $c$ and its value under the null hypothesis
increases. Moveover, a larger sample size also results in a higher
power. 
Table \ref{tab:realsimu2} shows the results from testing the
hypotheses $\H^2_{0, 1}$ to $\H^2_{0,
  3}$ that involve the linear combinations of two parameters.
The type I errors are well controlled when the sample size reaches
500 in all scenarios. 
Furthermore, the power increases when $c$ deviates from  $0$, and when
the sample size increases.  In summary,  the simulation studies demonstrate that HDFP
has satisfactory performance in testing linear hypotheses under both linear and logistic models.

\section{Real data analysis}\label{sec:realdata}
\subsection{Data preprocessing} \label{sec:preprocess}

We apply HDFP to analyze our AD
data, which includes $85$ AD patients and $61$ healthy subjects. 
The dataset includes the PSDs evaluated at 74 continuous frequencies of the MEG source time series at 68 ROIs,
where the detailed preprocessing procedure is presented in \ref{sec:realdatapre}
of the supplementary material. 
The outcome is the MMSE score that assesses cognitive impairment. The
lower the score, the worse the cognitive function \citep{arevalo2021mini}. 
In addition, we consider three baseline covariates, gender, age and number of years of education.

\subsection{Comparison with other methods} \label{sec:realcmp}

We apply the HDFP method to study the association between the PSDs and the
MMSE adjusted for baseline covariates
and test the null hypothesis  that $\H_0: \beta_j(s)=0$ for $j=1,
\ldots, 68$.
The significant level are set as $\alpha=0.05/68$ to account for the
multiple testing.
All implementation details are the same as in the Section
\ref{sec:conf}, except that $\lambda$ is chosen in the set of $\{0.1,
0.3, 0.5, 0.7, 1.1, 1.3\}$ through CV.

We include the bootstrap method proposed by \cite{xue2021hypothesis}
as the benchmark.  
Since the bootstrap method does not account for the baseline covariates, 
we employ a projection step to eliminate the influence of baseline covariates.
Specifically, consider $\widetilde{\Z}$  as an $n \times (p+1)$ matrix
where the first column consists of ones and the remaining columns
represent the baseline covariates. Let $\P = \I - \widetilde{\Z}
(\widetilde{\Z}\trans\widetilde{\Z})^{-1}\widetilde{\Z}\trans$ and $\Y
= (Y_i, i=1, \ldots, n)\trans$. We generate the new outcomes as $\P \Y$
and the new functional predictors as $\P \{\X_1(t), \ldots,
  \X_n(t)\}\trans$. Then we implement the bootstrap method considering 
the new outcomes and the new functional predictors. Here we remove the
covariate effect because $\P
\widetilde{\Z} =\0$.  


\begin{figure}[!t]
  \centering 
  \includegraphics[scale=0.12]{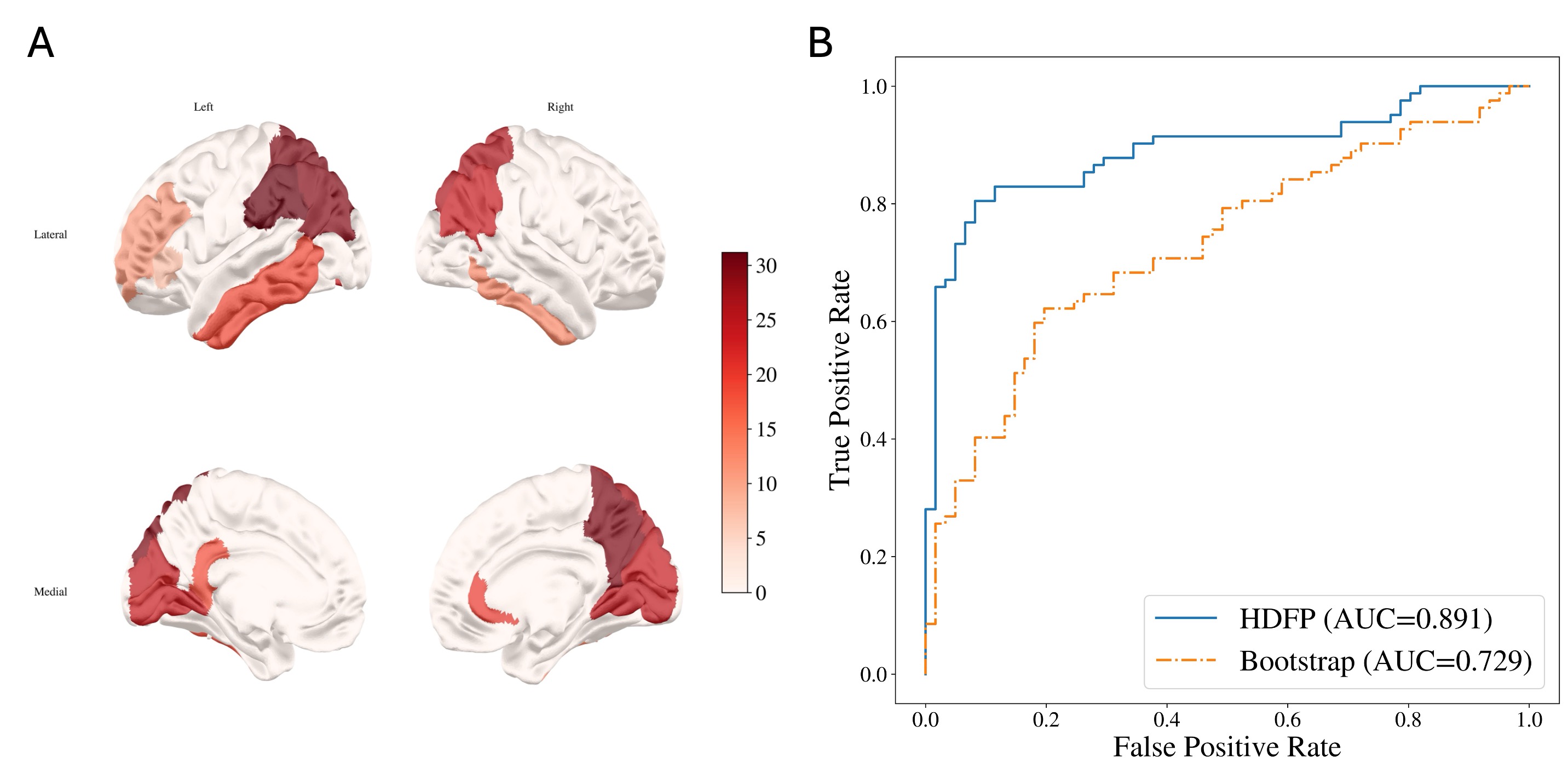}
  \caption{
    \textbf{A:} Negative log p-values from the significant ROIs selected via HDFP. 
    Only the significant regions with p-value less
      than 0.05/68 are shown.
    \textbf{B:} The receiver operating characteristic curves and corresponding AUCs 
    for HDFP and the bootstrap method in distinguishing healthy subjects with AD patients.
  }
  \label{fig:realcmp}
\end{figure}

We demonstrate significant ROIs selected via HDFP
with the p-values less than $0.05/68$ in Figure \ref{fig:realcmp}A.
We do not present the results from the bootstrap
method as no significant ROIs are  identified
after applying the Bonferroni correction. 
Overall, the significant ROIs identified by HDFP 
exhibit high symmetry between the left and right hemispheres, 
which is intuitive and consistent with the existing literature \citep{jagust2018imaging}.
Most of the significant ROIs are located 
in the inferior and posterior temporal cortices 
and the posterior parietal-occipital cortices, 
which reflects the similar regional distribution of 
the glucose hypometabolism and the high tau accumulation in the 
AD brain \citep{jagust2018imaging,jin2023}.
%
%

We further analyze the accuracies of HDFP and the bootstrap method in
classifying AD and healthy controls based on their chosen significant
ROIs that associated with the MMSE score.
In evaluating HDFP, we construct the logistic model between the disease indicator
and the PSDs from the significant ROIs, adjusting for the baseline
covariates.  We minimize (\ref{eq:lapHT})
for the parameter estimation and apply penalties to all the functional
predictors. In evaluating the bootstrap method, given that it fails to
identify any significant PSDs,   we proceed with a standard logistic regression between the disease indicator and the baseline covariates.
 We compare the area under the receiver operating
characteristic (ROC) curve (AUC) of both methods. Figure
\ref{fig:realcmp} illustrates that HDFP achieves better
classification accuracy, indicating that the PSDs from selected ROIs
indeed offer valuable information for distinguishing AD from
control samples.

\begin{figure}
  \centering 
  \includegraphics[scale=0.7]{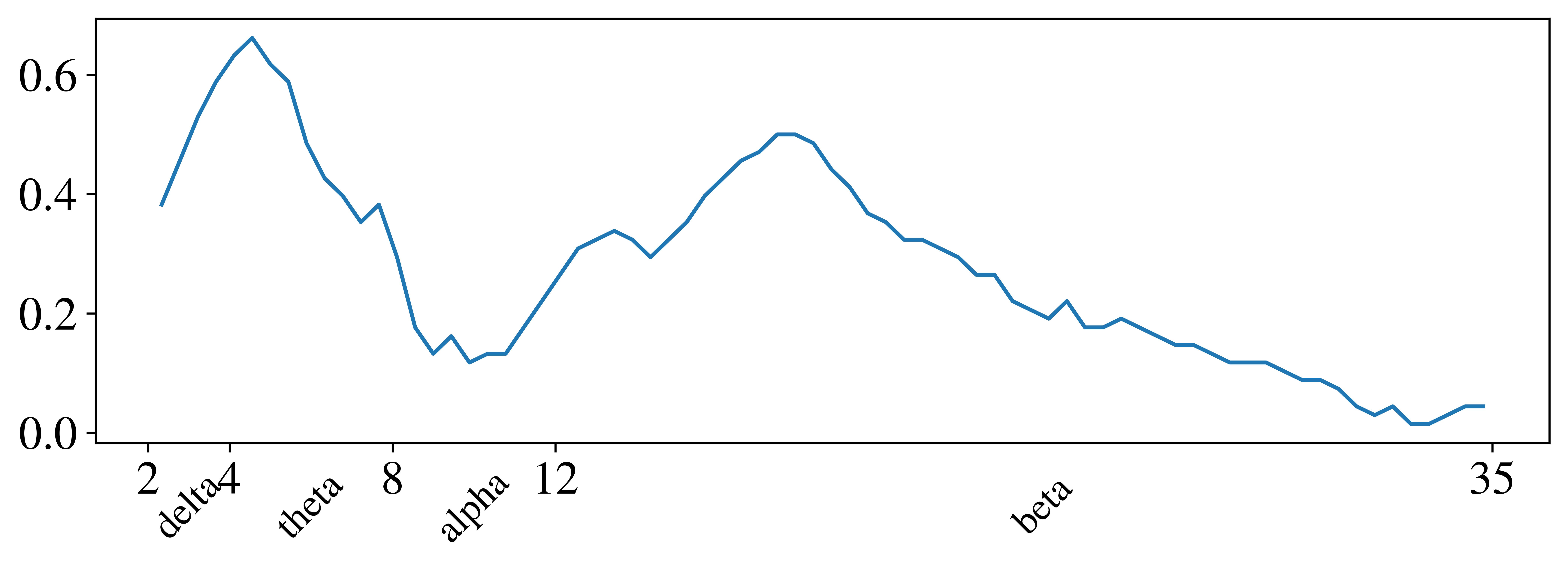}
  \caption{
  The percentage of the significant ROIs based on the point-wise $100(1-0.05/68)\%$ confidence intervals
  for each frequency point 
  among the $68$ ROIs via HDFP method. 
  }
  \label{fig:realinf}
\end{figure}

We further conduct subsequent analyses on the HDFP function
estimators. We establish point-wise 
CIs at the $100(1-0.05/68)\%$ level for each functional coefficient across the 68 ROIs.
If the CI for a specific frequency point does not cover zero, the functional
effect is considered as significantly different from zero at that frequency.  We then draw the percentage of the
significant ROIs in  across frequencies in Figure \ref{fig:realinf}. 
Figure \ref{fig:realinf}  shows that the majority of significant regions are concentrated in
the theta and low-beta ($[12, 20]$ Hz) frequency bands.  This suggests
the theta and low-beta bands could  play a crucial role in cognitive function, 
consistent with the findings in \cite{ranasinghe2022altered}.

Furthermore, as shown in Figure \ref{fig:realcmp}A, the
significant regions are not completely symmetric. This raise a
question of whether the functional pattern is asymmetric in
affecting the cognitive function. 
To explore this, we further examine the hypothesis regarding whether
the effects from the PSDs derived from the
left and right hemispheres are equal. 
We categorize the significant ROIs into two groups: Group 1 consists of the ROIs that are significant in both the left and right hemispheres, while Group 2 comprises ROIs that are significant in only one hemisphere.
We denote $\mathcal{G}_1$ 
be the set of indices of the ROIs in Group 1
and $\mathcal{G}_2$ be the set of indices of the ROIs in Group 2.
If $\beta_j(s)$ is the functional coefficient corresponding to the $j$th ROI
in one hemisphere,
we denote $\tilde{\beta}_j(s)$ as the functional coefficient corresponding to the $j$th ROI
in the other hemisphere.

We then test the two null hypotheses that 
$\H^r_{0, 1}: \tilde{\beta}_j(s) - \beta_j(s) = 0$ for all $j \in \mathcal{G}_1$
and 
$\H^r_{0, 2}: \tilde{\beta}_j(s) - \beta_j(s) = 0$ for all $j \in \mathcal{G}_2$.
All the implementation details are the same as in the Section~\ref{sec:realcmp}.
In $\H^r_{0, 1}$ where ROIs are significant in both hemispheres, 
we obtain test statistics $T=24.418$ with pvalue $0.418$, 
showing that no significant difference between the left and right hemispheres.
In $\H^r_{0, 2}$ where ROIs are significant in only one hemisphere, 
we obtain test statistics $T=60.248$ with pvalue $3.786 \times 10^{-4}$,
indicating a significant difference between the left and right hemispheres
in at least one ROI. This analysis suggests that there are
indeed asymmetry in the PSD effects from a subset of ROIs.

Finally, we summarize the effects from the baseline covariates on the
MMSE score in Table \ref{tab:alp}, which shows that the education
level is a strong predictor on the cognitive function,  with higher
education correlating to higher MMSE scores on average.
It indicates that an additional year of education is associated with an increase in the MMSE score, 
and thus lower probability of suffering from AD, 
which is consistent with the existing literature 
\citep{sando2008risk}.

\begin{table}[!thbp]
      \centering
      \caption{The estimated coefficients of the baseline covariates  
      and the corresponding $95\%$ confidence intervals after 
      applying the HDFP to the MEG data.
      }
      \renewcommand{\arraystretch}{0.6}
      \begin{tabular}{lcc}
      \toprule[1.5pt]
      Baseline covariate & Effect & $95\%$ confidence interval\\
        \midrule[1.5pt]
      Gender & $-0.502$ & $[-1.920, \phantom{-}0.916]$ \\ 
      Age & $\phantom{-}0.388$ & $[-0.241, \phantom{-}1.017]$ \\
      Years of education & $\phantom{-}0.417$ & $[\phantom{-}0.155, \phantom{-}0.679]$ \\
      \bottomrule[1.5pt]
      \end{tabular}
      \label{tab:alp}
    \end{table}

\section{Conclusion and discussion} \label{sec:discuss}
We introduce the novel HDFP method for testing linear hypotheses
regarding functional parameters under the generalized functional linear model. We
develop a regularization technique for estimating functional
parameters and validate the asymptotic properties of the
estimators. Moreover, we design a hypothesis testing procedure capable
of assessing local alternative hypotheses. 
We apply the proposed method to analyze an AD MEG study.  Our method
achieves higher interpretability and the prediction accuracy than the
existing method. HDFP  is generalizable to  study other functional data, such as 
fMRI data and
EEG data.  Further research along this line
is ongoing in our group.

When presenting a large number of functional predictors, one can reduce
the computational complexity by preprocessing the data using a
screening method \citep{fan2008sure,fan2001variable}. 
We have conducted preliminary studies on adopting a screening preprocessing step prior to implementing HDFP. 
These studies suggest that such preprocessing not only significantly
reduces computational complexity but also preserves the error rates in
the hypothesis testing.

The programming code for the simulation studies and the real data
analysis is included in the supplementary material.

\bibliographystyle{agsm}
\bibliography{hdfuninfer}

\makeatletter\@input{xx1.tex}\makeatother
\end{document}


\allowdisplaybreaks
\thispagestyle{empty}
\baselineskip=28pt
\vskip 5mm
\begin{center}
  {\LARGE{\bf Supplementary of ``High dimensional test for  functional covariates'' }}
\end{center}

\Appendix

\section{Preprocessing of real data}\label{sec:realdatapre}
In this section, we outline the preprocessing of real data in the MEG study 
(refer to Section~\ref{sec:realdata}).
This includes our methods for extracting the Power Spectral Density (PSD) from MEG recordings 
and procedures for outlier removal.
For information on acquiring MEG recordings under the DK atlas, 
please refer to \cite{jin2023}.

To obtain the PSD of the MEG recordings, 
we firstly filter the MEG data with a band-pass filter of $1-40$ Hz
and then calculate the PSD with multitaper method
\citep{das2020multitaper} with the \texttt{pmtm} function in MATLAB.
We uniformly sample $100$ points within the interval $[1, 40]$ Hz for each ROI.
After the PSD is obtained, we further use a convolve function in MATLAB to smooth the PSD.
We then select the PSD with $[2, 35]$ Hz as the functional predictors $\X(s)$ and 
convert the PSD to dB scale.
Note that the frequency range $[2, 35]$ Hz includes the delta, theta, alpha and beta bands, 
which are commonly investigated in the MEG data analysis \citep{mandal2018comprehensive}.
After selecting PSD within $[2, 35]$ Hz, 
the PSD functional predictors contain $74$ points for each ROI.
The PSD data in the dB scale within the $[2, 35]$ Hz from MEG recordings
should be typically greater than $0$ \citep{ranasinghe2022altered}, 
so we discard the data whose mean of PSD is less than $0$ as outliers
and remove them from the analysis.
After all the preprocessing, 
our dataset includes $61$ healthy subjects and $82$ patients with AD ($n=143$), 
with three baseline covariates and 
the PSD functional predictors includes $74$ points for each ROI.

\section{Additional notations} \label{sec:anotations}
Here we define some additional notations that are used in Appendix.
Some of them are not very conventional and depends on the context.

Let $\Q _{\mM \cup S, \mM \cup S} (\btheta)$, $\Q _{\mM \cup
  S, (\mM \cup S)^c} (\btheta)$,  $\Q _{(\mM \cup S)^c, \mM \cup S} (\btheta)$, $\Q _{(\mM \cup S)^c, (\mM \cup
  S)^c} (\btheta)$ be the top left $\{q + (m + k_0)N\}\times \{q +
(m + k_0)N\}$, bottom left $\{q + (m+ k_0)N\}  \times \{(d - m -
k_0)N\}$,  top right  $ \{(d- m - k_0)N\}\times \{q + (m + k_0)N\}$,
bottom right blocks of $\{(d- m - k_0)N\}\times \{(p -
k_0)N\}$. Moreover, we define $\wh{\Q}(\btheta) = \partial^2
  \mL(\btheta) /\partial \btheta \partial \btheta\trans$ and define $\wh{\Q} _{\mM \cup S, \mM \cup S} (\btheta)$, $\wh{\Q} _{\mM \cup
  S, (\mM \cup S)^c} (\btheta)$,  $\wh{\Q} _{(\mM \cup S)^c, \mM \cup S} (\btheta)$, $\wh{\Q} _{(\mM \cup S)^c, (\mM \cup
  S)^c} (\btheta)$ similarly.  For any vector $\a = (\a_1\trans, \a_{2j}\trans, j =
1, \ldots, d)\trans$, we define the subset of $\a$ 
as 
$\a_{\mF} \equiv (\a_1\trans, \a_{2 j}\trans, j \in \mF)\trans$ for an
arbitrary index set $\mF$, and $\a_{2\mF} =(\a_{2 j}\trans, j \in
\mF)\trans$.

%
%

\section{Algorithm and Details of CPRSM} \label{sec:cprsm}
The CPRSM algorithm is in Algorithm \ref{alg:PRSM} 
and more details of the algorithm are provided in the following.

\begin{algorithm}[!htbp]
  \caption{Contractive Peaceman-Rachford splitting method (CPRSM)}
  \label{alg:PRSM}
  \begin{algorithmic}[1] 
    \DATA Dataset $\{Y_i, \X_i(s), \Z_i\}_{i=1}^n$
    \REQUIRE 
    $\mM$: the set of indices for the functional covariates of interest;
    $\lambda$: the penalty parameter; 
    $N$: the number of B-spline basis functions;
    $\epsilon$: the stop tolerance; 
    $R$: the constraint constant; 
    $(\alpha,\beta)$: the relaxation and penalty parameters in CPRSM, $\alpha=0.9$ following \cite{he2014strictly}. 
    $(\bG^{(0)}, \brho^{(0)})$: the initial values of $\bG$ and $\brho$,
     typically $\bG^{(0)} = \0$ and $\brho^{(0)} = \0$, 
     $K$: the maximal iteration number.
    \ENSURE Estimators $\wh\ba$ and $\wh\bG$.
    \FOR{k = 0:$K$}
    \STATE Update $\btheta^{(k)}$ following (\ref{eq:updatetheta})--(\ref{eq:updatetheta1}) to 
    obtain $\btheta^{(k+1)}$.
    \STATE Update $\brho^{(k)}$ following (\ref{eq:updaterhohalf}) to obtain $\brho^{(k+1/2)}$.
    \STATE Update $\bG^{(k)}$ following (\ref{eq:updategam}) to obtain $\bG^{(k+1)}$.
    \STATE Update $\brho^{(k+1/2)}$ following (\ref{eq:updaterho}) to obtain $\brho^{(k+1)}$.
    \STATE $\Delta_{\brho}=\brho^{(k+1)}-\brho^{(k)}$, 
    $\Delta_{\btheta}=\btheta^{(k+1)}-\btheta^{(k)}$, 
    $\Delta_0=N^{1/2}\D \btheta^{(k+1)} - \vec(\bG^{(k+1)})$
    \STATE 
    $\Delta =\max(\|\Delta_{\brho}\|_2/\|\brho^{(k+1)}\|_2, \|\Delta_{\btheta}\|_2/\|\btheta^{(k+1)}\|_2,
    \|\Delta_0\|_2/\|\vec(\bG^{(k+1)})\|_2) $
    \IF {$\Delta< \epsilon$}
    \STATE BREAK
    \ENDIF
    \ENDFOR
    \STATE  $\wh\ba = \btheta_1^{(k+1)}$, $\wh\bG = \bG^{(k+1)}$.
  \end{algorithmic}
  \end{algorithm}

%
\begin{enumerate}
\item (\ref{eq:updatetheta}) is a simple quadratic minimization under the linear case 
      or using Newton-Raphson under the logistic case.
      Let 
      \bse
      g(\btheta) = f_1(\btheta) - \brho^{(k)\rm T} \{\D\btheta-N^{-1/2} \vec(\bG ^{(k)})\} +
       \frac{\beta}{2} \|\D\btheta- N^{-1/2} \vec(\bG ^{(k)})\|_2^2.
      \ese
      We have the following first and second derivatives, 
      \bse
      \frac{\partial g(\btheta)}{\partial \btheta}
      &=& 
      \frac{\partial f_1(\btheta)}{\partial \btheta} - \D^{\rm T}  \brho^{(k)}
      + \beta \left\{\D^{\rm T} \D \btheta - N^{-1/2}\D^{\rm T}\vec(\bG^{(k)})
      \right\} \\
      \frac{\partial^2 g(\btheta)}{\partial \btheta^{\rm T} \partial \btheta}
      &=&
      \frac{\partial^2 f_1(\btheta)}{\partial \btheta^{\rm T} \partial \btheta} + 
      \beta \D^{\rm T}\D.
      \ese
      Under the linear case, we can simply solve 
      \bse
      \frac{\partial g(\btheta)}{\partial \btheta} = \0.
      \ese
      Under the logistic case, we can use Newton-Raphson method with the update 
      \bse
      \btheta^{(k)(l+1)} = \btheta^{(k)(l)} - 
      \left\{\frac{\partial^2 g(\btheta^{(k)(l)})}{\partial \btheta^{\rm T} \partial \btheta}\right\}^{-1}
      \frac{\partial g(\btheta^{(k)(l)})}{\partial \btheta}.
      \ese
    
\item The projection (\ref{eq:proj}) can be achieved by adopting the linear programming method of
\citet{duchi2008}.

\item (\ref{eq:updategam}) is a quadratic minimization with SCAD penalization. 
We can write the fifth line as
\bse
\argmin_{\bG}  f_2(\bG) + g_2(\bG)
\ese
where $g_2(\bG) = - \brho^{(k+ 1/2)\rm T}\{\D\btheta^{(k +
                           1)}-N^{-1/2} \vec(\bG)\} + \frac{\beta}{2}
                         \|\D\btheta^{(k + 1)}-
                       N^{-1/2} \vec(\bG)\|_2^2$.
So
\bse
\frac{\partial g_2}{\partial N^{-1/2}\vec(\bG)} 
=  \brho^{(k+1/2)} + 
\beta \left\{
N^{-1/2}\vec(\bG) -  \D \btheta^{(k +1)}
\right\} = \0.
\ese
So 
\bse
N^{-1/2}\vec(\bG^{(k+1/2)}) = \D \btheta^{(k+1)} - \brho^{(k+1/2)}/\beta.
\ese
Then it is equevalent to solve 
\bse
\frac{\beta }{2}
\left\|
  N^{-1/2}\vec(\bG^{(k+1/2)}) - N^{-1/2}\vec(\bG)
\right\|^2_2 + 
\sum_{j = 1}^{T-1}\rho_{\lambda}(\|N^{-1/2}\bg_{j}\|_2)
\ese
We find $N^{-1/2}\bg_{j}^{(k+1)}$ as 
the projection of  $N^{-1/2}\bg_{j}^{(k+1/2)}$ to the ball 
\bse
 \mathbb{B}_2 \{(\|N^{-1/2}\bg_{j}^{(k+1/2)}\|_2
- \lambda/\beta)_+\}&\text{ if }&
\|N^{-1/2}\bg_{j}^{(k+1/2)}\|_2 \leq \lambda +\lambda/\beta\\
\mathbb{B}_2 [\{(a - 1)\beta \|N^{-1/2}\bg_{j}^{(k+1/2)}\|_2 - a
\lambda\}/( a\beta -\beta -1)]&\text{ if }& \lambda+\lambda/\beta
<\|N^{-1/2}\bg_{ j}^{(k+1/2)}\|_2 \leq a\lambda\\
\mathbb{B}_2 (\|N^{-1/2}\bg_{ j}^{(k+1/2)}\|_2)  &\text{ if }&
\|N^{-1/2}\bg_{ j}^{(k+1/2)}\|_2 > a\lambda. 
\ese
\end{enumerate}


\section{Amenable penalty} \label{sec:regcon}

We first list a set of assumptions on the univariate penalty function
$\rho_\lambda$ which are similar to those in \cite{loh2015,  loh2017}.
\begin{enumerate}[label=(A\arabic*)]
	\item The function $\rho_{\lambda}(t)$ satisfies $\rho_{\lambda}(0) =0$
	and is symmetric around zero. \label{it:a1}
	\item On the nonnegative real line $t\ge 0$, the function $\rho_{\lambda}(t)$ is
	nondecreasing. Furthermore, $\rho_{\lambda}(t)$ is subadditive, i.e.
	$\rho_{\lambda}(t_1 + t_2)\leq \rho_{\lambda}(t_1) +
	\rho_{\lambda}(t_2)$ for all $t_1, t_2 \geq 0$. \label{it:a2}

	\item For $t>0$, the function $\rho_{\lambda}(t)/t$ is non-increasing
	in t.  \label{it:a3}
	\item The function $\rho_{\lambda}(t)$ is differentiable at all $t\neq0$
	and sub-differentiable at $t=0$, with $\lim_{t \to 0+}
	\rho'_{\lambda}(t) = \lambda $, where $\rho'(t)$ denotes the derivative of
	$\rho(t)$. Together with the symmetric Condition in  \ref{it:a1}, this
	leads to  $\lim_{t \to 0-}  \rho'_{\lambda}(t) = -\lambda $.  
\label{it:a4}
	\item There exists $\mu >0$ so that $\rho_{\lambda}(t) + \mu t^2/2$ is
	convex. \label{it:a5} 
	\item There exists a $\gamma \in (0, +\infty)$ such that
	$\rho_{\lambda}'(t) = 0$ for all $t \geq \gamma\lambda$.\label{it:a6} 
\end{enumerate}
In Conditions \ref{it:a1}--\ref{it:a6}, the first three are general requirements for the penalty function 
\citep{zhang2012general}.
Condition \ref{it:a4} excludes the regularizers that are not differentiable at zero, 
for example, the lasso penalty.
Condition \ref{it:a5} is referred to as the week convexity condition \citep{vial1982strong,chen2014convergence}
which controls the level of nonconvexity of the penalty function.
Condition \ref{it:a6} is to remove the estimation bias via allow 
penalty to be zero if the estimator is $\gamma\lambda$ away from zero.

We also need some additional conditions on the model and the data
to facilitate the development of the theoretical properties.
We discuss these conditions here  but defer the detailed conditions to  Section~\ref{sec:regcon} in the supplementary material.
Condition~\ref{con:boundpsi} requires the boundedness of the function $\psi(t)$ in the genearlized 
linear model.
Condition~\ref{con:subexp} 
imposes the sub-Gaussian assumption on the functional covariates, baselines covariates 
as well as the error term.
Condition~\ref{con:bspline} restricts the smoothness of the functional coefficients 
$\beta_j(s)$,
and it also requires b-spline can approximate the true functional coefficients well enough.
Condition~\ref{con:npN} shows 
the relationship between the sample size $n$,
the number of B-spline basis $N$ 
and the number of functional covariates $d$ and baseline covariates $q$.
It requires that the sample size should be as least in the order of $\log(p)$ 
to ensure the consistency of the estimator.
Condition~\ref{con:secondorder} 
gaurantees  the boundedness and invertibility of the second order derivative of the log-likelihood function.

\section{Some useful lemmas} \label{sec:slemmas}

Define an auxiliary function $q_{\lambda}(t) = \lambda t -
      \rho_{\lambda}(t)$, where $q_{\lambda}(t) - \mu t^2/2$ is
      concave and everywhere differentiable in $t$ as shown in Lemma
      \ref{lem:fromlemm5support}. 
\begin{Lem}\label{lem:devore1} 
There is a constant $D_r > 0 $ such that for each spline $\sum_{k=
  1}^{N} c_k B_{k}(t)$, and for each $1\leq l \leq \infty$ 
\bse
D_r \|\c'\|_l \leq \left[\int_0^1 \left\{\sum_{k=
  1}^{N} c_k B_{k}(t)\right\}^l dt\right]^{1/l} \leq \|\c'\|_l, 
\ese
where $\c' = \{c_k \{(t_k - t_{k- b} )/ b\}^{1/l}, k = 1, \ldots,
N\}\trans$.
\end{Lem}
\noindent Proof: This is a direct consequence of Theorem 5.4.2 on page
145 in \cite{devore1993}. \qed

\begin{Lem}\label{lem:fromlemma4}
Conditions (A1)--(A4)
imply that $\rho_{\lambda}$ is $\lambda $-Lipschitz and  all
sub gradients and derivatives of $\rho_{\lambda}$ are bounded by
$\lambda $ in magnitude. 
Conditions (A1)--(A5) imply
\bse
\lambda  \|\bg_{ j}\|_2\leq \rho_\lambda(\|\bg_{j}\|_2) + \mu/2 \|\bg_{j}\|_2^2,
\ese
\end{Lem}
\noindent Proof: This lemma is a direct consequence of Lemma 4 in
\cite{loh2015}.

\begin{Lem}\label{lem:fromlemma5}
Suppose $\rho_\lambda$ satisfies Conditions (A1)--(A5). Let
$\V \in\mathbb{R}^{N\times d- m}$,  let $\mathcal{A}$ be the index set of $k$
largest elements of $\|\V_{\cdot j}\|_2, j = 1, \ldots, d-m$ in magnitude, and 
 let $\mathcal{A}^c$ be the index set of the remaining $d-m-k_0$
columns of $\V$. 
Suppose $\xi>0$ and satisfies 
\bse
\xi \sum_{j \in \mathcal{A}}\rho_{\lambda}( \|\V_{\cdot j }\|_2) -\sum_{j \in \mathcal{A}^c}
\rho_{\lambda}(\|\V_{\cdot j }\|_2)\geq 0. 
\ese
Then 
\bse
\xi \sum_{j \in \mathcal{A}}\rho_{\lambda}(\|\V_{\cdot j }\|_2) -\sum_{j \in \mathcal{A}^c}
\rho_{\lambda}(\|\V_{\cdot j }\|_2) \leq \lambda (\xi \sum_{j\in \mathcal{A}}\|\V_{\cdot j }\|_2- \sum_{j\in \mathcal{A}^c}\|\V_{\cdot j }\|_2).
\ese
Moreover, if $\bG_{0\cdot \mM^c} = (\|\bg_{0 j}\|_2, j \in \mM^c)\trans$ is
$k$-sparse, then suppose there are $\bG_{\cdot \mM^c}  = \{\bg_{j}, j \in \mM^c\}$ such that $\xi
\sum_{j \in \mathcal{M}^c}\left\{\rho_{\lambda}(\|\bg_{0 j} \|_2) -
\rho_{\lambda}(\|\bg_{j} \|_2) \right\}>0$ and $\xi \geq 1$, we have 
\bse
\sum_{j \in \mM^c }\xi \rho_{\lambda}(\|\bg_{0 j} \|_2) - \rho_{\lambda}(\|\bg_{j} \|_2) \leq \lambda \left(\xi
\sum_{j \in \mA }\|\V_{\cdot j}\|_2 - \sum_{j \in \mA^c }\|\V_{\cdot j}\|_2\right), 
\ese
where $\V= \bG_{\cdot \mM^c}- \bG_{0 \cdot \mM^c}$, $\mathcal{A}$ is the index set of the $k$
largest   $\|\V_{\cdot j}\|_2$,  and
$\mathcal{A}^c$ is the index set of the remaining $d-m-k$ columns  of $\V$.
\end{Lem}
\noindent Proof: This lemma is a direct consequence of Lemma 5 in
\cite{loh2015}. \qed

\begin{Lem}\label{lem:fromlemma4loh2017}
Suppose $\rho_{\lambda}$ is $(\mu, \gamma)$-amenable,
$\|\wh \bg_{ j}\|_2\geq 
\lambda \gamma $ for $j \in \mM \cup S$, then
$q_{\lambda}'(\|\wh \bg_{ j}\|_2) = \lambda$. 
\end{Lem}
\noindent Proof: Because $\rho_{\lambda}$ is $(\mu, \gamma)$-amenable,
$\rho'(\|\wh \bg_{ j}\|_{2}) = 0$ by Condition (A6) and (A1). Hence $q_{\lambda}'(\|\wh \bg_{j}\|_{2})  =\partial
\lambda \|\wh \bg_{j}\|_2/\partial
\|\wh \bg_{j}\|_2 = \lambda $. This proves the result. \qed

\begin{Lem}\label{lem:fromlemm5support}
Consider a $\mu$-amenable regularizer $\rho_{\lambda}$. Then \\
(a) $|\rho_{\lambda}'(t)| \leq \lambda$ for all $t\neq 0$. \\
(b) The function $q_{\lambda}(\|\t\|_2) - \mu/2 \|\t\|_2^2$ is concave
in $\|\t\|_2$ and
everywhere differentiable in $\t$, where $q_{\lambda}(\|\t\|_2) = \lambda\|\t\|_2 - \rho_{\lambda} (\|\t\|_2)$. 
\end{Lem}
\noindent Proof: This lemma is a direct consequence of Lemma 5 in
\cite{loh2017}. \qed

\begin{Lem}\label{lem:fromlemma11loh2017}
Let $\A, \B \in \mathbb{R}^{p\times p}$ be invertible. For any matrix
norm $\|\cdot\|$, we have 
\bse
\|\A^{-1} - \B^{-1}\| \leq \frac{\|\A^{-1}\|^2 \|\A - \B\|}{1
  -\|\A^{-1}\| \|\A - \B\| }. 
\ese
In particular, if $\|\A^{-1}\|\|\A - \B\| \leq 1/2$, then $\|\A^{-1} -
\B^{-1}\| = O(\|\A^{-1}\|^2\|\A - \B\|)$. 
\end{Lem}
\noindent Proof: This lemma is  Lemma 11 in
\cite{loh2017}. \qed

\begin{Lem}\label{lem:frombentkus}
Suppose $\X_1, \ldots, \X_n$ are independent $m$ dimensional random
vector which satisfies, $E(X_i)= 0$ and $\sumi\cov(\X_i) = \I_m$. Let
$\Z$ be a m-dimensional standard  multivariate normal vector, then 
\bse
\sup_{\mathcal C} |\Pr(\sumi \X_i \in {\mathcal C}) - \Pr(\Z\in
\mathcal{C})|= O(m^{1/4}\sumi E\|\X_i\|_2^3). 
\ese
\end{Lem}
\noindent Proof: The lemma  follows Theorem 1.1 in
\cite{bentkus2005}.

\section{Main proofs} \label{sec:sproofs}

Write $\btheta =
\{\ba\trans, N^{-1/2}\vec(\bG)\trans\}\trans$. For notation simplicity, we
denote $\mL(\btheta) \equiv \mL(\ba, \bG)$.  The derivative of $\mL(\btheta)$ with respect to $\btheta$ is 
\bse
\frac{\partial \mL(\btheta) }{ \partial \btheta}=  n^{-1}\sumi
\{\psi'_i(\ba, \bG) - Y_i\}
\left[\Z_i\trans, N^{1/2}\left\{\int_{0}^1\X_i (s) \otimes
  \B(s)  ds\right\}\trans\right]\trans
\ese
and the second derivative of $\mL(\ba, \bG)$ with respect to $\btheta$ is 
\bse
\frac{\partial^2 \mL(\btheta) }{\partial \btheta \partial \btheta\trans}
&=&  n^{-1}\sumi \psi''_i(\ba, \bG) \left(\left[\Z_i\trans, N^{1/2}\left\{\int_{0}^1\X_i (s) \otimes
  \B(s)  ds\right\}\trans\right]\trans\right)^{\otimes2}.  
\ese

Denote $\wt{\btheta} = \{\wt{\ba}\trans,
\vec(\wt{\bG})\trans\}\trans$
be a stationary points for (\ref{eq:lapHT}),
respectively, which satisfies the first order condition that
\be
\partial \mL(\wt{\btheta})/\partial \btheta\trans (\btheta -
\wt{\btheta}) + N^{-1/2}\sum_{j \in \mM^c } \partial
\rho_{\lambda}(\|\wt{\bg}_j\|_2)/\partial \wt{\bg}_{j}\trans
({\bg}_{ j} - \wt{\bg}_{ j} ) \geq 0\label{eq:firstordercond}
\ee

\begin{Lem}\label{lem:subGaussian}
  Assume Conditions \ref{con:subexp} and \ref{con:bspline} hold, we
  have for any given unit vector $\v = (v_{jk}, j = 1, \ldots, d, k = 1,
\ldots, N)$ with $\|\v\|_2 = 1$, $  N^{1/2}  \sum_{j = 1}^d \sum_{k = 1}^N
  v_{jk} \int_0^1 X_{ij}(s) B_{k}(s) ds$ is a sub-Gaussian
  random variable. Therefore,
  \bse \left[\Z_i\trans, N^{1/2} \left\{\int_{0}^1\X_i (s) \otimes
      \B(s)  ds\right\}\trans\right]\trans
  \ese
  is a sub-Gaussian random vector. 
\end{Lem}
\noindent Proof:
For a given unit vector $\v = (v_{jk}, j = 1, \ldots, d, k = 1,
\ldots, N)$ with $\|\v\|_2 = 1$. Denote $\v_{j\cdot} = (v_{j1},
\ldots, v_{jN})\trans$. 
\bse
&& \left\{\sum_{j = 1}^d \sum_{k = 1}^N v_{jk} \int_0^1 X_{ij}(s) N^{1/2}B_{k}(s) ds\right\}^l
\\
&\leq& \left[\sum_{j = 1}^d \sum_{k = 1}^N \left\{v_{jk}^2  \int_0^1
    X^2_{ij}(s)ds \int_0^1 N B^2_{k}(s) ds\right\}^{1/2}\right]^l\\
&\leq& \left[\sum_{j = 1}^d \sum_{k = 1}^N \left\{v_{jk}^2  \int_0^1
    X^2_{ij}(s)ds \int_0^1 N B_{k}(s) ds\right\}^{1/2}\right]^l\\
&\leq&  \left[\sum_{j = 1}^d \sum_{k = 1}^N \left\{v_{jk}^2  \int_0^1
    X^2_{ij}(s)ds \int_0^1 N  B_{k}(s) ds\right\}\right]^{l/2}\\
&=&  \left[\sum_{j = 1}^d \int_0^1 \sum_{k = 1}^N \left\{v_{jk}^2  \int_0^1
    X^2_{ij}(s)ds  N B_{k}(s) \right\} ds \right]^{l/2}\\
&\leq& \left[\sum_{j = 1}^d \sum_{k=1}^N v_{jk}^2 \int_0^1
  X^2_{ij}(s)ds N  (\tau_k - \tau_{k-b})/b\right]^{l/2}\\
&\leq& \left[\sum_{j = 1}^d \|\v_{j\cdot}\|_2 \int_0^1
    X^2_{ij}(s)ds \sum_{k=1}^N \frac{v_{jk}^2}{\|\v_{j\cdot}\|_2}
    H \right]^{l/2}\\
  &\leq&  \left[ H\sum_{j = 1}^d \|\v_{j\cdot}\|_2 \int_0^1
    X^2_{ij}(s)ds\right]^{l/2}, 
  \ese
  where the second inequality holds because $B_k(s) \in [0, 1]$, the sixth
 line holds by Lemma \ref{lem:devore1}, the last line holds because
by Condition \ref{con:bspline},  $\tau_k - \tau_{k-b} \leq h_{\max} b \leq
h_{\min} H b$ and fact that $h_{\min}N \leq 1$. Furthermore $\sum_{k=1}^d v_{jk}^2/\|\v_{j\cdot}\|_2 =
 \|\v_{j\cdot}\|_2\leq \|\v\|_2 = 1$. 
  Hence,
  \be\label{eq:expG}
  E\left\{N^{1/2}\sum_{j = 1}^d \sum_{k = 1}^N v_{jk} \int_0^1 X_{ij}(s)
    B_{k}(s) ds\right\}^l \leq E\left[ H\sum_{j = 1}^d \|\v_{j\cdot}\|_2 \int_0^1
    X^2_{ij}(s)ds\right]^{l/2}. 
  \ee
  Now because for any unit vector $\u$, $\sum_{j=1}^N u_j \int_0^1
    X^2_{ij}(s)ds$ is a sub-exponential distributed random variable by
    Condition \ref{con:subexp}, and $\{\|\v_{j\cdot}\|_2, j = 1, \ldots,
    d\}\trans$ is a unit vector, 
  by the definition of sub-exponential random variable in (5.15) in
  \citep{vershynin2010},   we have there is a constant $C<\infty$ such that 
    \bse
 \left( E\left[\sum_{j = 1}^d \|\v_{j\cdot}\|_2 \int_0^1
    X^2_{ij}(s)ds \right]^{l/2}\right)^{2/l} = C l/2
  \ese
  which implies
  \bse
  E\left[\sum_{j = 1}^d \|\v_{j\cdot}\|_2 \int_0^1
    X^2_{ij}(s)ds \right]^{l/2} = (l/2)^{l/2} C^{l/2}. 
  \ese
  Plug the above result to (\ref{eq:expG}), we have
  \bse
&& l^{-1/2} \left[ E\left\{N^{1/2}\sum_{j = 1}^d \sum_{k = 1}^N v_{jk} \int_0^1 X_{ij}(s)
    B_{k}(s) ds\right\}^l\right]^{1/l}  \\
&=& l^{-1/2}\{ (H   l/2)^{l/2}
  C^{l/2}\}^{1/l} = l^{-1/2} (l/2)^{1/2} H^{1/2}C^{1/2}= (HC/2)^{1/2} 
  \ese
 Therefore, we have $\|  N^{1/2} \sum_{j = 1}^d \sum_{k = 1}^N v_{jk} \int_0^1 X_{ij}(s)
B_{k}(s) ds\|_{\psi_2} <\infty$. By the defintion of the
    sub-Gaussian random variable as shown in (5.11) in  \citep{vershynin2010}, we have $  N^{1/2} \sum_{j = 1}^d \sum_{k = 1}^N v_{jk} \int_0^1 X_{ij}(s)
    B_{k}(s) ds$ is a sub-Gaussian random variable. Now because $\Z_i$
    is a fixed dimensional sub-Gaussian vector,  we have \bse
    \left[\Z_i\trans, N^{1/2} \left\{\int_{0}^1\X_i (s) \otimes
      \B(s)  ds\right\}\trans\right]\trans
  \ese
  is a sub-Gaussian random vector. This proves the result. 
  \qed

  \begin{Lem}\label{lem:convh}
    For any constant $s>1$, we have 
    \bse
    \mathbb{B}_1(\sqrt{s}) \cap \mathbb{B}_2({1}) \subseteq 
    {\rm cl}[{\rm conv}\{\mathbb{B}_0(s)\cap \mathbb{B}_2(3)\}], 
    \ese
    where the $l_k$ balls with radius $r$, $\mathbb{B}_k(r), k = 0, 1, 2$,
    are taken in p-dimensional space, and ${\rm cl} 
    (\cdot)$ and ${\rm conv}(\cdot)$ 
    denote the topological closure and convex hull, respectively.
    \end{Lem}
    \noindent Proof: 
    From  Lemma 11 in \cite{loh2012}, we get
    \bse
    \mathbb{B}_1(\sqrt{s}) \cap \mathbb{B}_2({1}) \subseteq 
    3{\rm cl}\{{\rm conv}\{\mathbb{B}_0(s)\cap \mathbb{B}_2(3)\}\}.
    \ese
    Here for a set $A$, $3A$ is defined as the set that
    satisfies
    $\sup_{\btheta\in 3A}<\btheta,\z>=3 \sup_{\btheta\in A}<\btheta,\z>$ for any
    $z$. Let $U$ be a subset of $\{1, \dots, p\}$ and $\z_U$ be the
    subvector of $\z$ with only the elements whose indices in $U$ retained.
    Now when $A={\rm cl}\{{\rm conv}\{\mathbb{B}_0(s)\cap
    \mathbb{B}_2(1)\}\}$, we get
    $\sup_{\btheta\in
      3A}<\btheta,\z>=3\max_{|U|=\lfloor s
      \rfloor}\sup_{\|\btheta_U\|_2\le1}<\btheta_U,\z_U>
    =\max_{|U|=\lfloor s
      \rfloor}\sup_{\|\btheta_U\|_2\le3}<\btheta_U,\z_U>
    =3\|\z_S\|_2$, 
    hence
    $3{\rm cl}\{{\rm conv}\{\mathbb{B}_0(s)\cap \mathbb{B}_2(1)\}\}
    ={\rm cl}\{{\rm conv}\{\mathbb{B}_0(s)\cap \mathbb{B}_2(3)\}\}$.
    Thus the results hold. This proves the Lemma.  
  \qed

  \begin{Lem}\label{lem:l2l1bounds}
    For a fixed matrix $\bOmega \in \mathbb{R}^{(q + Nd) \times (
        q + Nd)}$, parameter $s \geq 1$, and tolerance $\delta >0$. For $\V$ be
    a $N\times d$ dimensional matrix with $v_{ij}$ be its $i, j$th
    element. Furthermore, let $\W = (w_1, \ldots, w_q)\trans$ be a $q$
    dimensional vector.  Define
    $$\mathbb{K}(t) = \{(\W\trans, \V\trans)\trans \in \R^{q + N d},
    \|\{(\W\trans, \|\V_{\cdot j}\|_2, j =
    1, \ldots, d)\}\trans\|_0 \leq t, \|\{\W\trans, \vec(\V)\trans
    \}\trans\|_2 \leq 1\}. $$ Suppose the following
    condition  holdes
    \be
    \label{eq:l2l1boundscon}|\{\W\trans, \vec (\V)\trans\} \bOmega
    \{\W\trans, \vec (\V)\trans\}\trans| \leq \delta, ~\forall \{\W\trans, \vec(\V)\trans\}\trans \in
    \mathbb{K}(2s). 
    \ee
    Then 
    \bse
    && |\{\W\trans, \vec (\V)\trans\} \bOmega \{\W\trans, \vec (\V)\trans 
    \}\trans| \\
    &\leq&  27 \delta\left[\|\{\W\trans, \vec (\V)\trans\}\trans\|_2^2
    + s^{-1} \left\{ \|\W\|_1 + \sum_{j = 1}^d \|\V_{\cdot j}\|_2
    \right\}^2\right], \forall \W \in \mathbb{R}^q, \V \in
    \mathbb{R}^{N\times d}. 
    \ese
    \end{Lem}
    \noindent Proof:
    Define $$\mathbb{C}(s) \equiv \left[\{\W\trans,
      \vec(\V)\trans \}\trans: \|\W\|_1+ \sum_{j=1}^d\|\V_{\cdot j}\|_2 
    \leq \sqrt{s} \| \{\W\trans, \vec(\V)\trans\}\trans\|_2\right]. $$
    First note that $\| \{\W\trans, \vec(\V)\trans\}\trans\|_2 =
    \|(\W\trans, \|\V_{\cdot j}\|_2, j = 1, \ldots, d)\trans\|_2$. Therefore,   for
    $\{\W\trans, \vec(\V)\trans\}\trans \in \mathbb{C}(s)$,
    by Lemma \ref{lem:convh}, we have $(\W\trans, \|\V_{\cdot j}\|_2, j = 1,
    \ldots, d )\trans  \in
    3{\rm cl}\{{\rm conv}\{\mathbb{B}_0(s)\cap \mathbb{B}_2(3)\}$. Consider a weighted
    linear combination of the form $\{\W\trans, \vec(\V)\trans\}\trans = \sum_k a_k
    \{\W^{k\trans}, \vec(\V^k)\trans\}\trans$, with weights
    $a_k\geq 0$ such that $\sum_k a_k = 1$ and $ \|((\W^{k\trans}, \|\V^k_{\cdot
      j}\|_2, j = 1, \ldots, d)\trans\|_0 \leq s$ and $
    \|\{\W^{k\trans}, \vec(\V^k)\trans\}\trans\|_2
    \leq 3$ for each $k$. Then we can write
    \bse
    \{\W\trans, \vec (\V)\trans\} \bOmega \left\{\W\trans, \vec (\V)\trans
    \right\}\trans &=& \sum_k a_k \left\{\W^{k\trans}, \vec
      (\V^k)\trans\right\} \bOmega \sum_k a_k \left\{\W^{k\trans}, \vec (\V^k)\trans
    \right\}\trans \\
    &=&\sum_{k, l} a_k a_l \left\{\W^{k\trans}, \vec (\V^k)\trans \right\}
    \bOmega \left\{\W^{l\trans},  \vec (\V^l)\trans
    \right\}\trans. 
    \ese
    Now because $ 1/3 \left\{\W^{k\trans} , \vec (\V^k)\trans\right\}
    \trans$, $ 1/3 \left\{\W^{l\trans} , \vec (\V^l)\trans\right\}
    \trans$ and $$1/6 \left[\left\{\W^{k\trans} , \vec (\V^k)\trans\right\}+ 
      \left\{\W^{l\trans}, \vec (\V^l)\trans\right\} \right] $$ are in
    $\mathbb{K}(2s)$, follow (\ref{eq:l2l1boundscon}),  we can write
    \bse
    &&\bigg| \left\{\W^{k\trans} , \vec (\V^k)\trans\right\} \bOmega
    \left\{ \W^{l\trans},  \vec (\V^l)\trans
    \right\}\trans\bigg|\\
    &=& \frac{1}{2}\bigg| \left[\left\{\W^{k\trans}, \vec (\V^k)\trans\right\}+ 
      \left\{\W^{l\trans}, \vec (\V^l)\trans\right\} \right] \bOmega
    \left[\left\{\W^{k\trans} , \vec (\V^k)\trans\right\}+ 
      \left\{\W^{l\trans}, \vec (\V^l)\trans\right\} \right]\trans\\
    && - \left\{\W^{k\trans}, \vec (\V^k)\trans\right\} \bOmega
    \left\{\W^{k\trans},  \vec (\V^k)\trans
    \right\}\trans - \left\{\W^{l\trans}, \vec (\V^l)\trans\right\}
    \bOmega \left\{ \W^{l\trans}, \vec (\V^l)\trans
    \right\}\trans\bigg|\\
    &\leq&\frac{1}{2}(36\delta + 9\delta + 9\delta) = 27 \delta, 
    \ese
    for all $k, l$. Therefore, $|\{\W\trans, \vec (\V)\trans\} \bOmega
    \left\{\W\trans, \vec (\V)\trans
    \right\}\trans | \leq \sum_{k, l} a_k a_l (27 \delta) \leq  27\delta
    \|\a\|_2^2 = 27\delta$. Therefore,
    \be\label{eq:C1}
    \{\W\trans, \vec (\V)\trans\} \bOmega \left\{\W\trans, \vec (\V)\trans
    \right\}\trans  \leq 27 \delta,  \forall \{\W\trans, \vec (\V)\trans
    \}\trans \in \mathbb{C}(s). 
    \ee
    
    On the other hand for $\{\W\trans, \vec(\V)\trans\}\trans \not\in
    \mathbb{C}(s)$, we have
    \bse
    \frac{\{\W\trans , \vec (\V)\trans\} \bOmega \left\{\W\trans, \vec (\V)\trans
    \right\}}{ \left(\|\W\|_1 + \sum_{j = 1}^d \|\V_{\cdot j}\|_2 \right)^2} \leq
    \frac{1}{s} \sup_{ \u \in \mathbb{C}(s)} |\u\trans \bOmega \u| \leq
    \frac{27\delta}{s}, 
    \ese
    where $\u \equiv \{\W_1\trans, \vec(\V_1)\trans\}\trans \equiv
    \sqrt{s}  \{\W\trans, \vec (\V)\trans\}\trans/(\|\W\|_1+ \sum_{j = 1}^d
    \|\V_{\cdot j}\|_2)$. $\u\in
    \mathbb{C}(s)$ because $ \|\W\|_1+ \sum_{j=1}^d\|\V_{\cdot j}\|_2 
    \geq \sqrt{s} \| \{\W\trans, \vec(\V)\trans\}\trans\|_2$, 
    $\|\u\|_2\leq 1$ and $ \|\W_1\|_1 + \sum_{j=1}^d\|\V_{1 \cdot j}\|_2 =
      \sqrt{s}( \|\W\|_1+ \sum_{j = 1}^d
    \|\V_{\cdot j}\|_2) /(\|\W\|_1+ \sum_{j = 1}^d
    \|\V_{\cdot j}\|_2) = \sqrt{s}$. Combine with (\ref{eq:C1}), we have
    \bse
    \{\W\trans, \vec (\V)\trans\} \bOmega \left\{\W\trans, \vec (\V)\trans 
    \right\} \leq 27 \delta\left\{\|\{\W\trans, \vec (\V)\trans\} \trans\|_2^2
    + \frac{1}{s}\left( \|\W\|_1 + \sum_{j = 1}^d
    \|\V_{\cdot j}\|_2 \right)^2 \right\}
    \ese 
    \qed

    \begin{Lem}\label{lem:subexpbound}
      Assume Conditions \ref{con:boundpsi}, \ref{con:subexp} and
      \ref{con:bspline} hold, there is a constant $c >0$ such that for any
      given vector $\v$, 
      \bse
    &&  \Pr\left[\bigg| \v\trans \sumi \psi''_i(\ba, \bG) \left(\left[\Z_i\trans, N^{1/2}\left\{\int_{0}^1\X_i (s) \otimes
              \B(s)  ds\right\}\trans\right]\trans\right)^{\otimes2} \v\right.\\
        &&\left.- \v\trans E \left\{\sumi \psi''_i(\ba, \bG) \left(\left[
                \Z_i\trans, N^{1/2}\left\{\int_{0}^1\X_i (s) \otimes
      \B(s)  ds\right\}\trans\right]\trans\right)^{\otimes2}\right\} \v \bigg|
    \geq nt\right] \leq 2 \exp\left\{-c n\min\left(t^2, t\right)\right\}. 
    \ese
    \end{Lem}
    \noindent Proof:
    By Lemma \ref{lem:subGaussian} \bse \left[\Z_i, N^{1/2} \left\{\int_{0}^1\X_i (s) \otimes
          \B(s)  ds\right\}\trans\right]\trans
      \ese
      is a sub-Gaussian random vector. Furthermore $ 0<\psi''(\ba, \bG) < \infty$
      by Condition \ref{con:boundpsi}, we have \bse \psi''(\ba,
      \bG)^{1/2}\left[  \Z_i\trans,  N^{1/2} \left\{\int_{0}^1\X_i (s) \otimes
          \B(s)  ds\right\}\trans\right]\trans
      \ese is sub-Gaussian vector. Hence, follow Lemma 14 (G.1) in
      \citep{loh2012}, we have there is a constant $c>0$ such that 
      \bse
    &&  \Pr\left[\bigg|\v\trans\sumi \psi''_i(\ba, \bG) \left(\left[\Z_i\trans, N^{1/2}\left\{\int_{0}^1\X_i (s) \otimes
              \B(s)  ds\right\}\trans\right]\trans\right)^{\otimes2} \v\right.\\
        &&\left.-\v\trans E \left\{\sumi \psi''_i(\ba, \bG)
            \left(\left[\Z_i\trans, N^{1/2}\left\{\int_{0}^1\X_i (s) \otimes
      \B(s)  ds\right\}\trans\right]\trans \right)^{\otimes2}\right\} \v \bigg|
    \geq nt\right] \leq 2 \exp\left\{c n\min\left(t^2, t\right)\right\}. 
    \ese
    This proves the result. \qed

\begin{Lem}\label{lem:K2s}
  Assume Conditions \ref{con:boundpsi}, \ref{con:subexp} and
 \ref{con:bspline} hold, there is a constant $c_1 >0$ such that \bse
&& \Pr\left(\sup_{\v \in \mathbb{K}(2s)} \bigg|\sumi
\v\trans \left[\psi''_i(\ba, \bG) \left(\left[\Z_i\trans, N^{1/2}\left\{\int_{0}^1\X_i (s) \otimes
         \B(s)  ds\right\}\trans\right]\trans\right)^{\otimes2} \right.\right.\\
   &&\left.\left.- E \left\{\psi''_i(\ba, \bG)
         \left(\left[\Z_i\trans, N^{1/2}\left\{\int_{0}^1\X_i (s) \otimes
 \B(s)  ds\right\}\trans\right]\trans\right)^{\otimes2}\right\} \right] \v\bigg| > nt \right)\\
&\leq&   2 \exp\left\{-  c_1 n\min \left(t^2, t\right) + 2 s\log(9p)\right\}.
\ese
\end{Lem}
\noindent Proof:
For each subset ${\mathcal U} \subset (1, \ldots, p)$, we define
the set $S_{\mathcal U} $ as $S_{\mathcal U} = [\{\W\trans, \vec(\V)\trans\}\trans \in
{\mathbb R}^{p + Nd}, \|\{\W\trans, \vec(\V)\trans\}\trans\|_2\leq 1,
\supp\{(\W\trans, \|\V_{\cdot j}\|_2, j = 1, \ldots, d)\}\trans \subseteq  {\mathcal U}]$, and note that ${\mathbb K}(2 s) =
\cup_{ |{\mathcal U}| \leq 2s }S_{\mathcal U}$.  We define 
$\mathcal{A} = \{\u_1, \ldots, \u_m\}\subset S_{\mU}$   to be  a 1/3-cover of $S_{\mathcal
 U}$, if for every $\v \in S_{\mathcal U}$, there is some $\u_i \in
\mathcal{A} $ such that $\|\v - \u_i\|_2 \leq 1/3$. Define
$\Delta\v=\v-\u_j$ where $\u_j = \arg\min_{\u_i}\|\v -\u_i\|_2$. We have $\|\Delta\v\|_2  \leq
1/3$.
The same as those shown in Lemma 15 in \cite{loh2012}, by
\cite{ledoux2013}, we can construct $\mathcal{A}$ with $|\mathcal{A}|
< 9^{2sN + q}$. Define 
\bse
\Phi (\v_1, \v_2) &=& \v_1 \trans n^{-1}\sumi  \left[\psi''_i(\ba,
 \bG) \left(\left[\Z_i\trans,  N^{1/2}\left\{\int_{0}^1\X_i (s) \otimes
         \B(s)  ds\right\}\trans\right]\trans\right)^{\otimes2} \right.\\
   &&\left.- E \left\{\psi''_i(\ba, \bG) \left(\left[\Z_i\trans, N^{1/2}\left\{\int_{0}^1\X_i (s) \otimes
 \B(s)  ds\right\}\trans\right]\trans\right)^{\otimes2}\right\} \right]\v_2. 
\ese 
We have 
\bse
&& |\Phi (\v, \v)| \\
&= & |\Phi (\Delta \v + \u_j , \Delta \v
+ \u_j )|  \\
&\le&  \max_i  |\Phi (\u_i,
\u_i)| + \max_i  |\Phi (\Delta
\v,\u_i)  | + \max_i | \Phi (\u_i,\Delta
\v)|  +   | \Phi (\Delta
\v,\Delta
\v)  |\\
&\leq&  \max_i |\Phi (\u_i,
\u_i)| + 2 \max_i |\Phi (\Delta
\v,\u_i)  | + | \Phi (\Delta
\v,\Delta
\v)  |. 
\ese 
Hence, 
\bse
\sup_{\v \in S_{\mathcal U} } |\Phi (\v, \v)| 
\leq \max_i |\Phi (\u_i,
\u_i)| + 2 \sup_{\v \in S_{\mathcal U} } \max_i |\Phi (\Delta
\v,\u_i)  | + \sup_{\v \in S_{\mathcal U} } | \Phi (\Delta
\v,\Delta
\v)  |. 
\ese
Since $\|3\Delta\v\|_2 \leq 1$ and $\supp(3 \Delta\v)\subseteq  \mU$, 
$3\Delta\v  \in S_{\mathcal{U}}$. 
It follows that 
\bse
&&\sup_{\v \in S_{\mathcal U} }|\Phi (\v, \v)| \\
&\leq&
\max_i |\Phi (\u_i,
\u_i)| + 2/3 \sup_{\v \in S_{\mathcal U} } \max_i |\Phi (3\Delta
\v,\u_i)  | + 1/9\sup_{\v \in S_{\mathcal U} } | \Phi (3\Delta
\v,3\Delta
\v)  |\\
&\le&
\max_i |\Phi (\u_i,
\u_i)| + 2/3 \{\sup_{\v \in S_{\mathcal U} } |\Phi (3\Delta
\v, 3\Delta\v)  |\}^{1/2}\{ \max_i|\Phi (
\u_i,\u_i)  | \}^{1/2}
+ 1/9\sup_{\v \in S_{\mathcal U} } | \Phi (\v,
\v)  |\\
&\le&
\max_i |\Phi (\u_i,
\u_i)| + \sup_{\v \in S_{\mathcal U} } \{2/3 |\Phi (\v, \v)| + 1/9
|\Phi (\v, \v)|\}. 
\ese
Hence,  
$\sup_{\v \in S_{\mathcal U} }|\Phi (\v, \v)|\leq 9/2 \max_i |\Phi (\u_i,
\u_i)|$. By Lemma \ref{lem:subexpbound} and a union bound, we have 
\be\label{eq:unionbound}
&& \Pr\left(\sup_{\v \in S_{\mathcal U} } \bigg|\sumi \v\trans\left[\psi''_i(\ba, \bG) \left(\left[\Z_i\trans, N^{1/2}\left\{\int_{0}^1\X_i (s) \otimes
         \B(s)  ds\right\}\trans\right]\trans\right)^{\otimes2} \right.\right.\nonumber\\
   &&\left.\left.- E \left\{\psi''_i(\ba, \bG) \left(\left[\Z_i\trans, N^{1/2}\left\{\int_{0}^1\X_i (s) \otimes
 \B(s)  ds\right\}\trans\right]\trans\right)^{\otimes2}\right\} \right]\v\bigg| > 9/2 nt \right)\nonumber\\
&\leq&  9^{2sN + q}   2 \exp\left\{-  c n\min \left(t^2, t\right)\right\}.
\ee
Now replacing $t$ with $2/9 t$, we have there is a constant $c_1$
\bse
&& \Pr\left(\sup_{\v \in S_{\mathcal U} } \bigg|\sumi \v\trans\left[\psi''_i(\ba, \bG) \left(\left[\Z_i\trans, N^{1/2}\left\{\int_{0}^1\X_i (s) \otimes
         \B(s)  ds\right\}\trans\right]\trans\right)^{\otimes2} \right.\right.\\
   &&\left.\left.- E \left\{\psi''_i(\ba, \bG) \left(\left[\Z_i\trans, N^{1/2}\left\{\int_{0}^1\X_i (s) \otimes
 \B(s)  ds\right\}\trans\right]\trans\right)^{\otimes2}\right\}
\right] \v\bigg| >  nt \right)\\
&\leq&  9^{2sN + q}   2 \exp\left\{-  c_1 n\min \left(t^2, t\right)\right\}.
\ese
Finally, taking a union bound over the $p \choose 2s$ choices of
 $\mathcal{U}$, and noting that ${p \choose 2s} \le p^{2s} $,  we have 
\bse
&& \Pr\left(\sup_{\v \in \mathbb{K}(2s)} \bigg|\sumi \v\trans\left[\psi''_i(\ba, \bG) \left(\left[\Z_i\trans, N^{1/2}\left\{\int_{0}^1\X_i (s) \otimes
         \B(s)  ds\right\}\trans\right]\trans\right)^{\otimes2} \right.\right.\\
   &&\left.\left.- E \left\{\psi''_i(\ba, \bG) \left(\left[\Z_i\trans, N^{1/2}\left\{\int_{0}^1\X_i (s) \otimes
 \B(s)  ds\right\}\trans\right]\trans\right)^{\otimes2}\right\} \right] \v\bigg| > nt \right)\\
&\leq&   2 \exp\left\{-  c_1 n\min \left(t^2, t\right) + (2sN +
 q)\log(9) + 2s \log(p)\right\}.
\ese
This proves the result.  \qed

\begin{Lem}\label{lem:Re}
 Assume Conditions \ref{con:boundpsi}--\ref{con:secondorder} hold, $\forall \W \in \mathbb{R}^q, \V \in
\mathbb{R}^{N\times d}$, there is constant $c_1 >0$, such that  
\bse
&& \{\W\trans, \vec(\V)\trans\} n^{-1}\sumi 
 \psi''_i(\ba, \bG) \left(\left[\Z_i\trans, N^{1/2}\left\{\int_{0}^1\X_i (s) \otimes
         \B(s)
         ds\right\}\trans\right]\trans\right)^{\otimes2}\{\W\trans,
   \vec(\V)\trans\}\trans\\
&\geq&  c_{\psi}\alpha_{\min} \|\{\W\trans, \vec(\V)\trans\}\trans\|_2^2/2- \tau(n, p)  \left\{ \|\W\|_1 + \sum_{j = 1}^d \|\V_{\cdot j}\|_2
\right\}^2, 
\ese
and
\bse
&& \{\W\trans, \vec(\V)\trans\} n^{-1}\sumi 
 \psi''_i(\ba, \bG) \left(\left[\Z_i\trans, N^{1/2}\left\{\int_{0}^1\X_i (s) \otimes
         \B(s)
         ds\right\}\trans\right]\trans\right)^{\otimes2}\{\W\trans,
   \vec(\V)\trans\}\trans \\
   &\leq&  3c_{\psi}\alpha_{\max} \|\{\W\trans, \vec(\V)\trans\}\trans\|_2^2/2+  \tau(n, p)\left\{ \|\W\|_1 + \sum_{j = 1}^d \|\V_{\cdot j}\|_2
\right\}^2 
\ese
with probability greater than $1 - 2 \exp\left\{-  c_1 n\min \left(
   c_{\psi}^2\alpha_{\min}^2/54 ^2, 1\right)/2\right\}$, where $\tau(n,
p)=\tau_1\{log(p)+ N + q\}/n$, where  $\tau_1 = 5 c_{\psi}\alpha_{\min}/\{c_1\min \left(
 c_{\psi}^2\alpha_{\min}^2/54 ^2, 1\right)\}$. 
   \end{Lem}
   \noindent Proof: Let
   \bse
   \bOmega &=& n^{-1}\sumi \left[\psi''_i(\ba, \bG) \left(\left[\Z_i\trans, N^{1/2}\left\{\int_{0}^1\X_i (s) \otimes
         \B(s)  ds\right\}\trans\right]\trans\right)^{\otimes2} \right.\\
   &&\left.- E \left\{\psi''_i(\ba, \bG) \left(\left[\Z_i\trans, N^{1/2}\left\{\int_{0}^1\X_i (s) \otimes
 \B(s)  ds\right\}\trans\right]\trans\right)^{\otimes2}\right\}
\right]. 
\ese
Let $$t = c_{\psi}\alpha_{\min}\left[E \left\{\psi''_i(\ba, \bG) \left(\left[\Z_i\trans, N^{1/2}\left\{\int_{0}^1\X_i (s) \otimes
 \B(s)
 ds\right\}\trans\right]\trans\right)^{\otimes2}\right\}\right]/54
= c_{\psi}\alpha_{\min}/54, $$
by     Lemma (\ref{lem:K2s}), for $p\geq 9$,  we have 
\bse
&& \Pr\left(\sup_{\v \in S_{\mathcal U} } \bigg|\sumi
 \v\trans\left[\psi''_i(\ba, \bG) \left(\left[\Z_i\trans, N^{1/2}\left\{\int_{0}^1\X_i (s) \otimes
         \B(s)  ds\right\}\trans\right]\trans\right)^{\otimes2} \right.\right.\\
   &&\left.\left.- E \left\{\psi''_i(\ba, \bG)
         \left(\left[\Z_i\trans, N^{1/2}\left\{\int_{0}^1\X_i (s) \otimes
 \B(s)  ds\right\}\trans\right]\trans\right)^{\otimes2}\right\}
\right]\v \bigg| > n  c_{\psi}\alpha_{\min}/54 \right)\\
&\leq& \Pr\left(\sup_{\v \in S_{\mathcal U} } \bigg|\sumi
\v\trans  \left[\psi''_i(\ba, \bG) \left(\left[\Z_i\trans, N^{1/2}\left\{\int_{0}^1\X_i (s) \otimes
         \B(s)  ds\right\}\trans\right]\trans\right)^{\otimes2} \right.\right.\\
   &&\left.\left.- E \left\{\psi''_i(\ba, \bG)
         \left(\left[\Z_i\trans, N^{1/2}\left\{\int_{0}^1\X_i (s) \otimes
 \B(s)  ds\right\}\trans\right]\trans\right)^{\otimes2}\right\}
\right]\v \bigg| > n \min \left(c_{\psi} \alpha_{\min}/54, 1 \right) \right)\\
&\leq&   2 \exp\left\{-  c_1 n\min \left(c_{\psi}^2 \alpha_{\min}^2/54 ^2, 1\right)
 + (2 sN + q) \log 9 + 2s \log(p)\right\}\\
&\leq& 2 \exp\left\{-  c_1 n\min \left(c_{\psi}^2 \alpha_{\min}^2/54 ^2, 1\right)
 + 5  s (N + q + \log(p)\})\right\}. 
\ese
Let $s = c_1n\min \left( \alpha^2_{\min}/54^2, 1\right)/[10 \{N+ q +  \log(p)\}]$,
we obtain
\bse
&& \Pr\left(\sup_{\v \in S_{\mathcal U} } \bigg|\sumi
 \v\trans\left[\psi''_i(\ba, \bG) \left(\left[\Z_i\trans, N^{1/2}\left\{\int_{0}^1\X_i (s) \otimes
         \B(s)  ds\right\}\trans\right]\trans\right)^{\otimes2} \right.\right.\\
   &&\left.\left.- E \left\{\psi''_i(\ba, \bG)
         \left(\left[\Z_i\trans, N^{1/2}\left\{\int_{0}^1\X_i (s) \otimes
 \B(s)  ds\right\}\trans\right]\trans\right)^{\otimes2}\right\}
\right] \v \bigg| > n c_{\psi} \alpha_{\min}/54 \right) \\
&\leq& 2 \exp\left\{-  c_1
n \min \left( c_{\psi}^2\alpha_{\min}^2/54 ^2, 1\right)/2\right\}. 
\ese
Now by Lemma \ref{lem:l2l1bounds}, we have
\bse
&&\bigg|\{\W\trans, \vec(\V)\trans\}n^{-1}\sumi \left[
 \psi''_i(\ba, \bG) \left(\left[\Z_i\trans, N^{1/2}\left\{\int_{0}^1\X_i (s) \otimes
         \B(s)  ds\right\}\trans\right]\trans\right)^{\otimes2} \right.\\
   &&\left.- E \left\{\psi''_i(\ba, \bG)
         \left(\left[\Z_i\trans, N^{1/2}\left\{\int_{0}^1\X_i (s) \otimes
 \B(s)  ds\right\}\trans\right]\trans\right)^{\otimes2}\right\}\right]
\{\W\trans, \vec(\V)\trans\}\trans \bigg|\\
&\leq& c_{\psi}\alpha_{\min}/2 + s^{-1} \left\{ \|\W\|_1 + \sum_{j = 1}^d \|\V_{\cdot j}\|_2
\right\}^2, \forall \W \in \mathbb{R}^q, \V \in
\mathbb{R}^{N\times d}. 
\ese
This implies $\forall \W \in \mathbb{R}^q, \V \in
\mathbb{R}^{N\times d}$, 
\bse
&& \{\W\trans, \vec(\V)\trans\} n^{-1}\sumi 
 \psi''_i(\ba, \bG) \left(\left[\Z_i\trans, N^{1/2}\left\{\int_{0}^1\X_i (s) \otimes
         \B(s)
         ds\right\}\trans\right]\trans\right)^{\otimes2}\{\W\trans,
   \vec(\V)\trans\}\trans\\
&\geq&  c_{\psi}\alpha_{\min}\|\{\W\trans, \vec(\V)\trans\}\trans\|_2^2/2 - s^{-1} \left\{ \|\W\|_1 + \sum_{j = 1}^d \|\V_{\cdot j}\|_2
\right\}^2, 
\ese
and
\bse
&& \{\W\trans, \vec(\V)\trans\} n^{-1}\sumi 
 \psi''_i(\ba, \bG) \left(\left[\Z_i\trans, N^{1/2}\left\{\int_{0}^1\X_i (s) \otimes
         \B(s)
         ds\right\}\trans\right]\trans\right)^{\otimes2}\{\W\trans,
   \vec(\V)\trans\}\trans \\
   &\leq&  3 c_{\psi}\alpha_{\max}\|\{\W\trans, \vec(\V)\trans\}\trans\|_2^2/2 +  s^{-1} \left\{ \|\W\|_1 + \sum_{j = 1}^d \|\V_{\cdot j}\|_2
\right\}^2 
\ese
with probability greater than $1 - 2 \exp\left\{-  c_1 n\min \left(
  c_{\psi}^2 \alpha_{\min}^2/54 ^2, 1\right)/2 \right\}$.   Recall that we have choose $s = c_1n\min \left( c_{\psi}^2\alpha^2_{\min}/54^2, 1\right)/[10 \{N+ q +  \log(p)\}]$. Now select $\tau(n, p) = c_{\psi}\alpha_{\min}/(2s)$, we prove the
result. \qed
\begin{Lem}\label{lem:firstexpandsecond}
 Assume Conditions \ref{con:boundpsi}--\ref{con:npN} hold, there
 are constants $C>0$ such that 
By union bound over $k =
1, \ldots, q$ and $j = 1, \ldots, d$, we have
\bse
&& \max\left(\left\|n^{-1}\sumi \left[\psi'(\ba, \bG_0) -
     \psi'\{\ba,\bb(\cdot)\}\right] \Z_{i}\right\|_{\infty}, \right.\\
 &&\left.\left\|n^{-1}\sumi \left[\psi'(\ba, \bG_0) - \psi'\{\ba,\bb(\cdot)\}\right]
N^{1/2}\int_0^1 X_{ij}
(s) \B(s) ds \right\|_2, ~j = 1, \ldots, d\right) \\
&\leq& C k_0^{1/2} N^{-\omega} 
\ese
with probability $1 - 8p^{-1}$
\end{Lem}
\noindent Proof: First by Condition \ref{con:boundpsi} and
\ref{con:bspline}, we can show that there is a positive constant
$C_1$ such that 
\be\label{eq:psidiff}
&&|\sumi \left[\psi'(\ba, \bG_0) - \psi'\{\ba,\bb(\cdot)\}\right]Z_{ij}|\nonumber\\
&=& \sumi |\psi''\{\ba,\bb^*(\cdot)\}| \left\{\int_0^1
\|\bG_0 \trans \B(s)- \bb(s)\|_2^2 ds \right\}^{1/2} \left(\int_0^1
\X_{i\mS} (s)\trans \X_{i\mS} (s)ds \right)^{1/2} |Z_{ij}|\nonumber\\
&\leq& \sumi |\psi''\{\ba,\bb^*(\cdot)\}| \left\{\int_0^1
\|\bG_0 \trans \B(s)- \bb(s)\|_2^2 ds \right\}^{1/2} \left(\int_0^1
\X_{i\mS} \trans(s) \X_{i\mS} (s)ds \right)^{1/2} |Z_{ij}|\nonumber\\
&= & C_1 k_0^{1/2} 2 N^{-\omega} \sumi \left(\int_0^1
 \X_{i\mS} \trans(s) \X_{i\mS} (s)ds \right)^{1/2} |Z_{ij}|\nonumber\\
&\leq &  C_1 k_0^{1/2} N^{-\omega} \sumi \left\{\left(\int_0^1
 \X_{i\mS}\trans(s) \X_{i\mS} (s)ds \right) + |Z_{ij}|^2\right\}. 
\ee
Now by Condition \ref{con:subexp}, for any unit vector $\v$,
$\v\trans \int_0^1 \X_{i\mS} \X_{i\mS}(s)\trans ds \v $ is a sub-exponential random variable
and $Z_{ij}^2$ is a sub-exponential random variable, by Proposition
5.16 the Bernstein-type inequality in \cite{vershynin2010} and union
bounded as those used in (\ref{eq:unionbound}) and fact that $|\mS| = k_0$, we have
\bse
&&\Pr\left[|\sumi \left\{\left(\int_0^1
 \X_{i\mS}\trans(s) \X_{i\mS}(s) ds \right) - E\left(\int_0^1
 \X_{i\mS}\trans(s) \X_{i\mS}(s) ds \right)\right\}|\geq nt \right]
\\
&\leq&  2
\exp\left\{-c'_3n\min(t^2, t) +k_0 \log(9)\right\}\\
&\leq& 2
\exp\left\{-c'_3n\min(t^2, t) +c_0'\log\{\max(p, n)\}\log(9)\right\}
\ese
for some constant $c'_3>0$, where the last inequality holds by
Condition \ref{con:npN}. Therefore, by selecting $t =
B\sqrt{c_0\log\{\max(p, n)\}/n}$,  we have $$\frac{\sumi \int_0^1
 \X_i\trans(s) \X_i (s)ds }{n}\leq E\left(\int_0^1
 \X_i\trans (s)\X_i(s) ds \right) + B \sqrt{c_0 \log\{\max(p, n)\}/n}$$ with probability greater than $1 -
2\exp\left[-c'_3 n \min \left\{ B^2 c_0 \log\{\max(p, n)\}/n, B \sqrt{c_0
    \log\{\max(p, n)\}/n} \right\}\right], $ where $B$ satisfies $c_3' c_0 \min(B^2,
B) = c_0'\log(9)+ 2$ and $c_0$ is defined in Condition \ref{con:npN} so that $c_0 \log\{\max(p, n)\}/n
\leq 1$.   Similarly, we can
show that
$$\frac{\sumi Z_{ij}^2}{n}\leq  E\left(Z_{ij}^2\right)  +B_1\sqrt{c_0 \log\{\max(p, n)\}/n} $$ with probability greater than $1 -
2\exp\left[-2 \log\{\max(p, n)\} \right]$ for  some constant
$c'_3>0$ and $c_4' c_0 \min(B_1^2, B_1) = 2$. Plug in (\ref{eq:psidiff}), we have
\be\label{eq:psidiffZ}
&& |n^{-1}\sumi \left[\psi'(\ba, \bG_0) - \psi'\{\ba,\bb(\cdot)\}\right]Z_{ij}|\nonumber\\
&\leq&   C_1 k_0^{1/2} N^{-\omega} \left\{E\left(\int_0^1
 \X_i\trans(s) \X_i (s)ds \right)  + E\left(Z_{ij}\right) + (B + B_1) \sqrt{c_0 \log\{\max(p, n)\}/n}
\right\}, 
\ee
with probability greater than $1 - 4\exp[-2 \log\{\max(p, n)\}]$. 

Similar to (\ref{eq:psidiff})
\bse
&& \left\|\sumi \left[\psi'(\ba, \bG_0) - \psi'\{\ba,\bb(\cdot)\}\right]
N^{1/2}\int_0^1 X_{ij}
(s) \B(s) ds \right\|_2\\
&\leq&  C_2 k_0^{1/2} N^{-\omega} \sumi \left\{\left(\int_0^1
 \X_i\trans(s) \X_i(s) ds \right) + N \int_0^1 X_{ij}
(s) \B(s) \trans ds \int_0^1 X_{ij}
(s) \B(s)  ds \right\}. 
\ese
Also because $N \int_0^1 X_{ij}
(s) \B(s) \trans ds$ is a sub-Gaussian random vector as shown in Lemma
\ref{lem:subGaussian},  similar to (\ref{eq:psidiffZ}), there are
constants $B_3, B_4$ such that 
\bse
&& \left\|n^{-1}\sumi \left[\psi'(\ba, \bG_0) - \psi'\{\ba,\bb(\cdot)\}\right]
N^{1/2}\int_0^1 X_{ij}
(s) \B(s) ds \right\|_2\\
&\leq&  C_2 k_0^{1/2}N^{-\omega } \left[E\left(\int_0^1
   \X_i\trans (s)\X_i (s)ds \right) \right.\\
 && \left.+ E \left\{N \int_0^1 X_{ij}
(s) \B(s) \trans ds \int_0^1 X_{ij}
(s) \B(s)  ds \right\}+( B_3 + B_4) \sqrt{\log\{\max(p, n)\}/n} \right], 
\ese
with probability $1 - 4 \exp[-2 \log\{\max(p, n)\}]$. 

Now because for any unit vector $\v$, $\int_0^1 \v\trans \X_i
(s)\X_i(s)\trans \v ds$, $N \int_0^1 X_{ij}
(s) \B(s) \trans \v ds \int_0^1 X_{ij}
(s) \v\trans \B(s)  ds$, and $Z_{ij}^2$ are all sub-exponential random variable, their
expectations are uniformly bounded by Condition \ref{con:subexp} and
by Condition \ref{con:npN}, we have there is a constant $C>0$,  for any given
$j, k$, we have 
\bse
&& \max\left( \left|n^{-1}\sumi \left[\psi'(\ba, \bG_0) -
     \psi'\{\ba,\bb(\cdot)\}\right] Z_{ik}\right|, \right.\\
 &&\left.\left\|n^{-1}\sumi \left[\psi'(\ba, \bG_0) - \psi'\{\ba,\bb(\cdot)\}\right]
N^{1/2}\int_0^1 X_{ij}
(s) \B(s) ds \right\|_2\right) \\
&\leq& C k_0^{1/2} N^{-\omega}
\ese
with probability $1 - 8 \exp[-2 \log\{\max(p, n)\}]$. By union bound over $k =
1, \ldots, q$ and $j = 1, \ldots, d$, we have
\bse
&& \max\left(\left\|n^{-1}\sumi \left[\psi'(\ba, \bG_0) -
     \psi'\{\ba,\bb(\cdot)\}\right] \Z_{i}\right\|_{\infty}, \right.\\
 &&\left.\left\|n^{-1}\sumi \left[\psi'(\ba, \bG_0) - \psi'\{\ba,\bb(\cdot)\}\right]
N^{1/2}\int_0^1 X_{ij}
(s) \B(s) ds \right\|_2, j = 1, \ldots, d\right) \\
&\leq& C k_0^{1/2} N^{-\omega} 
\ese
with probability $1 - 8 \max(p, n)^{-1}$ \qed

\begin{Lem}\label{lem:firstinfinty}
 Assume Conditions \ref{con:boundpsi}--\ref{con:npN} hold,  there are constants $C_0, C>0$\bse
&& \max\left\{\|n^{-1}\sumi\{\psi'_i(\ba,\bG_0) - Y_i\}
 \Z_{i}\|_{\infty}, \right.\\
 &&\left.\|n^{-1}\sumi
 \{\psi'_i(\ba, \bG_0) - Y_i\} \left[N^{1/2}\left\{\int_{0}^1 X_{ij} (s) \otimes
   \B(s)  ds\right\}\trans\right]\trans\|_2, j = 1, \ldots,
d\right\} \\
&\leq&  C_0 \sqrt{\log\{\max(p, n)\}/n} + C k_0^{1/2} N^{-\omega}
\ese
with probability $1 - 20\max(p, n)^{-1}$. 
 \end{Lem}
\noindent Proof:
\be\label{eq:firstorderexpand}
&& n^{-1}\sumi
\{\psi'_i(\ba, \bG_0) - Y_i\}
\left[\Z_i\trans, N^{1/2}\left\{\int_{0}^1\X_i (s) \otimes
   \B(s)  ds\right\}\trans\right]\trans\nonumber\\
&=&  n^{-1}\sumi
[ \psi'_i\{\ba,\bb(\cdot)\} - Y_i]
\left[\Z_i\trans, N^{1/2}\left\{\int_{0}^1\X_i (s) \otimes
   \B(s)  ds\right\}\trans\right]\trans\nonumber\\
&& + n^{-1}\sumi
[\psi'_i(\ba, \bG_0) - \psi'_i\{\ba,\bb(\cdot)\} ]
\left[\Z_i\trans, N^{1/2}\left\{\int_{0}^1\X_i (s) \otimes
   \B(s)  ds\right\}\trans\right]\trans 
\ee
For the first term, follow Lemma 14 in \cite{loh2012}, we have 
\be\label{eq:firstgreatt}
&& \Pr\left\{\|n^{-1}\sumi
 \{\psi'_i\{\ba,\bb(\cdot)\} - Y_i\}\Z_{i}\|_{\infty} \geq t \right\}\\
&\leq&  6
q\exp\{-c_3 n \min(t^2, t)\}
\ee
for some constants $c_3 >0$.
Furthermore for any given $j$ and any unit vector $\v$, let $\U =
[\{\psi'_i(\ba, \bG) - Y_i\}, i = 1, \ldots, n$, $\V = \left(\v\trans \left[N^{1/2}\left\{\int_{0}^1 X_{ij} (s) \otimes
   \B(s)  ds\right\}\trans\right]\trans,  i = 1, \dots, n\right)$, 
\bse
&& |  n^{-1}\sumi
\{\psi'_i\{\ba,\bb(\cdot)\}- Y_i\} \v\trans \left[N^{1/2}\left\{\int_{0}^1 X_{ij} (s) \otimes
   \B(s)  ds\right\}\trans\right]\trans|\\
&=& n^{-1} |\U\trans\V|\\
&=& \frac{1}{2}\left\{\Phi(\U + \V) - \Phi(\U) - \Phi(\V)\right\}
\ese
where $\Phi(\U)\equiv \frac{\|\U\|^2_2}{n} -
E\left(\frac{\|\U\|^2_2}{n}\right)$. Since by Condition \ref{con:subexp}
and Lemma \ref{lem:subGaussian}, $\U$, $\V$ are both
sub-Gaussian,  we have
\bse
|  n^{-1}\sumi
\{\psi'_i\{\ba,\bb(\cdot)\}- Y_i\} \v\trans \left[N^{1/2}\left\{\int_{0}^1 X_{ij} (s) \otimes
   \B(s)  ds\right\}\trans\right]\trans| \leq 3t/2, 
\ese
with probability greater than $1 - 6\exp\{c_4n \min (t^2, t)\}$. Now
by the union bound, similar to (\ref{eq:unionbound}) for some constant $c_4>0$,  we have
\bse
&& \Pr \left\{\max_{j=1, \ldots, d} \|  n^{-1}\sumi
\{\psi'_i\{\ba,\bb(\cdot)\}- Y_i\} \left[N^{1/2}\left\{\int_{0}^1 X_{ij} (s) \otimes
   \B(s)  ds\right\}\trans\right]\trans\|_2 \geq t\right\}\\
&= &\Pr \left\{\max_{j=1, \ldots, d} \sup_{\v, \|\v\|_2 \leq 1} |  n^{-1}\sumi
\{\psi'_i\{\ba,\bb(\cdot)\}- Y_i\} \v\trans \left[N^{1/2}\left\{\int_{0}^1 X_{ij} (s) \otimes
   \B(s)  ds\right\}\trans\right]\trans| \geq t\right\}\\
&\leq& 
6 p \exp[- c_4n \min (t^2, t) + N\{\log (9)\}]. 
\ese

Combine with (\ref{eq:firstgreatt}), let  we have
\bse
&& \max\left\{\|n^{-1}\sumi\{\psi'_i\{\ba,\bb(\cdot)\} - Y_i\}
 \Z_{i}\|_{\infty}, \right.\\
 &&\left.\|n^{-1}\sumi
 \{\psi'_i\{\ba, \bb(\cdot)\} - Y_i\} \left[N^{1/2}\left\{\int_{0}^1 X_{ij} (s) \otimes
   \B(s)  ds\right\}\trans\right]\trans\|_2, j = 1, \ldots,
d\right\}\\
&\leq & t 
\ese
with probability greater than $$1 - 6 p \exp[- c_4n \min (t^2, t) +
N\{\log (9)\}] - 6
q\exp\{-c_3 n \min(t^2, t)\}$$ which is larger than $$1 - 12 p \exp[- \min(c_3, c_5)n \min (t^2, t) +
c_0'\log(p)\{\log (9)\}],  $$ where $c_0'$ is defined in Condition \ref{con:npN}. Now select $t = C_0 \sqrt{c_0 \log\{\max(p, n)\}/n}$
so that $c_0\min(c_3, c_5) \min(C_0, C_0^2) =  c_0'\log(9) + 2, $
where $c_0$ is defined in Condition \ref{con:npN}, we
have
\bse
&& \max\left\{\|n^{-1}\sumi\{\psi'_i\{\ba,\bb(\cdot)\} - Y_i\}
 \Z_{i}\|_{\infty}, \right.\\
 &&\left. \| n^{-1}\sumi
 \{\psi'_i\{\ba,\bb(\cdot)\} - Y_i\} \left[N^{1/2}\left\{\int_{0}^1 X_{ij} (s) \otimes
   \B(s)  ds\right\}\trans\right]\trans\|_2, j = 1, \ldots,
d\right\}\\
&\leq & C_0 \sqrt{\log\{\max(p, n)\}/n}
\ese
with probability greater than $1 - 12\max(p, n)^{-1}$. Now combine with
(\ref{eq:firstorderexpand}) and Lemma \ref{lem:firstexpandsecond}, we
have there is a constant $C_0>0$ 
\bse
&& \max\left\{\|n^{-1}\sumi\{\psi'_i(\ba,\bG_0) - Y_i\}
 \Z_{i}\|_{\infty}, \right.\\
 &&\left.\| n^{-1}\sumi
 \{\psi'_i(\ba, \bG_0) - Y_i\} \left[N^{1/2}\left\{\int_{0}^1 X_{ij} (s) \otimes
   \B(s)  ds\right\}\trans\right]\trans\|_2, j = 1, \ldots,
d\right\} \\
&\leq&  C_0 \sqrt{\log\{\max(p, n)\}/n} + C k_0^{1/2} N^{-\omega}
\ese
with probability $1 - 20\max(p, n)^{-1}$. \qed

\noindent {\bf Proof of Theorem~\ref{th:1}}:

First by Taylor
expansion of the first order derivative,
\bse
\left\{\partial \mL(\wh{\btheta}) /\partial \btheta\trans- \partial
 \mL(\btheta_0) /\partial \btheta\trans \right\} \wh{\v} =\wh{\v}\trans
\partial^2 \mL(\btheta^*) /\partial \btheta \btheta\trans \wh{\v}, 
\ese
where $\btheta^*$ is a point on the line between $\wh{\btheta}$ and
$\btheta_0$, and hence is in the feasible set. Now by Lemma
\ref{lem:Re}, we obtain 
\be\label{eq:firsttosecond}
\left\{\partial \mL(\wh{\btheta}) /\partial \btheta\trans- \partial
 \mL(\btheta_0) /\partial \btheta\trans \right\} \wh{\v}  \geq
c_{\psi}\alpha_{\min} \|\wh{\v}\|_2^2/2- \tau_1 \log(p)/n  \left\{ \|\wh{\v}_1\|_1 + \sum_{j = 1}^d \|\wh{\v}_{2 j}\|_2
\right\}^2. 
\ee
Since function $\rho_{\lambda}(\|\bg_{ j}\|_2) +
\mu/2 \|\bg_{ j}\|_2^2$ is convex in $\|\bg_{
  j}\|_2$ by Condition (A5), and $L_2$ norm is convex, we have $\rho_{\lambda}(\|\bg_{ j}\|_2) +
\mu/2 \|\bg_{j}\|_2^2$ is a convex function
on $\bg_{ j}$. 

Therefore, we have 
\bse
\rho_{\lambda}(\| \bg_{0 j}\|_2)  +  \mu/2 \|\bG_{0 \cdot
 j}\|_2^2 - \rho_{\lambda}(\|\wh \bg_{ j}\|_2)  -  \mu/2 \|\wh\bg_{
 j}\|_2^2 \geq \{\partial \rho_{\lambda} (\|\wh \bg_{
 j}\|_2)/\partial \bg_{
 j}\trans + \mu \wh\bg_{
 j}\trans  \} ({\bg}_{0 j} - \wh{\bg}_{ j} ), 
\ese
which implies
\bse
&& \rho_{\lambda}(\| \bG_{0 \cdot j}\|_2) - \rho_{\lambda}(\|\wh
\bg_{ j}\|_2) + \mu/2 \|\wh \bg_{ j} - \bg_{0 j}
\|^2_2\\
&\geq& \partial \rho_{\lambda} (\|\wh \bg_{
 j}\|_2)/\partial \bg_{
 j}\trans  ({\bg}_{0  j} - \wh{\bg}_{ j} ). 
\ese
Therefore, combine with (\ref{eq:firstordercond}), we have
\bse
N^{-1/2} \sum_{j \in \mM^c }\left\{\rho_{\lambda}(\|\bG_{0 \cdot j}\|_2) - \rho_{\lambda}(\|\wh
\bg_{ j}\|_2) + \mu/2 \|\wh \bg_{j} - \bG_{0 \cdot j}
\|^2_2 \right\}\geq \partial \mL(\wh{\btheta})/\partial \btheta\trans
(\wh{\btheta} - {\btheta_0}). 
\ese
Combining with (\ref{eq:firsttosecond}), we have
\bse
&&c_{\psi}\alpha_{\min}\|\wh{\v}\|_2^2/2- \tau_1 \log(p)/n  \left\{ \|\wh{\v}_1\|_1 + \sum_{j
   = 1}^d \|\wh{\v}_{2 j}\|_2\right\}^2\\
 &\leq&  - \partial
 \mL({\btheta}_0)/\partial \btheta\trans \wh{\v} +N^{-1/2} \sum_{j \in \mM^c
 }\left\{\rho_{\lambda}(\| \bG_{0 \cdot j}\|_2) - \rho_{\lambda}(\|\wh
\bg_{j}\|_2) + \mu/2 \|\wh \bg_{ j} - \bG_{0 \cdot j}
\|^2_2 \right\}\\
&\leq&  - \partial
 \mL({\btheta}_0)/\partial \btheta\trans \wh{\v} +N^{-1/2} \sum_{j \in \mM^c
 }\left\{\rho_{\lambda}(\| \bG_{0 \cdot j}\|_2) - \rho_{\lambda}(\|\wh
\bg_{j}\|_2) + \mu/2 \|\wh{\v}\|^2_2\right\}, 
\ese
with probability greater than $1 - 2 \exp\left\{-  c_1 n\min \left(
c_{\psi}^2   \alpha_{\min}^2/54 ^2, 1\right)/2\right\}$. 
This implies
\be\label{eq:alphamu}
&& (c_{\psi}\alpha_{\min} /2-\mu/2 ) \|\wh{\v}\|_2^2\nonumber\\
&\leq& \tau_1 \log(p)/n  \left\{ \|\wh{\v}_1\|_1 + \sum_{j
   = 1}^d \|\wh{\v}_{2 j}\|_2\right\}^2 - \partial
 \mL({\btheta}_0)/\partial \btheta\trans \wh{\v} + N^{-1/2}\sum_{j \in \mM^c
 }\left\{\rho_{\lambda}(\| \bG_{0 \cdot j}\|_2) - \rho_{\lambda}(\|\wh
 \bg_{ j}\|_2)\right\} \nonumber\\
 &\leq&  \tau_1 \log(p)/n  \left\{ \|\wh{\v}_1\|_1 + \sum_{j
   = 1}^d \|\wh{\v}_{2 j}\|_2\right\}^2 -  \partial
 \mL({\btheta}_0)/\partial \ba\trans \wh{\v}_1 - \sum_{j=1}^d \partial
 \mL({\btheta}_0)/\partial \bg_{j}\trans \wh{\v}_{2j} \nonumber \\
 && + N^{-1/2}\sum_{j \in \mM^c
 }\left\{\rho_{\lambda}(\| \bG_{0 \cdot j}\|_2) - \rho_{\lambda}(\|\wh
 \bg_{ j}\|_2) \right\}\nonumber\\
 &\leq& \tau_1 \log(p)/n  \left\{ \|\wh{\v}_1\|_1 + \sum_{j
   = 1}^d \|\wh{\v}_{2 j}\|_2\right\}^2  +\max_{j = 1, \ldots, d}\left\{\| \partial
 \mL({\btheta}_0)/\partial \ba\|_{\infty}, \|\partial
 \mL({\btheta}_0)/\partial \bg_{ j}\|_2\right\} \left\{ \|\wh{\v}_1\|_1 + \sum_{j
   = 1}^d \|\wh{\v}_{2 j}\|_2\right\} \nonumber\\
&& + N^{-1/2}\sum_{j \in \mM^c
 }\left\{\rho_{\lambda}(\| \bG_{0 \cdot j}\|_2) - \rho_{\lambda}(\|\wh
 \bg_{ j}\|_2)\right\} \nonumber\\
 &\leq& \left[2 \tau_1 R \log(p)/n  + \max_{j = 1, \ldots, d}\left\{\| \partial
 \mL({\btheta}_0)/\partial \ba\|_{\infty}, \|\partial
 \mL({\btheta}_0)/\partial \bg_{ j}\|_2\right\}\right]\left\{ \|\wh{\v}_1\|_1 + \sum_{j
 = 1}^d \|\wh{\v}_{2 j}\|_2\right\} \nonumber\\
&& +N^{-1/2} \sum_{j \in \mM^c
 }\left\{\rho_{\lambda}(\| \bG_{0 \cdot j}\|_2) - \rho_{\lambda}(\|\wh
 \bg_{j}\|_2)\right\}\nonumber\\
 &\leq& \lambda/2 \left\{ \|\wh{\v}_1\|_1 + \sum_{j
 = 1}^d \|\wh{\v}_{2 j}\|_2\right\} + N^{-1/2}\sum_{j \in \mM^c
 }\{\rho_{\lambda}(\| \bG_{0 \cdot j}\|_2) - \rho_{\lambda}(\|\wh
 \bg_{j}\|_2) \}\nonumber\\
 &=& \lambda/2 \left\{ \|\wh{\v}_1\|_1 +
\sum_{j\in \mM}   \|\wh{\v}_{2 j}\|_2 +\sum_{j\in \mM^c}
\|\wh{\v}_{2 j}\|_2   \right\} + N^{-1/2}\sum_{j \in \mM^c
 }\{\rho_{\lambda}(\| \bG_{0 \cdot j}\|_2) - \rho_{\lambda}(\|\wh
 \bg_{ j}\|_2) \}\nonumber\\
 &\leq& \lambda/2 \left\{ \|\wh{\v}_1\|_1 +\sum_{j\in \mM}
   \|\wh{\v}_{2 j}\|_2 + 
\sum_{j\in \mM^c}  \rho_{\lambda}(\|\wh{\v}_{2 j}\|_2)/\lambda + \mu/(2\lambda) \|\wh{\v}_{2 j}\|_2^2\right\} +N^{-1/2} \sum_{j \in \mM^c
 }\{\rho_{\lambda}(\| \bG_{0 \cdot j}\|_2) - \rho_{\lambda}(\|\wh
 \bg_{ j}\|_2) \}\nonumber\\
 &\leq& \lambda/2 \left(\|\wh{\v}_1\|_1  + \sum_{j \in \mM}
   \|\wh{\v}_{2j}\|_2\right)+ \frac{\sum_{j\in \mM^c}   \left\{
     \rho_{\lambda}(\|\wh{\bg}_{ j}\|_2) +
     \rho_{\lambda}(\|{\bG}_{0 \cdot j}\|_2)\right\}}{2} + \mu/4
 \|\wh{\v}\|_2^2 \nonumber\\
 && +N^{-1/2} \sum_{j \in \mM^c
 }\{\rho_{\lambda}(\| \bG_{0 \cdot j}\|_2) - \rho_{\lambda}(\|\wh
 \bg_{ j}\|_2) \}\nonumber\\
 &=& \lambda/2 \left(\|\wh{\v}_1\|_1  + \sum_{j \in \mM}
   \|\wh{\v}_{2j}\|_2\right) + \mu/4
 \|\wh{\v}\|_2^2 + N^{-1/2} /2 \sum_{j \in \mM^c
 }\{3\rho_{\lambda}(\| \bG_{0 \cdot j}\|_2) - \rho_{\lambda}(\|\wh
 \bg_{ j}\|_2) \}, 
 \ee
 with probability greater than $1 - 2 \exp\left\{-  c_1 n\min \left(
 c_{\psi}^2  \alpha_{\min}^2/54 ^2, 1\right)/2\right\} - 20\max(p, n)^{-1}$. 
 Here the sixth line holds by the assumption that $$\lambda \geq
 \max\left\{2 \tau_1 R \log\{\max(p)\}/n,  C_0 \sqrt{\log\{\max(p, n)\}/n} +
   C k_0^{1/2} N^{-\omega}\right\}/4, $$ and the fact that
 \bse
 \max_{j = 1, \ldots, d}\left\{\| \partial
 \mL({\btheta}_0)/\partial \ba\|_{\infty}, \|\partial
 \mL({\btheta}_0)/\partial \bg_{ j}\|_2 \right\}\leq C_0 \sqrt{\log\{\max(p, n)\}/n} +
   C k_0^{1/2} N^{-\omega}, 
   \ese
   with probability greater than $1 - 20\max(p, n)^{-1}$ by Lemma \ref{lem:firstinfinty}. 
 The
eighth line holds by Lemma \ref{lem:fromlemma4}. Now if
\bse
\sum_{j \in \mM^c
 }\{3\rho_{\lambda}(\| \bG_{0 \cdot j}\|_2) - \rho_{\lambda}(\|\wh
 \bg_{j}\|_2) \} \leq 0, 
 \ese
 by (\ref{eq:alphamu}), we have
 \bse
 (c_{\psi}\alpha_{\min} /2-3\mu/4 ) \|\wh{\v}\|_2^2 \leq \lambda/2 \left(\|\wh{\v}_1\|_1  + \sum_{j \in \mM}
   \|\wh{\v}_{2j}\|_2\right) \leq \lambda/2 \sqrt{q + m}
 \|\wh{\v}\|_2, 
 \ese
 which implies
 \be\label{eq:v21}
\|\wh{\v}\|_2 \leq \lambda \sqrt{q + m}/ (c_{\psi}\alpha_{\min} -3\mu/2 ). 
 \ee
On the other hand if \bse
\sum_{j \in \mM^c
 }\{3\rho_{\lambda}(\| \bG_{0 \cdot j}\|_2) - \rho_{\lambda}(\|\wh
 \bg_{ j}\|_2) \} >  0. 
 \ese  by Lemma (\ref{lem:fromlemma5}), we have
 \be\label{eq:rhoineq}
 0 \leq N^{-1/2} \sum_{j \in \mM^c
 }\{3\rho_{\lambda}(\| \bG_{0 \cdot j}\|_2) - \rho_{\lambda}(\|\wh
 \bg_{j}\|_2) \} \leq \lambda \left( 3 \sum_{j\in \mA}
   \|\wh{\v}_{2j}\|_2 - \sum_{j\in \mA^c} \|\wh{\v}_{2j}\|_2\right), 
 \ee
 where $\mA$ is the index set of the $k_0$ largest
 $\|\wh{\v}_{2j}\|_2$, $j\in \mM^c$, and $\mA^c$ is the index set of the remaining
 $d - m - k_0$ of 
 $\wh{\v}_{2j}, j \in \mM^c$, which implies
 \bse
\sum_{j\in \mA^c} \|\wh{\v}_{2j}\|_2  \leq 3 \sum_{j\in \mA} \|\wh{\v}_{2j}\|_2
 \ese
 and hence
 \bse
 \sum_{j \in \mM^c } \|\wh{\v}_{2j}\|_2 \leq 4 \sum_{j\in \mA}
 \|\wh{\v}_{2j}\|_2. 
 \ese
 Plug (\ref{eq:rhoineq}) into (\ref{eq:alphamu}), we have
 \bse
 (c_{\psi}\alpha_{\min} /2-3\mu/4 ) \|\wh{\v}\|_2^2 &\leq& \lambda/2 \left(\|\wh{\v}_1\|_1  + \sum_{j \in \mM}
   \|\wh{\v}_{2j}\|_2\right) + \lambda /2 \left( 3 \sum_{j\in \mA}
   \|\wh{\v}_{2j}\|_2 - \sum_{j\in \mA^c} \|\wh{\v}_{2j}\|_2\right)\\
 &\leq& \lambda/2 \left(\|\wh{\v}_1\|_1  + \sum_{j \in \mM}
   \|\wh{\v}_{2j}\|_2\right) + \lambda /2 \left( 3 \sum_{j\in \mA}
   \|\wh{\v}_{2j}\|_2 \right)\\
 &\leq& 3 \lambda/2 \left(\|\wh{\v}_1\|_1  + \sum_{j \in \mM}
   \|\wh{\v}_{2j}\|_2 + \sum_{j\in \mA}
   \|\wh{\v}_{2j}\|_2 \right)\\
 &\leq& 3 \lambda/2 \sqrt{q + m + k_0} \left\{\|\wh{\v}_1\|_2^2 + \sum_{j \in
   \mM \cup \mA }
 \|\wh{\v}_{2j}\|_2^2\right\}^{1/2}\\
&\leq& 3\lambda \sqrt{q + m + k_0}/ 2 \|\wh{\v}\|_2. 
\ese
This gives 
\be \label{eq:v22}
\|\wh{\v}\|_2  \leq  3\lambda \sqrt{q + m + k_0}/ (c_{\psi}\alpha_{\min}
-3\mu/2 ). 
\ee
Finally, since \bse
\sum_{j\in \mA^c} \|\wh{\v}_{2j}\|_2  \leq 3 \sum_{j\in \mA} \|\wh{\v}_{2j}\|_2
\ese
we have 
\be\label{eq:v23}
&& \|\wh{\v}_1\|_1 + \sum_{j = 1}^d \|\wh{\v}_{2j}\|_2\nonumber\\
&=& \|\wh{\v}_1\|_1 + \sum_{j \in \mM } \|\wh{\v}_{2j}\|_2 + \sum_{j
 \in \mA } \|\wh{\v}_{2j}\|_2  + \sum_{j
 \in \mA^c} \|\wh{\v}_{2j}\|_2\nonumber\\
&\leq&  \|\wh{\v}_1\|_1 + \sum_{j \in \mM } \|\wh{\v}_{2j}\|_2 + 4\sum_{j
 \in \mA } \|\wh{\v}_{2j}\|_2 \nonumber \\
&\leq& 4 \sqrt{q + m + k_0} \|\wh{\v}\|_2\nonumber\\
&\leq& 12 \lambda (q + m + k_0)/ (c_{\psi} \alpha_{\min}
-3\mu/2 ). 
\ee
Combine (\ref{eq:v21}), (\ref{eq:v22}) and (\ref{eq:v23}), we prove
the result. \qed

\begin{Lem}\label{lem:convexF}
 Let $\x^* = (\x_1^{*\trans}, \x_{2j}^{*\trans}, j = 1, \ldots, d)\trans$ is feasible for
 the program
 \be\label{eq:B0}
 \min_{\x \equiv (\x_1\trans, \x_{2j}\trans, j = 1 ,\ldots, d)\trans}
 \left\{f(\x) -   \sum_{j\in \mM^c} g(\|\x_j\|_2) +\lambda
    \sum_{j\in \mM^c} \|\x_j\|_2 \right\} \text{ such that }
 \|\x_1\|_1 +  \sum_{j=1}^d \|\x_{2j}\|_2 \leq R, 
 \ee
 where $f (\x)\in C^2$, $g(\|\x\|)$ is differentiable in $\x$,  and $g(\|\x\|_2) - \kappa/2
 \|\x\|_2^2$ is concave in $\|\x_{2j}\|_2$. Assume there are
 $\mu_1^* >0$, $\v_j^*
 \in \partial \|\x_{2j}^*\|_2$ for $j\in \mM^c$ and $\v_j^* = 0$ for
 $j\in \mM$,  $\v^* = (\0_q\trans, \v_j^{*\trans}, j =
 1, \ldots, d)\trans$, and  $\w^* \in \partial \{\|\x_1^*\|_1 + 
 \sum_{j = 1}^d \|\x_{2j}^*\|_2\}$ such that
 \be
 \mu_1^* \left\{R - \|\x_1\|_1 -   \sum_{j=1}^d \|\x_{2j}\|_2 \right\}
 = 0 \label{eq:1}\\
 \frac{\partial f(\x^*)}{\partial \x^*} -
  \frac{\partial  \sum_{j \in \mM^c}  g (\|\x_{2j}^*\|_2)}{\partial \x^*} +  \lambda \v^* + \mu_1^* \w^* = 0\label{eq:2}\\
 \s\trans \frac{\partial^2 f(\x)}{\partial \x\partial \x\trans }\s
 >\kappa , \forall \s \in G^*, \label{eq:3}
 \ee
 where
 \bse
 G^* \equiv \left\{\mystrut \s \equiv (\s_1\trans, \s_{2j}\trans, j = 1,
   \ldots, d)\trans \in \mathbb{R}^{q + Nd}: \|\s\|_2 = 1;\right. \\
\sup_{\w
     \in \partial \{\|\x^*_{1}\|_1 +  \sum_{j=1}^d\|\x^*_{2j}\|_2\}}
   \s\trans \w\leq 0 \text{ if } \|\x^*_1\|_1  +  \sum_{j=1}^d\|\x^*_{2j}\|_2
   \leq R; \\
\sup_{(\v_j \in \partial \|\x_{2j}^*\|_1, j \in \mM^c)} \s\trans
 \left(\frac{\partial f(\x^*)}{\partial \x^*} -  
   \frac{\partial \sum_{j \in \mM^c} g(\|\x_{2j}^*\|_2)}{\partial \x^*}\right) +
 \lambda  \sum_{j\in \mM^c}\s_{2j}\trans \v_j = 0; \\
\left.    \mu_1^* \sup_{\w
     \in \partial \{\|\x^*_{1}\|_1 +  \sum_{j=1}^d\|\x^*_{2j}\|_2\}}
   \s\trans \w=  0 \right\}. 
 \ese
 Then $\x^*$ is an isolated local minimum of the program (\ref{eq:B0})
\end{Lem}
\noindent Proof: Suppose $\x^*$ is not an isolated local minimum. Then
there is a sequence $\{\x^k \equiv (\x_1^{k\trans},
\x_{2j}^{k\trans}, j = 1, \ldots, d)\trans\}$, so that $\x^k\to \x^*$ and
\bse
\phi(\x^k) \leq \phi(\x^*), 
\ese
where $\phi(\x)  = f(\x) -  \sum_{j\in \mM^c} g(\|\x_j\|_2) +\lambda
    \sum_{j\in \mM^c} \|\x_j\|_2$. Let $\s^k \equiv (\x^k -
   \x^*)/\|\x^k - \x^*\|_2$, so $(\s^k)$ is a set of feasible
   directions. Since $(\s^k) \subset \mathbb{B}_2(1)$, where
   $\mathbb{B}_2(1)$ is the $L_2$ ball with radius 1, the set must
   possess a point of accumulation $\s\in \mathbb{B}_2(1)$, and we
   can extract a convergence subsequence such that $(\s^k) \to
   \s$. With a slight abuse of notation, we stil use $(\s^k)$ to
   denote the subsequence. We will show that $\s \in G^*$.

   First because the feasible region is closed, $\s$ is also in the
   feasible direction of $\x^*$. If $\|\x^*_{1}\|_1 + 
   \sum_{j=1}^d\|\x^*_{2j}\|_2 = R$, by the sub-gradient of convexity
   function of $\|\cdot\|_1$ and $\|\cdot\|_2$, we have
   \bse
   0&\geq& \|\x^k_{1}\|_1 +
    \sum_{j=1}^d\|\x^k_{2j}\|_2- \|\x^*_{1}\|_1 - 
    \sum_{j=1}^d\|\x^*_{2j}\|_2\\
   &=& \|\x_1^* + \|\x^k -
\x^*\|_2\s^k_1\|_2 - \|\x_1^* \|_2 +  \sum_{j=1}^d \left\{\|\x_{2j}^* + \|\x^k -
\x^*\|_2\s^k_{2j}\|_2 - \|\x_{2j}^* \|_2\right\} \\
&\geq& \|\x^k - \x^*\|_2
\s_1^{k\trans} \w_1 +  \|\x^k - \x^*\|_2 \sum_{j = 1}^d
\s_{2j}^{k\trans}\w_{2j}\\
&=& \|\x^k - \x^*\|_2 \s^{k\trans} \w, 
\ese
where $\w = (\w_1\trans, \w_{2j}\trans, j = 1, \ldots, d)\trans \in \partial
(\|\x_1^*\|_1 +  \sum_{j = 1}^d \|\x_{2j}^*\|_2)$. When $k\to \infty$,
this leads to 
\be\label{eq:sw}
\sup_{\w
     \in \partial \{\|\x^*_{1}\|_1 +  \sum_{j=1}^d\|\x^*_{2j}\|_2\}}
   \s\trans \w\leq 0. 
\ee
Furthermore, (\ref{eq:1}) also implies that if $ \|\x_1^*\|_1 +
\sum_{j = 1}^d \|\x_{2j}^*\|_2 \neq R$, then $\mu_1^* = 0$. Since
$\mu_1^*\geq 0$, we have 
\be\label{eq:usw}
\mu_1^* \sup_{\w
     \in \partial \{\|\x^*_{1}\|_1 +  \sum_{j=1}^d\|\x^*_{2j}\|_2\}}
   \s\trans \w\leq 0. 
\ee
Moreover, by (\ref{eq:2}), we have 
\bse
\s\trans \frac{\partial f(\x^*)}{\partial \x^*} -  \sum_{j\in \mM^c}\s_{2j}\trans 
\frac{\partial g(\|\x_{2j}^*\|_2)}{\partial \x_{2j} } +  \lambda
\sum_{j \in \mM^c}\s_{2j}\trans \v_{j}^*  + \mu_1^* \s\trans \w^*= 0. 
\ese
Together with (\ref{eq:usw}), we have
\be\label{eq:sfg}
\s\trans \frac{\partial f(\x^*)}{\partial \x^*} -  \sum_{j\in \mM^c}\s_{2j}\trans 
\frac{\partial g(\|\x_{2j}^*\|_2)}{\partial \x_{2j} } +  \lambda
\sum_{j \in \mM^c}\s_{2j}\trans \v_{j}^* = - \mu_1^* \s\trans \w^*
\geq 0. 
\ee
By the definition of sub-gradient, we have
\bse
\|\x^*_{2j} + \|\x^k - \x^*\|_2 \s_{2j}^k\|_2 -
\|\x^*_{2j}  \|_2 \geq \|\x^k - \x^*\|_2\s^{k\trans}_{2j}\v_j
\ese
for all $\v_j \in \partial \|\x_{2j}^*\|_2, j \in \mM^c$ and  all
$k$. Furthermore because
\bse
\|\x^*_{2j} + \|\x^k - \x^*\|_2 \s_{2j}^k\|_2 -
\|\x^*_{2j}  \|_2  = \|\x_{2j}^k\|_2 - \|\x_{2j}^*\|_2, 
\ese
we have
\be\label{eq:sv}
\s_{2j}\trans \v_j = \lim_{k\to \infty }\s^{k\trans}_{2j}\v_j \leq
\lim_{k \to \infty} \frac{\|\x_{2j}^k\|_2 - \|\x_{2j}^*\|_2}{\|\x^k
 - \x^*\|_2}. 
\ee
Furthermore,
\bse
&& \s\trans \frac{\partial f(\x^*)}{\partial \x^*} - 
\sum_{j \in \mM^c}\s_{2j}\trans\frac{\partial g(\|\x_{2j}^*\|_2)}{\partial \x_{2j}^*} \\
&=&\lim_{k \to \infty} \s^{k\trans} \frac{\partial f(\x^*)}{\partial \x^*} - 
\sum_{j \in \mM^c} \s_{2j}^{k\trans}\frac{\partial g(\|\x_{2j}^*\|_2)}{\partial \x_{2j}^*}
\\
&=& \lim_{k \to \infty} \frac{\langle \x^k - \x^*, \frac{\partial
   f(\x^*)}{\partial \x^*}\rangle -  \sum_{j \in \mM^c} \langle
 \x_{2j}^k - \x_{2j}^*, \frac{\partial g(\|\x_{2j}^*\|_2)}{\partial
   \x_{2j}^*} \rangle}{\|\x^k - \x^*\|_2}\\
&=& \lim_{k \to \infty} \frac{f(\x^k) - \sum_{j\in \mM^c}
 g(\|\x_{2j}^k\|_2)-  f(\x^*) +  \sum_{j\in \mM^c}
 g(\|\x_{2j}^*\|_2)}{\|\x^k - \x^*\|_2}
\ese
since $\x^k \to \x^*$ and $f(\x^k) -  \sum_{j \in \mM^c}
g(\|\x^k_{2j}\|_2) $ is differentiable in $\x^k$. Combine with (\ref{eq:sv}), we have
\bse
&& \sup_{(\v_j \in \partial \|\x^*_{2j}\|_2, j \in \mM^c)} \s\trans
\frac{\partial f(\x^*)}{\partial \x^*} -  \sum_{j \in \mM^c}\s_{2j}\trans \frac{\partial
g(\|\x_{2j}^*\|_2)}{\partial \x_{2j}^*}  +  \lambda \sum_{j \in \mM^c}\s_{2j}\trans\v_j\\
&\leq& \lim_{k\to \infty} \frac{\phi(\x^k) - \phi(\x^*)}{\|\x^k - \x^*\|_2}\leq 0. 
\ese
Combine with (\ref{eq:sfg}), we have
\be\label{eq:sfg0}
\sup_{(\v_j \in \partial \|\x^*_{2j}\|_2, j \in \mM^c)} \s\trans
\frac{\partial f(\x^*)}{\partial \x^*} - \sum_{j\in \mM^c}\s_{2j}\trans \frac{\partial
g(\|\x_{2j}^*\|_2)}{\partial \x_{2j}^*}  + \lambda \sum_{j\in \mM^c}\s_{2j}\trans\v_j =
\bf 0. 
\ee
Now by (\ref{eq:2}), we have
\bse
\mu_1^* \s\trans \w^* = 0. 
\ese
Together with (\ref{eq:usw}), we have
\bse
\mu_1^* \sup_{\w
     \in \partial \{\|\x^*_{1}\|_1 + \sum_{j=1}^d\|\x^*_{2j}\|_2\}}
   \s\trans \w=  0. 
   \ese
Combine with  (\ref{eq:sw}) and (\ref{eq:sfg0}), we have $\s \in
G^*$.

By the convexity of $L_1$, $L_2$ norms, we have
\bse
\|\x_{1}^k\|_1 - \|\x_1^*\|_1 + \sum_{j=1}^d\{\|\x_{2j}^k\|_2 -
\|\x_{2j}^*\|_2\} \geq \x_1^{k\trans} \w_1 - \|\x_1^*\|_1 +
\sum_{j=1}^d(\x_{2j}^{k\trans} \w_{2j} - \|\x_{2j}^*\|_2)
\ese
for $\w = (\w_1\trans, \w_{2j}\trans, j = 1, \ldots, d)\trans \in
\partial (\|\x^*_{1}\|_1 + \sum_{j=1}^d\|\x^*_{2j}\|_2)$. Therefore, $
\x_1^{k\trans} \w_1 + \sum_{j=1}^d\x_{2j}^{k\trans} \w_{2j}  \leq
\|\x_1^{k}\|_1 + \sum_{j=1}^d\|\x_{2j}^{k}\|_2 \leq  R$.
Therefore,
\bse
\phi(\x^k) &=&  f(\x^k) - \sum_{j\in \mM^c} g(\|\x_{2j}^k\|_2)
+\lambda \sum_{j\in \mM^c}  \|\x_{2j}^k\|_2\\
&\geq& f(\x^k) - \sum_{j\in \mM^c} g(\|\x_{2j}^k\|_2) + \lambda 
\sum_{j\in \mM^c}\x_{2j}^{k\trans}\v^*_j + \mu_1^*\left(
\x_1^{k\trans} \w^*_1 + \sum_{j=1}^d\x_{2j}^{k\trans} \w^*_{2j}  - R
\right), 
\ese
for all $\v^* = (\0_q, \v^{*\trans}_j, j = 1, \ldots, d)$, $\v^*_j \in
\partial\|\x_{2j}^*\|_2$ for $j\in \mM^c$ and $\v^*_j = 0$ for $j\in
\mM$.
And
\bse
\phi(\x^*) &=&  f(\x^*) - \sum_{j\in \mM^c} g(\|\x_{2j}^*\|_2)
+\lambda \sum_{j\in \mM^c}  \|\x_{2j}^*\|_2\\
& = & f(\x^*) - \sum_{j\in \mM^c} g(\|\x_{2j}^*\|_2) + \lambda 
\sum_{j\in \mM^c}\x_{2j}^{*\trans}\v^*_j + \mu^*_1\left(
\x_1^{*\trans} \w^*_1 + \sum_{j=1}^d\x_{2j}^{*\trans} \w^*_{2j}  - R
\right). 
\ese
The equality holds because $\mu^*_1\left(
\x_1^{*\trans} \w^*_1 + \sum_{j=1}^d\x_{2j}^{*\trans} \w^*_{2j}  - R
\right) = 0$, and $\x_{2j}^{*\trans}\v^*_j =
\|\x_{2j}^{*\trans}\|_2$.
Therefore, we can write
\be\label{eq:phipshi}
&& \lim_{k\to \infty} \left\{\phi(\x^k) - \phi(\x^*)\right\}/\|\x^k -
\x^*\|_2^2\nonumber\\
&\geq& \lim_{k\to \infty} \left\{f(\x^k) - f(\x^*) -\sum_{j\in \mM^c}
 g(\|\x^k_{2j}\|_2) + \sum_{j\in \mM^c} g(\|\x^*_{2j}\|_2) + \langle
 \lambda  \v^* + \mu_1^*\w^*, \x^k - \x^*\rangle\right\}\nonumber\\
&& /\|\x^k -
\x^*\|_2^2\nonumber\\
&=&  \lim_{k\to \infty} \left\{f(\x^k) - f(\x^*) -\sum_{j\in \mM^c}
 g(\|\x^k_{2j}\|_2) + \sum_{j\in \mM^c} g(\|\x^*_{2j}\|_2) \right\}/\|\x^k -
\x^*\|_2^2\nonumber \\
&& - \left\langle \frac{\partial f(\x^*)}{\partial \x^*} -
 \frac{\partial \sum_{j\in \mM^c }g(\|\x_{2j}^*\|_2)}{\partial \x^*},
 \x^k - \x^*\right\rangle/\|\x^k - \x^*\|_2^2\nonumber\\
&=& \lim_{k \to \infty}\left\{ f(\x^k) - f(\x^*) - \langle
 \frac{\partial f(\x^*)}{\partial \x^*}, \x^k -
 \x^*\rangle\right\}/\|\x^k - \x^*\|_2^2\nonumber \\
&& - \lim_{k \to \infty}\left\{ \sum_{j\in \mM^c}g(
     \|\x_{2j}^k\|_2) - \sum_{j\in \mM^c}g( \|\x_{2j}^*\|_2) -
    \left \langle  \frac{\partial \sum_{j\in \mM^c
       }g(\|\x_{2j}^*\|_2)}{\partial \x^*}, \x^k -
     \x^*\right\rangle\right\}\nonumber\\
 &&/\|\x^k - \x^*\|_2^2
 \ee
 Now because $g(\|\x\|_2) - \kappa/2 \|\x\|_2^2$ is concave in
 $\|\x\|_2$, and hence $-g(\|\x\|_2) + \kappa/2 \|\x\|_2^2$ is
 convex in $\x$. Therefore, we have
 \bse
 -\sum_{j\in \mM^c} g(\|\x_{2j}^k\|_2) +  \sum_{j\in \mM^c}
 g(\|\x_{2j}^*\|_2) + \left\langle\frac{\partial \sum_{j\in \mM^c}
     g(\|\x^*_{2j}\|_2)}{\partial \x^*}, \x^k - \x^*\right\rangle
 \geq - k/2\|\x^k - \x^*\|_2^2. 
 \ese
 Furthermore note that $\phi(\x^k) - \phi(\x^*) \leq 0$, combine with
 (\ref{eq:phipshi}), we have
 \bse
\lim_{k \to \infty}\left\{ f(\x^k) - f(\x^*) - \langle
 \frac{\partial f(\x^*)}{\partial \x^*}, \x^k -
 \x^*\rangle\right\}/\|\x^k - \x^*\|_2^2 - k/2 \leq 0, 
\ese
which by Talyor expansion implies
\bse
\s\trans \frac{\partial^2 f(\x^*)}{\partial \x\partial \x\trans }\s &=&
\lim_{k \to \infty } (\x^k - \x^*)\trans \frac{\partial^2
 f(\x^*)}{\partial \x \partial \x\trans} (\x^k - \x^*)/\|\x^k -
\x^*\|_2^2\\
&\leq& \lim_{k\to \infty} \left\{\kappa\|(\x^k - \x^*)\|_2^2 +
 o(\|\x^k - \x^*\|_2^2)\right\}/\|\x^k - \x^*\|_2^2\\
&\leq& \kappa, 
\ese
which contradicts with (\ref{eq:3}). Hence $\x^*$ must be an isolated
local minimum. This proves the result. \qed

\begin{Lem}\label{eq:minandstation}
 Assume Conditions (A1)--(A6) and \ref{con:boundpsi} --
 \ref{con:secondorder} hold. Assume $\mu <c_{\psi}\alpha_{\min}/2$, $\delta \in
 [4R\tau_1\log(p)/n/\lambda, 1]$, $\max_{j \in (\mM\cup
   S)^c}\|\wh{\z}_j\|_2\leq 1- \delta$ for $\wh{\z}_j \in
 \partial \|N^{-1/2}\wh{\bg}_{j}\|_2$. Furthermore, assume that $n\geq 8\tau_1 (\log(p) + N + q)\{\sqrt{q} + \sqrt{m} + (2/\delta + 2)
\sqrt{k_0}\}^2/ (c_{\psi}\alpha_{\min} /2-
\mu)$. Suppose $\wt{\btheta} \equiv \{\wt{\ba}\trans,
N^{-1/2}\vec(\wt{\bG})\}\trans$ is a stationary point of program
\ref{eq:lapHT} and $\wh{\btheta} \equiv \{\wh{\ba}\trans,
N^{-1/2}\vec(\wh{\bG})\}\trans$ is the interior local minimizer of
\ref{eq:lapHT} and satisfies $\supp\{\|\wh{\bg}_{ j}\|_2, j = 1, \ldots,
d\} \subset \mM \cup S$. Then there is a constant $c_1>0$ such that $\supp\{\|\wt{\bg}_{ j}\|_2, j = 1, \ldots,
d\} \subset \mM \cup S$ with probability $1 - 2 \exp\left\{-  c_1 n\min \left(
 c_{\psi}^2  \alpha_{\min}^2/54 ^2, 1\right)/2 \right\}$. 
\end{Lem}
\noindent Proof: Let $\wt{\v} \equiv (\wt{\v}_1\trans,
\wt{\v}_{2j}\trans, j = 1, \ldots, d)\trans \equiv  \wt{\btheta} - \wh{\btheta}
= \{\wt{\ba}\trans - \wh{\ba}\trans, N^{-1/2}\vec(\wt{\bG})\trans -
N^{-1/2}\vec(\wh{\bG})\trans\}\trans$ by the Taylor expansion of the first order
derivative we have
\bse
\left\{\partial \mL(\wt{\btheta})/\partial \btheta\trans - \partial
 \mL(\wh{\btheta})/\partial \btheta\trans\right\}\wt{\v} =
\wt{\v}\trans\partial^2 \mL({\btheta}^*)/\partial \btheta \partial
\btheta\trans \wt{\v}, 
\ese
where $\btheta^*$ is a point on the line connecting $\wh{\btheta}$ and
$\wt{\btheta}$, and hence in the feasible set. Therefore, by Lemma
\ref{lem:Re}, we have
\bse
\left\{\partial \mL(\wt{\btheta})/\partial \btheta\trans - \partial
 \mL(\wh{\btheta})/\partial \btheta\trans\right\}\wt{\v} \geq
c_{\psi}\alpha_{\min} /2\|\wt{\v}\|_2^2 -  \tau_1 \{\log(p) + N +
q\}/n\left\{\|\wt{\v}_1\|_1 +\sum_{j = 1, \ldots, d}\|\wt{\v}_{2j}
 \|_2\right\}^2, 
\ese
with probability $1 - 2 \exp\left\{-  c_1 n\min \left(
c_{\psi}^2   \alpha_{\min}^2/54 ^2, 1\right)/2 \right\}$. 
Now note that
\bse
&& \mL(\btheta) - N^{-1/2}\sum_{j\in \mM^c }q_{\lambda}(\|\bg_{
 j}\|_2)\\
& =&
\mL(\btheta) - N^{-1/2}\sum_{j\in \mM^c } \mu \|\bg_{j}\|_2^2/2 +
N^{-1/2}\sum_{j\in \mM^c } \mu \|\bg_{
 j}\|_2^2/2 - N^{-1/2}\sum_{j\in \mM^c }q_{\lambda}(\|\bg_{j}\|_2), 
\ese
and $\sum_{j\in \mM^c } \mu \|\bg_{
 j}\|_2^2/2 - \sum_{j\in \mM^c }q_{\lambda}(\|\bg_{ j}\|_2)$ is
convex by Lemma \ref{lem:fromlemm5support}, and hence for any $\btheta$ in the feasible set,
we have
\bse
&& \wt{\v}\trans \frac{\partial\left[ \mL(\btheta) -N^{-1/2}\sum_{j \in \mM^c}
 q_{\lambda} \{\|\bg_{j}\|_2\}\right]}{\partial \btheta \partial
 \btheta\trans}\wt{\v} \nonumber\\
&\geq& \wt{\v}\trans \frac{\partial \mL(\btheta)}{\partial \btheta \partial
 \btheta\trans}\wt{\v}  - \mu \sum_{j\in \mM^c}\wt{\v}_{2j}\trans
\wt{\v}_{2j}\nonumber\\
&\geq& (c_{\psi}\alpha_{\min} /2- \mu)\|\wt{\v}\|_2^2 - \tau_1 \{\log(p) + N +
q\}/n\left\{\|\wt{\v}_1\|_1 +\sum_{j = 1, \ldots, d}\|\wt{\v}_{2j}
 \|_2\right\}^2. \nonumber
\ese
Furthermore, the first order condition implies
\bse
\left\{\partial \mL(\wt{\btheta})/\partial \btheta + N^{-1/2}\partial
 \sum_{j\in \mM^c} \rho_{\lambda}(\|\bg_{ j}\|_2)/\partial
\btheta\right\}
\trans (\wh{\btheta} - \wt{\btheta}) \geq 0. 
\ese
Therefore, we have
\bse
0\leq \left\{\frac{\partial \mL(\wt{\btheta})}{\partial \btheta\trans}
- N^{-1/2}\frac{\partial \sum_{j \in \mM^c} q_{\lambda} (\|\bg_{
   j}\|_2)}{\partial \btheta\trans}\right\} (\wh{\btheta} -
\wt{\btheta}) + \lambda N^{-1/2}\sum_{j \in \mM^c} \wt{\z}_j\trans
(\wh{\bg}_{ j} -\wt{\bg}_{ j}  ), 
\ese
where $\wt{\z}_j \in \partial \|N^{-1/2}\wt{\bg}_{ j}\|_2$ for $j\in
\mM^c$ and $\wt{\z}_j = \0$ for $j\in \mM$. And define $\wt{\z} = (\0_q\trans,
\wt{\z}_j\trans, j = 1, \ldots, d)\trans.$ Furthermore,
because $\wh{\btheta}$ is the interior local minimum, for $\wh{\z}_j
\in \partial \|N^{-1/2}\wt{\bg}_{j}\|_2$ for $j\in \mM^c$ and
$\wh{\z}_j = \0$ for $j \in \mM$, and let $\wh{\z} = (\0_q\trans,
\wh{\z}_j\trans, j = 1, \ldots, d)\trans$,  we have
\bse
\0 &=& \partial \mL(\wh{\btheta})/\partial \btheta + N^{-1/2}\partial \sum_{j \in \mM^c}
\rho_{\lambda}(\|\wh{\bg}_{ j}\|_2)/\partial \btheta \\
&=&  \partial \mL(\wh{\btheta})/\partial \btheta - N^{-1/2}\partial \sum_{j \in \mM^c}
q_{\lambda}(\|\wh{\bg}_{j}\|_2)/\partial \btheta + \lambda 
\wh{\z}, 
\ese
which leads to
\bse
\left\{\partial \mL(\wh{\btheta})/\partial \btheta\trans + N^{-1/2}\partial \sum_{j \in \mM^c}
\rho_{\lambda}(\|\wh{\bg}_{ j}\|_2)/\partial \btheta\trans
\right\}\wt{\v} + \lambda \wh{\z}\trans \wt{\v} = 0. 
\ese
Hence
\bse
0&\leq&  \left\{\partial \mL(\wh{\btheta})/\partial \btheta\trans + N^{-1/2}\partial \sum_{j \in \mM^c}
\rho_{\lambda}(\|\wh{\bg}_{ j}\|_2)/\partial \btheta\trans
\right\}\wt{\v} - \left\{\partial \mL(\wt{\btheta})/\partial \btheta\trans + N^{-1/2}\partial \sum_{j \in \mM^c}
\rho_{\lambda}(\|\wt{\bg}_{ j}\|_2)/\partial \btheta\trans
\right\}\wt{\v} \\
&&+ \lambda \wh{\z}\trans (\wt{\btheta} - \wh{\btheta} ) - \lambda \wt{\z}\trans (\wt{\btheta} - \wh{\btheta} ) \\
&=&  \left\{\partial \mL(\wh{\btheta})/\partial \btheta\trans + N^{-1/2}\partial \sum_{j \in \mM^c}
\rho_{\lambda}(\|\wh{\bg}_{ j}\|_2)/\partial \btheta\trans
\right\}\wt{\v} - \left\{\partial \mL(\wt{\btheta})/\partial \btheta\trans + N^{-1/2}\partial \sum_{j \in \mM^c}
\rho_{\lambda}(\|\wt{\bg}_{ j}\|_2)/\partial \btheta\trans
\right\}\wt{\v} \\
&& + \lambda N^{-1/2}\sum_{j \in \mM^c}\wh{\z}_j\trans \wt{\bG}_{\dot
 j} - \lambda N^{-1/2}\sum_{j \in \mM^c} \|\wh{\bg}_{ j}\|_2 + \lambda
N^{-1/2}\sum_{j \in \mM^c}  \wt{\z}_j\trans \wh{\bg}_{ j}, 
\ese
which implies
\be\label{eq:llg}
&& \lambda N^{-1/2}\sum_{j \in \mM^c}\|\wt{\bg}_{j}\|_2 - \lambda N^{-1/2}\sum_{j \in \mM^c}\wh{\z}_j\trans \wt{\bg}_{
 j} \nonumber\\
&\leq& \left\{\partial \mL(\wh{\btheta})/\partial \btheta\trans +N^{-1/2} \partial \sum_{j \in \mM^c}
\rho_{\lambda}(\|\wh{\bg}_{j}\|_2)/\partial \btheta\trans
\right\}\wt{\v} - \left\{\partial \mL(\wt{\btheta})/\partial \btheta\trans + N^{-1/2}\partial \sum_{j \in \mM^c}
\rho_{\lambda}(\|\wt{\bg}_{ j}\|_2)/\partial \btheta\trans
\right\}\wt{\v} \nonumber\\
&& - \lambda N^{-1/2}\sum_{j\in \mM^c}\|\wh{\bg}_{ j}\|_2 +
\lambda N^{-1/2}\sum_{j \in \mM^c} \wt{\z}_j\trans \wh{\bg}_{ j} \nonumber\\
&\leq& \left\{\partial \mL(\wh{\btheta})/\partial \btheta\trans + N^{-1/2}\partial \sum_{j \in \mM^c}
\rho_{\lambda}(\|\wh{\bg}_{ j}\|_2)/\partial \btheta\trans
\right\}\wt{\v} - \left\{\partial \mL(\wt{\btheta})/\partial \btheta\trans + N^{-1/2}\partial \sum_{j \in \mM^c}
\rho_{\lambda}(\|\wt{\bg}_{ j}\|_2)/\partial \btheta\trans
\right\}\wt{\v}  \nonumber\\
&\leq&  - (c_{\psi}\alpha_{\min} /2- \mu)\|\wt{\v}\|_2^2 + \tau_1 \{\log(p) + N +
q\}/n\left\{\|\wt{\v}_1\|_1 +\sum_{j = 1, \ldots, d}\|\wt{\v}_{2j}
 \|_2\right\}^2. 
\ee
The second inequality holds by the definition of the $\partial
\|\cdot\|_2$, and the last line holds by the Taylor expansion. Now
suppose the following relation holds, 
\bse
\|\wt{\v}_1\|_1 + \sum_{j=1}^d\|\wt{\v}_{2j}\|_2& \leq& 2 \{\sqrt{q} + \sqrt{m} + (2/\delta + 2)
\sqrt{k_0}\}\|\wt{\v}\|_2.
\ese
(\ref{eq:llg}) implies 
\bse
&&\lambda N^{-1/2}\sum_{j \in \mM^c}\|\wt{\bg}_{ j}\|_2 - \lambda N^{-1/2}\sum_{j \in \mM^c}\wh{\z}_j\trans \wt{\bg}_{  j}  \\
&\leq& \left\{4 \tau_1 (\log(p) + N + q)\{\sqrt{q} + \sqrt{m} + (2/\delta + 2)
\sqrt{k_0}\}^2 n^{-1} - (c_{\psi}\alpha_{\min} /2-
\mu)\right\}\|\wt{\v}\|_2^2. 
\ese
Now because $n\geq 8\tau_1 (\log(p) + N + q)\{\sqrt{q} + \sqrt{m} + (2/\delta + 2)
\sqrt{k_0}\}^2/ (c_{\psi}\alpha_{\min}/2 -
\mu)$, we have 
\bse
\0 &=& \lambda N^{-1/2}\sum_{j \in \mM^c}\|\wt{\bg}_{j}\|_2  - \lambda N^{-1/2}
\sum_{j \in \mM^c}\|\wt{\bg}_{j}\|_2 \leq \lambda N^{-1/2}\sum_{j \in \mM^c}\|\wt{\bg}_{ j}\|_2 - \lambda N^{-1/2}\sum_{j \in \mM^c}\wh{\z}_j\trans \wt{\bg}_{
 j} \\
&\leq& - (c_{\psi}\alpha_{\min} /2-
\mu) \|\wt{\v}\|_2^2 /2. 
\ese
The first inequality holds by the fact that $\|\wt{\z}_j\|_2\leq
1$. Therefore, we have 
\bse
\lambda \sum_{j \in \mM^c}\|N^{-1/2}\wt{\bg}_{j}\|_2   = \lambda \sum_{j \in \mM^c}\wh{\z}_j\trans N^{-1/2}\wt{\bg}_{
 j}. 
\ese
Now because $\max_{j\in (\mM\cup S)^c} \|\wh{\z}_j\|_2 <1$, we
conclude that $\wt{\bg}_{
 j} = \0$ for $j\in (\mM\cup S)^c$. Hence $\supp\{\|\wt{\bg}_{
 j}\|_2, j = 1, \ldots, d\} \subset \mM\cup S$. This would prove the
claim in the statement.

Thus, we only need to show that \bse
\|\wt{\v}_1\|_1 + \sum_{j=1}^d\|\wt{\v}_{2j}\|_2& \leq& 2 \{\sqrt{q} + \sqrt{m} + (2/\delta + 2)
\sqrt{k_0}\}\|\wt{\v}\|_2.
\ese From
(\ref{eq:llg}), we have
\be\label{eq:B23}
&& (c_{\psi}\alpha_{\min} /2- \mu)\|\wt{\v}\|_2^2 - \tau_1 \{\log(p) + N +
q\}/n\left\{\|\wt{\v}_1\|_1 +  \sum_{j = 1, \ldots, d}\|\wt{\v}_{2j}
 \|_2\right\}^2 \nonumber\\
&\leq& \left\{\partial \mL(\wt{\btheta})/\partial \btheta\trans + N^{-1/2}\partial \sum_{j \in \mM^c}
\rho_{\lambda}(\|\wt{\bg}_{ j}\|_2)/\partial \btheta\trans
\right\}\wt{\v}  -  \left\{\partial \mL(\wh{\btheta})/\partial \btheta\trans + N^{-1/2}\partial \sum_{j \in \mM^c}
\rho_{\lambda}(\|\wh{\bg}_{j}\|_2)/\partial \btheta\trans
\right\}\wt{\v} \nonumber\\
&\leq& \lambda N^{-1/2}\sum_{j \in \mM^c} \wh{\z}_j\trans
\wt{\bg}_{j} - \lambda N^{-1/2}\sum_{j \in \mM^c}  \|\wh{\bg}_{
 j}\|_2 - \lambda N^{-1/2}\sum_{j \in \mM^c}  \|\wt{\bg}_{j}\|_2  +  \lambda N^{-1/2}\sum_{j \in \mM^c} \wt{\z}_j\trans
\wh{\bg}_{ j} \nonumber\\
&=& \lambda N^{-1/2}\sum_{j \in \mM^c} \wt{\z}_j\trans
\wh{\bg}_{ j} - \lambda N^{-1/2}\sum_{j \in \mM^c}
\|\wt{\bg}_{ j}\|_2  + \lambda \wh{\z}\trans \wt{\v}. 
\ee
Now because $\supp(\{\|\wh{\bg}_{, j}\|_2, j = 1, \ldots, d\}) \subset
\mM\cup S$, we have
\be\label{eq:B24}
&& \lambda \sum_{j \in \mM^c} \wt{\z}_j\trans
\wh{\bg}_{ j} - \lambda \sum_{j \in \mM^c}
\|\wt{\bg}_{ j}\|_2 \nonumber\\
&\leq& \lambda \sum_{j \in \mM^c}
\|\wh {\bg}_{ j}\|_2 - \lambda \sum_{j \in \mM^c}
\|\wt{\bg}_{ j}\|_2 \nonumber\\
&=& \lambda \sum_{j \in S} \|\wh {\bg}_{ j}\|_2 - \lambda \sum_{j
 \in S} \|\wt {\bg}_{ j}\|_2 -\lambda \sum_{j \in (\mM\cup S)^c}
\|\wt{\bg}_{ j}\|_2 \nonumber\\
&=&  \lambda \sum_{j \in S} \|\wh {\bg}_{ j}\|_2 - \lambda \sum_{j
 \in S} \|\wt {\bg}_{ j}\|_2 -(\lambda \sum_{j \in (\mM\cup S)^c}
\|\wt{\bg}_{ j}\|_2 -\lambda \sum_{j \in (\mM\cup S)^c}
\|\wh{\bg}_{ j}\|_2 ) \nonumber \\
&=& \lambda \sum_{j \in S} \|\wt {\bg}_{ j} - \wh {\bg}_{ j}
- \wt {\bg}_{ j} \|_2 - \lambda \sum_{j
 \in S} \|\wt {\bg}_{ j}\|_2 -(\lambda \sum_{j \in (\mM\cup S)^c}
\|\wt{\bg}_{ j}\|_2 -\lambda \sum_{j \in (\mM\cup S)^c}
\|\wt {\bg}_{ j} + \wh{\bg}_{ j} - \wt {\bg}_{ j} \|_2
) \nonumber\\
&\leq& \lambda N^{1/2}\sum_{j \in S} \|\wt{\v}_j\|_2 -  \lambda N^{1/2} \sum_{j \in
 (\mM\cup S)^c } \|\wt{\v}_j\|_2. 
\ee
In addition,
\bse
\lambda \wh{\z}\trans \wt{\v} &=& \lambda \sum_{j \in S}\wh{\z}_j\trans
\wt{\v}_{2j} + \lambda \sum_{j \in (\mM\cup S)^c}\wh{\z}_j\trans
\wt{\v}_{2j}\\
&\leq& \lambda \max_{j \in S} \|\wh{\z}_j\|_2\sum_{j\in
 S}\|\wt{\v}_{2j}\|_2 + \lambda \max_{j \in (\mM\cup S)^c} \|\wh{\z}_j\|_2\sum_{j\in
 (\mM\cup S)^c}\|\wt{\v}_{2j}\|_2 \\
&\leq& \lambda \sum_{j\in
 S}\|\wt{\v}_{2j}\|_2 +  \lambda (1 - \delta)\sum_{j\in
 (\mM\cup S)^c}\|\wt{\v}_j\|_2. 
\ese
Combine with (\ref{eq:B23}) and (\ref{eq:B24}), we have
\bse
&&-\tau_1 \{\log(p) + N +
q\}/n\left\{\|\wt{\v}_1\|_1 +  \sum_{j = 1, \ldots, d}\|\wt{\v}_{2j}
 \|_2\right\}^2 \\
&\leq& (c_{\psi}\alpha_{\min} /2- \mu)\|\wt{\v}\|_2^2 - \tau_1 \{\log(p) + N +
q\}/n\left\{\|\wt{\v}_1\|_1 +  \sum_{j = 1, \ldots, d}\|\wt{\v}_{2j}
 \|_2\right\}^2\\
&\leq& \lambda \sum_{j \in S} \|\wt{\v}_{2j}\|_2 -  \lambda \sum_{j \in
 (\mM\cup S)^c } \|\wt{\v}_{2j}\|_2 +  \lambda \sum_{j\in
 S}\|\wt{\v}_{2j}\|_2 +  \lambda (1 - \delta)\sum_{j\in
 (\mM\cup S)^c}\|\wt{\v}_{2j}\|_2\\
&=& 2 \lambda \sum_{j \in S} \|\wt{\v}_{2j}\|_2- \lambda \delta \sum_{j\in
 (\mM\cup S)^c}\|\wt{\v}_{2j}\|_2. 
\ese
Now because $\delta\lambda/2 \geq 2 \tau_1 R\{\log(p) + N +
q\}/n \geq \tau_1 \{\log(p) + N +
q\}/n\left\{\|\wt{\v}_1\|_1 +  \sum_{j=1}^d\|\wt{\v}_{2j}
 \|_2\right\}$, the above display implies
\bse
-2^{-1} \delta \lambda \left\{\|\wt{\v}_1\|_1 +  \sum_{j=1}^d\|\wt{\v}_{2j}
 \|_2\right\} \leq 2 \lambda \sum_{j \in S} \|\wt{\v}_{2j}\|_2- \lambda \delta \sum_{j\in
 (\mM\cup S)^c}\|\wt{\v}_{2j}\|_2, 
\ese
which implies
\bse
2^{-1} \delta \lambda
\sum_{j\in (\mM \cup S)^c}\|\wt{\v}_{2j}\|_2  
&\leq& 2^{-1} \delta \lambda \|\wt{\v}_1\|_1 +  2^{-1} \delta \lambda
\sum_{j\in \mM}\|\wt{\v}_{2j}\|_2 +  (2 + 2^{-1} \delta) \lambda
\sum_{j\in S}\|\wt{\v}_{2j}\|_2 . 
\ese
Therefore,
\bse
\|\wt{\v}_1\|_1 + \sum_{j=1}^d\|\wt{\v}_{2j}\|_2& \leq& 
\|\wt{\v}_1\|_1 + \sum_{j\in \mM} \|\wt{\v}_{2j}\|_2 + \sum_{j\in S}
\|\wt{\v}_{2j}\|_2 + \sum_{j\in (\mM\cup S)^c}
\|\wt{\v}_{2j}\|_2\\
&\leq&  \|\wt{\v}_1\|_1 + \sum_{j\in \mM} \|\wt{\v}_{2j}\|_2 + \sum_{j\in S}
\|\wt{\v}_{2j}\|_2  + \|\wt{\v}_1\|_1 + \sum_{j\in \mM}
\|\wt{\v}_{2j}\|_2 +(4 + \delta)/\delta \sum_{j\in
  S}\|\wt{\v}_{2j}\|_2\\
&\leq& 2 \{\sqrt{q} + \sqrt{m} + (2/\delta + 2)
\sqrt{k_0}\}\|\wt{\v}\|_2. 
\ese
This completes the proof. \qed

\begin{Lem}\label{lem:strict}
  Assume Conditions (A1)--(A6) and and \ref{con:boundpsi} --
 \ref{con:secondorder} hold, assume $\mu <c_{\psi}\alpha_{\min}/2$ and
  $ n > \tau_1 \{\log(p) + N +
q\} (q + m + k_0 )   /(c_{\psi}\alpha_{\min} /2- \mu)  $. Then the function $
\mL(\btheta) + N^{-1/2}\sum_{j\in \mM^c}q_{\lambda}(\|\bg_{ j}\|_2) $
and $ \mL(\btheta) - \mu N^{-1/2}\sum_{j\in \mM^c} \|\bg_{ j}\|_2/2$ are strictly convexity over the set $[\btheta: \btheta = \{\ba\trans,
N^{-1/2}\vec(\bG)\trans\}\trans;   \bg_{  j} = \0 ~\forall j \in
(\mM\cup S)^c]$. 
\end{Lem}
\noindent Proof: First define a vector $\v = (\v_1\trans, \v_{2j}\trans, j = 1,
\ldots, d)\trans$ with $\|\v_{2j}\|_2 >0$ for $j \in \mM\cup S$ and
$\|\v_{2j}\|_2 = 0$, otherwise. By Lemma \ref{lem:Re}, we have for
$\btheta$ in the feasible set of program (\ref{eq:or}), we have
\bse
\v\trans \partial \mL(\btheta)/\partial \btheta\partial
\btheta\trans\v \geq c_{\psi}\alpha_{\min} \|\v\|_2^2 - \tau_1 \{\log(p) + N +
q\} /n \left\{\|\v_1\|_1 + \sum_{j=1}^d\|\v_{2j}\|_2^2\right\}^2. 
\ese
Therefore, we have
\bse
\v\trans \partial \mL(\btheta)/\partial \btheta\partial
\btheta\trans\v \geq \left\{c_{\psi}\alpha_{\min} - \tau_1 \{\log(p) + N +
q\} (q + m + k_0 ) /n \right\}\|\v_{2j}\|_2^2, 
\ese
and
\bse
\v\trans \partial \mL(\btheta)/\partial \btheta\partial
\btheta\trans\v  - \mu \|\v\|_2^2 \geq \left\{c_{\psi}\alpha_{\min} - \mu - \tau_1 \{\log(p) + N +
q\} (q + m + k_0 ) /n \right\}|\v_{2j}\|_2^2. 
\ese
Now because $\left\{c_{\psi}\alpha_{\min} - \mu - \tau_1 \{\log(p) + N +
q\} (q + m + k_0 ) /n \right\} >0$, $\mL(\btheta) - \mu \sum_{j\in \mM^c}\|\bg_{ j}\|_2^2/2
$ is strictly convex on the set that $[\btheta: \btheta = \{\ba\trans,
\vec(\bG)\trans\}\trans;   \bg_{ j} = \0 ~\forall j \in
(\mM\cup S)^c]$.
Finally, because
\bse
&& \mL(\btheta) - N^{-1/2}\sum_{j\in \mM^c}q_{\lambda}(\|\bg_{
  j}\|_2) \\
&=& 
\mL(\btheta) -\mu N^{-1/2} \sum_{j\in \mM^c}\|\bg_{ j}\|_2^2/2 +
\left\{\mu N^{-1/2}\sum_{j\in \mM^c}\|\bg_{ j}\|_2^2/2 - N^{-1/2} \sum_{j\in
  \mM^c}q_{\lambda}(\|\bg_{ j}\|_2)\right\}, 
\ese
where the second part is convex by Lemma \ref{lem:fromlemm5support}. Therefore, $
\mL(\btheta) -N^{-1/2} \sum_{j\in \mM^c}q_{\lambda}(\|\bg_{ j}\|_2) $
and $ \mL(\btheta) +N^{-1/2}  \sum_{j\in \mM^c}\rho_{\lambda}(\|\bg_{
  j}\|_2)  = \mL(\btheta) - N^{-1/2}\sum_{j\in \mM^c}q_{\lambda}(\|\bg_{
  j}\|_2)  + \lambda N^{-1/2}\sum_{j\in \mM^c}\|\bg_{
  j}\|_2$ are strictly convex over the set $[\btheta: \btheta = \{\ba\trans,
\vec(\bG)\trans\}\trans;   \bg_{ j} = \0 ~\forall j \in
(\mM\cup S)^c]$. This proves the result. \qed

Suppose the indices of  nonzero $\bb_{j}(\cdot), j \in \mM^c$ are given, we
minimize
\be\label{eq:or}
&& \mL
(\ba,  \bG) + N^{-1/2}\sum_{j \in \mM^c} \rho_{\lambda}(\|\bg_{
  j}\|_2), \nonumber\\
&& \text{ such that } \|\ba\|_1 + N^{-1/2}\sum_{j=1}^d \|\bg_{
  j}\|_2 \leq R, \text{ and } \bg_{ j} = \0 \text{ for } j \in
(\mM\cup S)^c, 
\ee
with respect to $\ba, \bG$ to obtain the estimator, which named oracle
estimator.

In Theorem~\ref{th:2}, we show that the minimizer of (\ref{eq:lapHT})  
is unique under some conditions. Moreover, it approaches the oracle estimator, 
i.e., 
the estimator of (\ref{eq:or}) when the indices of nonzero $\bb_{j}(\cdot), j \in \mM^c$ are given.

\begin{Th} \label{th:2}Assume Conditions (A1)--(A6) and
  and \ref{con:boundpsi} --
   \ref{con:secondorder} hold. Further assume that $\mu
   <c_{\psi}\alpha_{\min}/2$ and  $24 \lambda (q + m + k_0)/  (c_{\psi}\alpha_{\min}
 -3\mu/2 ) \leq R$, $$\lambda \geq
   \max\left\{2 \tau_1 R \log(p)/n,  C_0 \sqrt{\log\{\max(p, n)\}/n} +
     C k_0^{1/2} N^{-\omega}\right\}/4. $$ 
  Let $\wh{\btheta}_{\mM\cup S} = \{\wh\ba\trans,  N^{-1/2}\wh\bg_{
      j}\trans, j \in \mM \cup S\}$ be the minimizer for
    (\ref{eq:or}). Without loss of generality, we define $\wh{\btheta} = \{\wh\ba\trans,  N^{-1/2}\wh\bg_{
      j}\trans, j \in \mM\cup S, \0_{d - m -
      k_0}\trans\}\trans$, where we assume the first $q+
    (m + k_0)N$ elements are nonzero,  and satisfies
 \be\label{eq:construct}
 \frac{\partial \mL(\wh{\btheta})}{\partial \btheta} - N^{-1/2}\frac{\partial
  \sum_{j\in \mM^c} q_{\lambda}(\|\wh{\bG}_{\cdot j}\|_2)}{\partial
  \btheta} +\lambda \wh{\z} = \0, 
 \ee
 where $\wh{\z} =(\0_q, \wh{\z}_j\trans, j = 1, \ldots, d)\trans$, and  $\wh{\z}_j \in
   \partial \|N^{-1/2}\wh{\bG}_{\cdot j}\|_2$
 Assume there is a $\delta \in
   [4R\tau_1\log(p)/n/\lambda, 1]$, and $\max_{j \in (\mM\cup
     S)^c}\|\wh{\z}_j\|_2\leq 1- \delta$ and   \bse
   n&\geq& \max\left[8\tau_1 (\log(p) + N + q)\{\sqrt{q} + \sqrt{m} + (2/\delta + 2)
  \sqrt{k_0}\}^2/ (c_{\psi}\alpha_{\min} /2-
  \mu),\right.\\
  &&\left.\tau_1 \{\log(p) + N +
  q\} (q + m + k_0 )   /(c_{\psi}\alpha_{\min} /2- \mu) \right].
  \ese 
 Then (\ref{eq:lapHT}) has a unique minimizer $\wh \btheta$. 
  \end{Th}

\noindent {\bf Proof:}  {\bf \noindent Step 1:} By Lemma \ref{lem:strict}, $\wh{\btheta}_{\mM \cup S}$ be
the minimizer for (\ref{eq:or}). It is easy to see that the condition in
Theorem \ref{th:1} holds and $\sum_{j \in (\mM\cup
  S)^c}\|\wh{\bG}_{\cdot j} - \bg_{0 j}\|_2 \leq 3 \sum_{j \in
  \mM^c\cap S}\|\wh{\bG}_{\cdot j} - \bg_{0 j}\|_2$ because $\sum_{j \in (\mM\cup
  S)^c}\|\wh{\bG}_{\cdot j} - \bg_{0j}\|_2 = 0$. Therefore,
applying Theorem \ref{th:1}, we have
\bse
\|\wh{\ba}\|_1 +  N^{-1/2} \sum_{j = 1}^d \|\wh{\bG}_{\cdot j}\|_2
&\leq& 12 \lambda (q + m + k_0)/  (c_{\psi}\alpha_{\min}
-3\mu/2 ) + \|\ba_0\|_1 +  N^{-1/2}\sum_{j = 1}^d \|{\bG}_{0\cdot j}\|_2\\
&\leq& R, 
\ese
and hence $\wh{\btheta}_{\mM\cup S}$ is in the feasible region. The
second inequality holds by the assumption that $24 \lambda (q + m + k_0)/  (c_{\psi}\alpha_{\min}
-3\mu/2 ) \leq R$ in the theorem statement.

{\bf \noindent Step 2:} We show that $\wh{\btheta}$ is a local minimum
 (\ref{eq:lapHT}) by verifying the conditions in Lemma
 \ref{lem:convexF} in the supplementary material. 
Because $$\mL(\btheta) +N^{-1/2}\sum_{j \in \mM^c} \rho_{\lambda}\|\bg_{
 j}\|_2 = \mL(\btheta) -N^{-1/2}\sum_{j \in \mM^c} q_{\lambda}\|\bg_{
 j}\|_2 + \lambda N^{-1/2}\sum_{j \in \mM^c} \|\bg_{ j}\|_2, $$
we can write $f = \mL$, $g = q_{\lambda}$ and $(\x^*, \v^*, \w^*,
\mu_1^*) = (\wh{\btheta}, \wh{\z}, \wh{\z}_1, 0)$, where $\wh{\z} =
(\0_q\trans, \wh{\z}_j\trans, j
= 1, \ldots, d)\trans$ and $\wh{\z}_j\in \partial \|N^{-1/2}\wh{\bG}_{\cdot
 j}\|_2$ and $\wh{\z}_1 \in \partial \{\|\wh{\ba}\|_1 + N^{-1/2}\sum_{j =
 1}^d \|\wh{\bg}_{ j}\|_2\}$. Lemma \ref{lem:fromlemm5support} in the supplementary
material ensures the concavity and differentiability of $g(\|\x\|_2)
- \mu/2\|\x\|_2^2$ in $\|\x\|_2$ and $\x$, respectively. Furthermore,
since $\mu_1^*= 0$, (\ref{eq:1}) is satisfied. (\ref{eq:2}) is
satisfied by our construction in (\ref{eq:construct}). Therefore, it remains to
verify (\ref{eq:3}). We first show that if $\s \in G^*$, $\s_{2j} =
0$ for $j \in (\mM \cup S)^c$, and therefore (\ref{eq:2}) only need to
be satisfied for the vectors in the form of $\s = (\s_1\trans,
\s_{2j}\trans, j \in \mM\cup S, \0\trans_{d - m - k_0})\trans$.  Suppose it does not hold so that there is a $\bnu
\in G^*$ such that $\bnu = (\bnu_1\trans,
\bnu_{2j}\trans,  j \in \mM\cup S, \0\trans_{d - m - k_0})\trans$.  We
define $\z' = (\0_q\trans, \z_{2j}^{'\rm T}, j = 1,
\ldots, d)$ such that $\z'_{2k} = \wh{\z}_{2k}$ for  $k\neq j$, and
$\z'_{2k} = \partial \|\bnu_{2k}\|_2$ for $k = j$. Clearly $\z'_{2k} \in \partial
\|N^{-1/2}\wh{\bg}_{ k}\|_2$, $k = 1, \ldots, d$, $\|\z'_{2j}\| = 1$ because $\bnu_{2j}>\0$,  and
\bse
\lambda \bnu\trans \z' >\lambda \bnu\trans \wh{\z}, 
\ese
because $\max_{j \in (\mM \cup S)^c}\|\wh{\z}_{2j}\|_2 <1$. Therefore,
\bse
&& \bnu\trans \left[\frac{\partial \mL(\wh{\btheta})}{\partial \btheta} -
 N^{-1/2}\frac{\partial \sum_{j \in \mM^c} q_{\lambda}(\|\bg_{
     j}\|_2)}{\partial \btheta}\right] +\lambda \bnu\trans \z'\\
&>&\bnu\trans \left[\frac{\partial \mL(\wh{\btheta})}{\partial \btheta} -
N^{-1/2}  \frac{\partial \sum_{j \in \mM^c} q_{\lambda}(\|\bg_{
     j}\|_2)}{\partial \btheta}\right] + \lambda \bnu\trans \wh{\z}\\
&=& 0, 
\ese
which contradicts with the requirement of $G^*$ that
\bse
\sup_{(\v_j \in \partial \|\wh{\bg}_{j}\|_1, j \in \mM^c)} \bnu\trans \left[\frac{\partial \mL(\wh{\btheta})}{\partial \btheta} -
 N^{-1/2} \frac{\partial \sum_{j \in \mM^c} q_{\lambda}(\|\bg_{
     j}\|_2)}{\partial \btheta}\right] + \lambda  \sum_{j \n \mM^c}
\bnu_{2j}\trans \v_j= 0. 
\ese
Therefore, we conclude that  if $\s \in G^*$, $\s_{2j} =
\0$ for $j \in (\mM \cup S)^c$. Now by Lemma \ref{lem:strict} in the
supplementary material and the
fact that $\wh{\bg}_{ j} = \0$ for $j\in (\mM\cup S)^c$, we
conclude that (\ref{eq:3}) holds by setting $\kappa = \mu$. Hence all
conditions of Lemma \ref{lem:convexF} in the supplementary material
are satisfied, so we conclude $\wh{\btheta}$ is an isolated local
minimzer of (\ref{eq:lapHT}).

Now by Lemma \ref{eq:minandstation}, because $\wh{\bg}_{j}= \0$ for $j \in (\mM
\cup S)^c$ and $\wh{\btheta}$ is an interior minimizer,
$\wt{\bg}_{ j}= \0$ for $j \in (\mM
\cup S)^c$
for any given stationary point $\wt{\btheta}$  of (\ref{eq:lapHT}). Hence, we write any stationary point in the form of
$\wt{\btheta} = (\wt{\ba}_1\trans, N^{-1/2}\wt{\bg}_{ j}\trans, j \in
\mM\cup S, \0_{d - m - k_0}\trans)\trans$, where
$\wt{\btheta}_{\mM\cup S} = (\wt{\ba}_1\trans,  N^{-1/2}\wt{\bg}_{ j}\trans, j \in
\mM\cup S)\trans$ is a stationary point for
(\ref{eq:or}). Furthermore, (\ref{eq:or}) is strictly convex by Lemma
\ref{lem:strict}, and $\wt{\btheta}_{\mM\cup S} $ is unique in the
feasible set and $\wh{\btheta}_{\mM\cup S}$ and $\wt{\btheta}_{\mM\cup
 S}$ are unique. Therefore, $\wh{\btheta}$ is the unique local
minimum. This proves the result.

Let $\btheta^*$ is the point on the line connecting $\wh{\btheta}$ and
$\btheta_0$. We write
\be\label{eq:first0}
\wh{\Q} (\btheta^*)= \left\{\begin{array}{cc}\wh{\Q}_{(\mM \cup S)(\mM \cup S)
                  }(\btheta^*)&\wh{\Q}_{(\mM \cup S)  (\mM \cup S) ^c}(\btheta^*)\\
\wh{\Q}_{(\mM \cup S)^c  (\mM \cup S)}(\btheta^*)& \wh{\Q}_{(\mM \cup S)^c
                                              (\mM \cup S)^c}
                                              (\btheta^*)\end{array}\right\}. 
                                          \ee

Then by the construction in (\ref{th:2}), let $\wh{\btheta} =
(\wh{\btheta}_{\mM \cup S}\trans,{\bf 0}_{(\mM \cup S)^c}\trans)\trans $, $\wh{\btheta}_{\mM \cup S}$ is the minimizer for
(\ref{eq:or}), then  we have
\bse
&& \wh{\Q} (\btheta^*)(\wh{\btheta} - \btheta_0)+ \left[\begin{array}{l}\left\{\frac{\partial
                              \mL(\btheta_0)}{\partial \btheta}\right\}_{\mM \cup S} -
                              \left\{\frac{\partial \sum_{j \in \mM^c}
                              q_{\lambda}(\|\bg_{0 j}\|_2)}{\partial
                              \btheta }\right\}_{\mM \cup S}\\
\left\{\frac{\partial
                              \mL(\btheta_0)}{\partial \btheta}\right\}_{2(\mM \cup S)^c} -
                              \left\{\frac{\partial \sum_{j \in \mM^c}
                              q_{\lambda}(\|\bg_{0 j}\|_2)}{\partial
                              \btheta }\right\}_{2(\mM \cup S)^c}\end{array}\right]\\
&& +\lambda \left\{
                              \begin{array}{l}\wh{\z}_{\mM
                                \cup S }\\ \wh{\z}_{2(\mM
                                \cup S)^c} \end{array}\right\} = {\bf
                            0}.
                          \ese
                          Taking the upper $q + m+k_0$ non-zero component, we get
\be\label{eq:diffbb1} 
&& \wh{\btheta}_{\mM \cup S} -  \btheta_{ 0\mM \cup S}\nonumber\\
&=& \{\wh{\Q}_{\mM \cup S,
 \mM\cup S}(\btheta^*)\}^{-1}\left[- \left\{\frac{\partial
                              \mL(\btheta_0)}{\partial \btheta}\right\}_{\mM
                            \cup S} + \left\{\frac{\partial
                           \sum_{j \in \mM^c}   q_{\lambda}(\|\wh{\bg}_{ j}\|_2)}{\partial
                              \btheta}\right\}_{\mM \cup S}
                          -\lambda \wh{\z}_{\mM\cup
                            S} \mystrut\right],
\ee
while taking the lower $d-m-k_0$ components, this leads to
\bse
\wh{\z}_{(\mM \cup S)^c} &=&  \lambda^{-1}  \left[ \left\{\frac{\partial
                            \sum_{j \in \mM^c}
                            q_{\lambda}(\|\wh{\bg}_{j}\|_2)}{\partial
                              \btheta}\right\}_{2(\mM \cup
                                    S)^c} - \left\{\frac{\partial
                              \mL(\btheta_0)}{\partial \btheta_0}\right\}_{2(\mM
                            \cup S)^c} \right] \\
&&- \lambda^{-1}\wh{\Q}_{(\mM \cup
                          S)^c  (\mM \cup S)}(\btheta^*) (\wh{\btheta}_{\mM \cup S} -
                        \btheta_{ \mM\cup S}) \\
&=& \lambda^{-1}  \left[ \left\{ \frac{\partial
                           \sum_{j \in \mM^c}
                            q_{\lambda}(\|\wh{\bg}_{ j}\|_2)}{\partial
                              \btheta}\right\}_{2(\mM \cup
                                    S)^c} - \left\{\frac{\partial
                              \mL(\btheta_0)}{\partial \btheta}\right\}_{2(\mM
                            \cup S)^c} \right] \\
&&+ \lambda^{-1} \wh{\Q}_{(\mM \cup
                          S)^c  (\mM \cup S)}(\btheta^*)\{\wh{\Q}_{(\mM \cup
                          S)(\mM \cup S) }(\btheta^*)\}^{-1}\\
&&\times \left(\left[\left\{\frac{\partial
                              \mL(\btheta_0)}{\partial \btheta}\right\}_{\mM \cup S} -
                             \left\{\frac{\partial \sum_{j \in \mM^c} 
                              q_{\lambda}(\|\wh{\bg}_{ j}\|_2)}{\partial
                              \btheta}\right\}_{\mM \cup
                            S}\right] +
                          \lambda \wh{\z}_{\mM \cup
                            S} \right).
\ese
Further by Condition (A4), we have
\bse
\left\{ \frac{\partial
                              \sum_{j \in \mM^c} q_{\lambda}(\|\wh{\bg}_{ j}\|_2)}{\partial
                              \btheta}\right\}_{2(\mM \cup
                                    S)^c}
&=& \left[\left\{ \frac{\partial \lambda \|\wh{\bg}_{ j}\|_2}{\partial
                              \bg_{ j}\trans }\right\} -  \left\{\frac{\partial
                              \rho_{\lambda}(\|\wh{\bg}_{ j}\|_2) }{\partial
                              \bg_{ j}\trans}\right\}, j \in (\mM \cup
                          S)^c \right ]\trans\\
&=& \bf 0.
\ese

Therefore, we have
\bse
\wh{\z}_{2(\mM \cup S)^c} &=& \lambda^{-1}  \left[ - \left\{\frac{\partial
                              \mL(\btheta_0)}{\partial \btheta}\right\}_{2(\mM
                            \cup S)^c} \right] + \lambda^{-1} \wh{\Q}_{(\mM \cup
                          S)^c  (\mM \cup S)}(\btheta^*)\{\wh{\Q}_{(\mM \cup
                          S)(\mM \cup S) }(\btheta^*)\}^{-1}\\
&&\times \left(\left[\left\{\frac{\partial
                              \mL(\btheta_0)}{\partial \btheta}\right\}_{\mM \cup S} -
                            \left\{ \frac{\partial
                             \sum_{j \in \mM^c}
                             q_{\lambda}(\|\wh{\bg}_{ j}\|_2)}{\partial
                              \btheta}\right\}_{\mM \cup
                            S}\right] +
                          \lambda \wh{\z}_{\mM \cup S} \right).
\ese
Hence, to use Theorem \ref{th:1}, we must show that $\max_{j \in
 (\mM\cup S)^c}\|\wh{\z}_j\|_2 <1$.
\qed

\noindent {\bf Proof of Theorem \ref{th:3}:} 
First of all $\wh{\btheta}_{2(\mM \cup S)^c } = \0$ by
construction. So for any unit vector $\v = (\v_1, \v_{2j}\trans, j =
1, \ldots, d)$, $\w = (\w_1\trans, \w_{2j}\trans, j = 1, \ldots,
d)\trans$, recall that
\bse
\v \frac{\partial^2 \mL(\btheta)}{\partial \btheta\partial
 \btheta\trans}\w = \v\trans\sumi \psi''_i(\ba, \bG) \left(\left[\Z_i\trans, N^{1/2}\left\{\int_{0}^1\X_i (s) \otimes
 \B(s)  ds\right\}\trans\right]\trans\right)^{\otimes2}\w
\ese
Denote $\nabla ^3 \mL(\bb)$ to be the third order
gradient of $\mL$, we have
\bse
&& \v \nabla ^3 \mL(\bb)\w\\
&=&n^{-1}\v\trans\sumi \psi''_i(\ba, \bG) \left(\left[\Z_i\trans, N^{1/2}\left\{\int_{0}^1\X_i (s) \otimes
     \B(s)  ds\right\}\trans\right]\trans\right)^{\otimes2}\w \\
&&\times \left(\left[\Z_i\trans, N^{1/2}\left\{\int_{0}^1\X_i (s) \otimes
 \B(s)  ds\right\}\trans\right]\trans\right). 
\ese
Here we consider vector $\v, \w$ such that $\|\v_{2j}\|_2 >0,
\|\w_{2j}\|_2 >0$ for $j \in \mM\cup S$ and $\|\v_{2j}\|_2 = 0,
\|\w_{2j}\|_2 = 0$  otherwise. First by Theorem \ref{th:1}, we have
\bse
\|\wh{\btheta} - \btheta_0\|_2  \leq  3\lambda \sqrt{q + m + k_0}/ (c_{\psi}\alpha_{\min}
-3\mu/2 ). 
\ese
Let $\mathbb{K} \equiv \{\v = (\v_1\trans, \v\trans_{2j}, j = 1,
\ldots, d)\trans, \v_{2j} = \0 \text{ for } j \in (\mM\cup S)^c,
\|\v\|_2\leq 1\}$, then there are constants $a_1>0, a_2>0$ such that 
\be\label{eq:Qupp}
&& \sup_{\v, \w \in \mathbb{K}} |\v\trans \left\{\wh{\Q}(\btheta^*) - \frac{\partial^2 \mL(\btheta_0)}{\partial \btheta\partial
   \btheta\trans} \right\}\w|\nonumber\\
&\leq& \sup_{\v, \w \in \mathbb{K}} |n^{-1}\v\trans\sumi \psi'''_i(\ba^{**}, \bG^{**}) \left(\left[\Z_i\trans, N^{1/2}\left\{\int_{0}^1\X_i (s) \otimes
     \B(s)  ds\right\}\trans\right]\trans\right)^{\otimes2}\w \nonumber\\
&&\times \left(\left[\Z_i\trans, N^{1/2}\left\{\int_{0}^1\X_i (s) \otimes
     \B(s)  ds\right\}\trans\right]\right) (\wh{\btheta} - \btheta_0)
|\nonumber\\
&\leq& \{3\lambda \sqrt{q + m + k_0}/ (c_{\psi}\alpha_{\min}
-3\mu/2 )\}\nonumber\\
&&\times \sup_{\v, \w \in \mathbb{K}}\bigg |n^{-1}\v\trans\sumi |\psi'''_i(\ba^{**}, \bG^{**})| \left(\left[\Z_i\trans, N^{1/2}\left\{\int_{0}^1\X_i (s) \otimes
     \B(s)  ds\right\}\trans\right]\trans\right)^{\otimes2}\w \nonumber\\
&&\times |\left(\left[\Z_i\trans, N^{1/2}\left\{\int_{0}^1\X_i (s) \otimes
     \B(s)  ds\right\}\trans\right]\right) (\wh{\btheta} - \btheta_0) / \{3\lambda \sqrt{q + m + k_0}/ (c_{\psi}\alpha_{\min}
-3\mu/2 )\}|\bigg|\nonumber\\
&\leq &(a_1 + \log(p)\sqrt{q + m + k_0}/n) \{3\lambda \sqrt{q + m + k_0}/ (c_{\psi}\alpha_{\min}
-3\mu/2 )\} \nonumber\\
&= & O_p\left(\sqrt{q + m + k_0} [\sqrt{\log\{\max(p, n)\}/n}
 +k_0^{1/2} N^{-\omega} ]\right)
\ee
with probability greater than $1 - 2 \exp\left\{-  a_2 n\min \left(
  c_{\psi}^2 \alpha_{\min}^2/54 ^2, 1\right)/2\right\} - 20\max(p, n)^{-1}$,
where $\btheta^{**} \equiv (\ba^{**\trans}, \vec(\bG^{**})\trans)\trans$ is
on the line connecting $\btheta^*$ and $\btheta_0$. Here the second to the last line holds by using the same arguments as
those lead to Theorem \ref{th:2} and the fact that $|\psi_i'''(\ba,
\bG)|>0$ by Condition \ref{con:boundpsi}, and $|\left(\left[\Z_i\trans, N^{1/2}\left\{\int_{0}^1\X_i (s) \otimes
     \B(s)  ds\right\}\trans\right]\right) (\wh{\btheta} - \btheta_0) / \{3\lambda \sqrt{q + m + k_0}/ (c_{\psi}\alpha_{\min}
-3\mu/2 )\}| = O_p(1)$. Furthermore, by Lemma \ref{lem:K2s}, we have 
\be\label{eq:Qupptrue}
&& \sup_{\v, \w \in \mathbb{K}} \v\trans\left\{\frac{\partial^2 \mL(\btheta_0)}{\partial \btheta\partial
   \btheta\trans}  - \Q(\btheta_0)\right\}\w\nonumber\\
&=& \sqrt{\log(\max(p, n))/n}\nonumber\\
&=& o_p(\sqrt{q + m + k_0} [\sqrt{\log\{\max(p, n)\}/n}
 +k_0^{1/2} N^{-\omega} ]), 
\ee
with probability greater than $1 - a_3\max(p, n)^{-1}$ for
some constants $a_3 >0$. Now define $\mathbb{K}_1 \equiv \{\v = (\v_1\trans, \v\trans_{2j}, j = 1,
\ldots, d)\trans, \|\v_{2j}\|_2 > \0 \text{ for only one $j$ },
\|\v\|_2\leq 1\}$, similar to (\ref{eq:Qupp}) there are constants $a_4, a_5 >0$ such that
\be\label{eq:Qlower}
&& \sup_{\v\in \mathbb{K}_1, \w \in \mathbb{K}} |\v\trans \left\{\wh{\Q}(\btheta^*) - \frac{\partial^2 \mL(\btheta_0)}{\partial \btheta\partial
   \btheta\trans} \right\}\w|\nonumber\\
&\leq& \sup_{\v\in \mathbb{K}_1, \w \in \mathbb{K}} |n^{-1}\v_1\trans\sumi \psi'''_i(\ba^{**}, \bG^{**}) \left(\left[\Z_i\trans, N^{1/2}\left\{\int_{0}^1\X_i (s) \otimes
     \B(s)  ds\right\}\trans\right]\trans\right)^{\otimes2}\w \nonumber\\
&&\times \left(\left[\Z_i\trans, N^{1/2}\left\{\int_{0}^1\X_i (s) \otimes
     \B(s)  ds\right\}\trans\right]\right) (\wh{\btheta} - \btheta_0)
|\nonumber\\
&\leq& \{3\lambda \sqrt{q + m + k_0}/ (c_{\psi}\alpha_{\min}
-3\mu/2 )\}\nonumber\\
&&\times \sup_{\v\in \mathbb{K}_1, \w \in \mathbb{K}}\bigg |n^{-1}\v\trans\sumi |\psi'''_i(\ba^{**}, \bG^{**})| \left(\left[\Z_i\trans, N^{1/2}\left\{\int_{0}^1\X_i (s) \otimes
     \B(s)  ds\right\}\trans\right]\trans\right)^{\otimes2}\w \nonumber \\
&&\times |\left(\left[\Z_i\trans, N^{1/2}\left\{\int_{0}^1\X_i (s) \otimes
     \B(s)  ds\right\}\trans\right]\right) (\wh{\btheta} - \btheta_0) / \{3\lambda \sqrt{q + m + k_0}/ (c_{\psi}\alpha_{\min}
-3\mu/2 )\}|\bigg|\nonumber\\
&\leq &\{3\lambda \sqrt{q + m + k_0}/ (c_{\psi}\alpha_{\min}
-3\mu/2 )\}/4 \nonumber\\
&&\times \sup_{\v\in \mathbb{K}_1, \w \in \mathbb{K}}\bigg |n^{-1}(\v +
\w) \trans\sumi |\psi'''_i(\ba^{**}, \bG^{**})| \left(\left[\Z_i\trans, N^{1/2}\left\{\int_{0}^1\X_i (s) \otimes
     \B(s)  ds\right\}\trans\right]\trans\right)^{\otimes2}(\v + \w) \nonumber \\
&&\times |\left(\left[\Z_i\trans, N^{1/2}\left\{\int_{0}^1\X_i (s) \otimes
     \B(s)  ds\right\}\trans\right]\right) (\wh{\btheta} - \btheta_0) / \{3\lambda \sqrt{q + m + k_0}/ (c_{\psi}\alpha_{\min}
-3\mu/2 )\}|\bigg|\nonumber\\
&& + \{3\lambda \sqrt{q + m + k_0}/ (c_{\psi}\alpha_{\min}
-3\mu/2 )\}/4 \nonumber\\
&&\times \sup_{\v\in \mathbb{K}_1, \w \in \mathbb{K}}\bigg |n^{-1}(\v -
\w) \trans\sumi |\psi'''_i(\ba^{**}, \bG^{**})| \left(\left[\Z_i\trans, N^{1/2}\left\{\int_{0}^1\X_i (s) \otimes
     \B(s)  ds\right\}\trans\right]\trans\right)^{\otimes2}(\v -\w) \nonumber \\
&&\times |\left(\left[\Z_i\trans, N^{1/2}\left\{\int_{0}^1\X_i (s) \otimes
     \B(s)  ds\right\}\trans\right]\right) (\wh{\btheta} - \btheta_0) / \{3\lambda \sqrt{q + m + k_0}/ (c_{\psi}\alpha_{\min}
-3\mu/2 )\}|\bigg|\nonumber\\
&\leq& [a_4 + \{\log(p) + q + N\}\sqrt{q + m + k_0 + 1}/n]\{3\lambda \sqrt{q + m + k_0}/ (c_{\psi}\alpha_{\min}
-3\mu/2 )\} \nonumber\\
&=& O_p\left(\sqrt{q + m + k_0} [\sqrt{\log\{\max(p, n)\}/n}
 +k_0^{1/2} N^{-\omega} ]\right), 
\ee
with probability $1 - 2 \exp\left\{-  a_5 n\min \left(
  c_{\psi}^2 \alpha_{\min}^2/54 ^2, 1\right)/2\right\} - 20\max(p,
n)^{-1}$. Similar to (\ref{eq:Qupptrue})
\be\label{eq:Qlowertrue}
&& \sup_{\v \in \mathbb{K}_1, \w \in \mathbb{K}} \v\trans\left\{\frac{\partial^2 \mL(\btheta_0)}{\partial \btheta\partial
   \btheta\trans}  - \Q(\btheta_0)\right\}\w\nonumber\\
&=& o_p(\sqrt{q + m + k_0} [\sqrt{\log\{\max(p, n)\}/n}
 +k_0^{1/2} N^{-\omega} ]). 
 \ee
 Combine (\ref{eq:Qupp}) and (\ref{eq:Qupptrue}) and Lemma \ref{lem:fromlemma11loh2017} in
 the supplementary material, we have
 \be\label{eq:QQL2diff}
 && \|\wh{\Q}_{\mM\cup S, \mM\cup S}(\btheta^*)^{-1} - {\Q}_{\mM\cup
   S, \mM\cup S}(\btheta_0)^{-1}\|_2\nonumber\\
 &\leq& O_p(\wh{\Q}_{\mM\cup S, \mM\cup S}(\btheta^*)^{-1} - {\Q}_{\mM\cup
   S, \mM\cup S}(\btheta_0)^{-1})\nonumber\\
 &=& O_p\left(\sqrt{q + m + k_0} [\sqrt{\log\{\max(p, n)\}/n}
 +k_0^{1/2} N^{-\omega} ]\right), 
\ee
with probability $1 - \exp\left\{- O(n)\right\} - O\{\max(p,
n)^{-1}\}$, and therefore
\be\label{eq:supQ}
&& \sup_{j \in \mM\cup S} \sum_{k \in S} \|\E_j\trans[\{\wh{\Q}_{\mM \cup S,
 \mM\cup S}(\btheta^*)\}^{-1}  - \{{\Q}_{\mM \cup S,
 \mM\cup S}(\btheta^*)\}^{-1}] \E_k \|_2 \nonumber\\
&=&  O_p\left\{(q + m + k_0)[\sqrt{\log\{\max(p, n)\}/n}
 +k_0^{1/2} N^{-\omega} ]\right\}\nonumber\\
&=& o_p(1), 
\ee
where $\E_j$ is 
a matrix so that $\E_j\trans(\wh{\btheta} - \btheta_0) = N^{-1/2}\wh{\bg}_{ j}
-N^{-1/2} {\bG}_{0\cdot j}$. 

Furthermore, let $\mathbb{K}_2 \equiv \{\v = (\v_1\trans, \v\trans_{2j}, j = 1,
\ldots, d)\trans, \|\v_{2j}\|_2 > \0 \text{ for only one $j \in
 \mM\cup S$ }, \|\v_{2j}\|_2 = \0 \text{ for $j \in   (\mM\cup S)^c$ } 
\|\v\|_2\leq 1\}, $ then 
\be
&& \sup_{\v \in \mathbb{K}_1}\bigg|\v_{\mM\cup S}\trans\wh \Q_{(\mM \cup S)^c, \mM \cup S} (\btheta^*) \{\wh \Q_{(\mM \cup S), \mM \cup
 S}(\btheta^*)\}^{-1}\left\{\frac{\partial \mL(\btheta_0)}{\partial \btheta}\right\}_{\mM \cup S}\bigg|\nonumber\\
&\le& \sup_{\v \in \mathbb{K}_1} \bigg|\v_{\mM\cup S}\trans[\wh\Q_{(\mM \cup S)^c, \mM \cup S} (\btheta^*)- \Q_{(\mM \cup S)^c, \mM \cup S}(\btheta_0)]\{\wh \Q_{(\mM \cup S), \mM \cup
 S}(\btheta^*)\}^{-1} \left\{\frac{\partial \mL(\btheta_0)}{\partial
 \btheta}\right\}_{\mM \cup S}\bigg|\nonumber\\
&& +\sup_{\v \in \mathbb{K}_1}\bigg|\v_{\mM\cup S}\trans\Q_{(\mM \cup S)^c,
   \mM \cup S}(\btheta_0)(\{\wh \Q_{(\mM \cup S), \mM \cup
 S}(\btheta^*)\}^{-1} \nonumber\\
&&- \{\Q_{(\mM \cup S), \mM \cup S}(\btheta_0)\}^{-1})\left\{\frac{\partial \mL(\btheta_0)}{\partial
 \btheta}\right\}_{\mM \cup S}\bigg|\nonumber\\
&& +\sup_{\v \in \mathbb{K}_1} \bigg|\v_{\mM\cup S}\trans\Q_{(\mM \cup S)^c, \mM \cup S} (\btheta_0)\{\Q_{(\mM \cup S), \mM \cup S}(\btheta_0)\}^{-1} \left\{\frac{\partial \mL(\btheta_0)}{\partial
 \btheta}\right\}_{\mM \cup S}\bigg|\nonumber\\
&\leq& \sup_{\v\in {\mathbb K}_1, \w \in {\mathbb K}}
\bigg |\v_{1 (\mM \cup S)^c} \trans [\wh\Q_{(\mM \cup S)^c, \mM \cup S} (\btheta^*)-
\Q_{(\mM \cup S)^c, \mM \cup
 S}(\btheta)]\w_{\mM \cup S}\bigg |\nonumber\\
&&\times  \|\{\wh \Q_{(\mM \cup S), \mM \cup
 S}(\btheta^*)\}^{-1}\|_2 \bigg\|\left\{\frac{\partial \mL(\btheta_0)}{\partial
 \btheta}\right\}_{\mM \cup S}\bigg\|_2\nonumber\\
&& + \|\Q_{ (\mM \cup S)^c, \mM \cup S} (\btheta_0)\|_2 \bigg\|\{\wh \Q_{(\mM \cup S), \mM \cup
 S}(\btheta^*)\}^{-1} - \{\Q_{(\mM \cup S), \mM \cup
 S}(\btheta_0)\}^{-1}\bigg\|_2 \bigg\|\left\{\frac{\partial \mL(\btheta_0)}{\partial
 \bb}\right\}_{\mM \cup S}\bigg\|_2\nonumber\\
&& + \sup_{\v\in \mathbb{K}_1}\bigg|\v_{\mM\cup S}\trans\Q_{(\mM \cup S)^c, \mM \cup S}(\btheta_0) \{\Q_{(\mM \cup S), \mM
 \cup S}(\btheta_0)\}^{-1} \left\{\frac{\partial \mL(\btheta_0)}{\partial
 \btheta}\right\}_{\mM \cup S}\bigg|\nonumber\\
&\leq& O_p\left( (q + m + k_0) [\sqrt{\log\{\max(p, n)\}/n}
 +k_0^{1/2} N^{-\omega} ]^2\right)\label{eq:ineq1}\\
&& + \sup_{\v\in \mathbb{K}_1}\bigg|\v\trans\Q_{(\mM \cup S)^c, \mM \cup S} (\btheta)\{\Q_{(\mM \cup S), \mM
 \cup S}(\btheta)\}^{-1} \left\{\frac{\partial \mL(\btheta_0)}{\partial
   \btheta}\right\}_{\mM \cup S}\bigg| \nonumber\\
&=& O_p\left( (q + m + k_0) [\sqrt{\log\{\max(p, n)\}/n}
 +k_0^{1/2} N^{-\omega} ]^2\right.\nonumber\\
 &&\left.+ \sqrt{q + m + k_0} [\sqrt{\log\{\max(p, n)\}/n}
   +k_0^{1/2} N^{-\omega} ]\right)\label{eq:ineq2}\\
 &=& O_p\left( \sqrt{q + m + k_0} [\sqrt{\log\{\max(p, n)\}/n}
 +k_0^{1/2} N^{-\omega} ]\right)\label{eq:ineq3}, 
\ee
with probability  greater than $1 - \exp\left\{- O(n)\right\} - O\{\max(p,
n)^{-1}\}$, 
(\ref{eq:ineq1}) is obtained by using (\ref{eq:Qlower}) and
(\ref{eq:Qlowertrue}) and (\ref{eq:QQL2diff}) and the fact that
\bse
&&\|\{\wh \Q_{(\mM \cup S), \mM \cup
 S}(\bb^*)\}^{-1} \|_2\\
&\leq& \|\{\wh \Q_{(\mM \cup S), \mM \cup
 S}(\bb^*)\}^{-1} -[\Q_{ (\mM \cup S), \mM \cup S}(\bb)]^{-1}\|_2 +\| \{\Q_{(\mM \cup S), \mM \cup S}(\bb)\}^{-1}\|_2\\
&\leq& O_p(1).
\ese
And (\ref{eq:ineq2}) holds by using the same argument as
those lead to Lemma \ref{lem:firstinfinty}. (\ref{eq:ineq3}) holds by
the assumption that $(q + m + k_0) \sqrt{\log\{\max(p, n)\}/n} \to
0$ and $k_0
 N^{-2\omega} = O[\log\{\max(p, n)\}/n]$. By the assumption that $k_0
 N^{-2\omega} = O[\log\{\max(p, n)\}/n]$,  $\lambda =
 O_p ( \left[\log\{\max(p, n)\}/n\right]^{1/4})$, and the assumption
 that we have 
\be\label{eq:llambda}
\sup_{\v \in \mathbb{K}_1}\bigg|\v_{\mM\cup S}\trans\wh \Q_{(\mM \cup S)^c, \mM \cup S} (\btheta^*) \{\wh \Q_{(\mM \cup S), \mM \cup
 S}(\btheta^*)\}^{-1}\left\{\frac{\partial \mL(\btheta_0)}{\partial
   \btheta}\right\}_{\mM \cup S}\bigg| = o_p(\lambda). 
\ee
Therefore, 
\be\label{eq:zmax}
&& \max_{j \in (\mM \cup S)^c} \|\wh{\z}_j\|_2\\
&\leq&  \lambda^{-1}
\sup_{\v \in \mathbb{K}_1}\bigg|\v_{2(\mM\cup S)^c}\trans \left\{\frac{\partial
                              \mL(\btheta_0)}{\partial \btheta}\right\}_{2(\mM
                            \cup S)^c} \bigg|\nonumber\\
                          && + \lambda^{-1} \sup_{\v \in
                            \mathbb{K}_1} \bigg|\v_{2(\mM\cup S)^c}\trans \wh{\Q}_{(\mM \cup
                          S)^c  (\mM \cup S)}(\btheta^*)\{\wh{\Q}_{(\mM \cup
                          S)(\mM \cup S) }(\btheta^*)\}^{-1}\left\{\frac{\partial
                              \mL(\btheta_0)}{\partial
                              \btheta}\right\}_{\mM \cup S}\bigg| \nonumber\\
                          && - \sup_{\v \in
                            \mathbb{K}_1} \bigg|\v_{2(\mM\cup S)^c}\trans \wh{\Q}_{(\mM \cup
                          S)^c  (\mM \cup S)}(\btheta^*)\{\wh{\Q}_{(\mM \cup
                          S)(\mM \cup S) }(\btheta^*)\}^{-1} \nonumber\\
                        &&\times \left[\left\{ \frac{\partial
                             \sum_{j \in \mM^c}
                             q_{\lambda}(\|\wh{\bg}_{ j}\|_2)}{\partial
                              \btheta}\right\}_{\mM \cup
                            S} - 
                          \lambda \wh{\z}_{\mM \cup S}\right]\bigg|\nonumber\\
                        &\leq&- \sup_{\v \in
                            \mathbb{K}_1} \bigg|\v_{2(\mM\cup S)^c}\trans \wh{\Q}_{(\mM \cup
                          S)^c  (\mM \cup S)}(\btheta^*)\{\wh{\Q}_{(\mM \cup
                          S)(\mM \cup S) }(\btheta^*)\}^{-1}\nonumber \\
                        &&\times \left[\left\{ \frac{\partial
                             \sum_{j \in \mM^c}
                             q_{\lambda}(\|\wh{\bg}_{ j}\|_2)}{\partial
                              \btheta}\right\}_{\mM \cup
                            S} - 
                          \lambda \wh{\z}_{\mM \cup S}\right]  \bigg|+
                        o_p(\lambda) \nonumber
                        \ee
                        where the last inequality holds by using the
                        same arguments leads to Lemma 
                        \ref{lem:firstinfinty} and
                        (\ref{eq:llambda}). 
Also by (\ref{eq:diffbb1}) we have for $j \in \mM\cup S$, recall that
$\E_j$ is 
a matrix so that $\E_j\trans(\wh{\btheta} - \btheta_0) = N^{-1/2}\wh{\bg}_{ j}
-N^{-1/2} {\bG}_{0\cdot j}$
\be\label{eq:diffbb}
&& \sup_{j \in  S}N^{-1/2}\|\wh{\bg}_{ j} -  {\bG}_{0 \cdot j}\|_2\nonumber\\
&\leq&\sup_{j \in S} \|\E_j\trans\{\wh{\Q}_{\mM \cup S,
 \mM\cup S}(\btheta^*)\}^{-1} \left\{\frac{\partial
                              \mL(\btheta_0)}{\partial \btheta}\right\}_{\mM
                            \cup S} \|_2\nonumber\\
&& + \sup_{j \in S} \|\E_j\trans\{\wh{\Q}_{\mM \cup S,
 \mM\cup S}(\btheta^*)\}^{-1}  \left[\left\{\frac{\partial
                           \sum_{j \in \mM^c}   q_{\lambda}(\|\wh{\bg}_{ j}\|_2)}{\partial
                              \btheta}\right\}_{\mM \cup S}
                          -\lambda \wh{\z}_{\mM\cup
                            S} \mystrut\right]\bigg\|\nonumber\\
                        &=& \sup_{j \in S} |\E_j\trans\{\wh{\Q}_{\mM \cup S,
 \mM\cup S}(\btheta^*)\}^{-1} \left\{\frac{\partial
                              \mL(\btheta_0)}{\partial \btheta}\right\}_{\mM
                            \cup S} |\nonumber\\
&& + \sup_{j \in \mM\cup S} \|\sum_{k \in S}\E_j\trans\{\wh{\Q}_{\mM \cup S,
 \mM\cup S}(\btheta^*)\}^{-1} \E_k  \left[\frac{\partial
                            q_{\lambda}(\|\wh{\bg}_{ k}\|_2)}{\partial
                            \wh{\bg}_{ k}  }
                          -\lambda
                          \wh{\z}_k\mystrut\right]\bigg\|\nonumber\\
                        &=& o_p(\lambda) +\lambda  \sup_{j \in S} \sum_{k \in S} \|\E_j\trans\{\wh{\Q}_{\mM \cup S,
                          \mM\cup S}(\btheta^*)\}^{-1} \E_k \|_2\nonumber\\
                        &\leq& o_p(\lambda) + 2\lambda c_{\infty}\nonumber\\
                        &\leq& 3\lambda c_{\infty}, 
                        \ee
                        where the third to the last line holds by 
                        Lemma \ref{lem:fromlemm5support} that $\sup_{k \in  S}\left\|\frac{\partial
                            q_{\lambda}(\|\wh{\bg}_{ k}\|_2)}{\partial
                            \wh{\bg}_{ k}  }
                          -\lambda
                          \wh{\z}_k\mystrut\right\|_2 \leq \lambda$. The
                        second to the last line holds because
                        \bse
                  &&  \sup_{j \in S} \sum_{k \in S} \|\E_j\trans\{\wh{\Q}_{\mM \cup S,
                     \mM\cup S}(\btheta^*)\}^{-1} \E_k \|_2 \\
                   &\leq& \sup_{j \in S} \sum_{k \in S} \|\E_j\trans[\{\wh{\Q}_{\mM \cup S,
                          \mM\cup S}(\btheta^*)\}^{-1} - \{{\Q}_{\mM \cup S,
                          \mM\cup S}(\btheta^*)\}^{-1}]\E_k \|_2\\
                        && +   \sup_{j \in S} \sum_{k \in S} \|\E_j\trans\{{\Q}_{\mM \cup S,
                          \mM\cup S}(\btheta^*)\}^{-1} \E_k \|_2\\
                        &\leq& 2c_{\infty}, 
                        \ese
                        by the assumption that  $\sup_{j S} \sum_{k \in S} \|\E_j\trans\{{\Q}_{\mM \cup S,
                          \mM\cup S}(\btheta^*)\}^{-1} \E_k \|_2 \leq
                        c_{\infty}$ and
                        (\ref{eq:supQ}).
                        Now because $\min_{j \in S}(\|\bg_{ j}\|_2) \geq
                        \lambda (\gamma + 3 c_{\infty})$, by
                        (\ref{eq:diffbb}), we have
                        \bse
                        \min_{j \in S}(\|\wh{\bg}_{ j}\|_2) \geq
                        \lambda \gamma, 
                        \ese
                        by Lemma \ref{lem:fromlemma4loh2017} in the in the supplementary
                        material, we have
                        \be\label{eq:qlambda}
                        \sup_{k \in  S}\left[\frac{\partial
                            q_{\lambda}(\|\wh{\bg}_{ k}\|_2)}{\partial
                            \wh{\bg}_{ k}  }
                          -\lambda
                          \wh{\z}_k\mystrut\right]  = \0. 
                        \ee
                       Now plug this result in (\ref{eq:zmax}), we
                       have
                       \bse
                       \max_{j \in (\mM\cup S)^c} \|\wh{\z}_j\|_2 =
                       o_p(1), 
                       \ese
                       and therefore,     $\max_{j \in (\mM\cup S)^c}
                       \|\wh{\z}_j\|_2 <1$ with probability greater
                       than $1 - \exp\{O(n)\} - O\{
\max(p, n)^{-1}\}$. Therefore, Theorem \ref{th:2} applies so that
$\wh{\btheta} = \{\wh{\btheta}_{\mM\cup S}\trans, \0\trans\}\trans$ is the unique
solution for (\ref{eq:lapHT}) and also plug (\ref{eq:qlambda}) in to (\ref{eq:diffbb})
\bse
&& \wh{\btheta}_{\mM \cup S} -  \btheta_{ 0\mM \cup S}\nonumber\\
&=& - \{{\Q}_{\mM \cup S,
 \mM\cup S}(\btheta_0)\}^{-1}\left\{\frac{\partial
                              \mL(\btheta_0)}{\partial \btheta}\right\}_{\mM
                            \cup S}  + o_p(1)\\
                          &=& - \{{\Q}_{\mM \cup S,
 \mM\cup S}\{\ba_0, \bb(\cdot)\}\}^{-1}\left\{n^{-1}\sumi
\{\psi'_i\{\ba_0, \bb(s)\} - Y_i\}
\left[\Z_i\trans, N^{1/2}\left\{\int_{0}^1\X_i (s) \otimes
 \B(s)  ds\right\}\trans\right]\trans\right\}_{\mM
\cup S}\\
&&\times \{1+ o_p(1)\}, 
\ese
with probability greater than $1 - \exp\{-(n)\} -O \{\max(p,
n) ^{-1}\}$, where the last inequality holds by using the same
arguments as those lead to Lemma \ref{lem:firstinfinty}, and the fact
$\sup_{s \in [0,
   1]}|\B\trans (s) \bg_{0j} - \beta_j(s)| = O_p(N^{-\omega})$ in
 Condition \ref{con:bspline} and the condition the assumption 
that $k_0
N^{-2\omega} = o[\log\{\max(p, n)\}/n]$.
This proves the result. \qed

\noindent {\bf Proof of Lemma~\ref{lem:T0}:}
\bse
&& \bPsi^{-1/2}(\bSig, \Q, \bb) \bomega_n \\
&=& -  \sqrt{n}  \bPsi^{-1/2}(\bSig, \Q, \bb) \A \Q^{-1}_{\mM\cup S,
 \mM\cup S}\{\ba_0, \bb(\cdot)\}_{\mM\cup S,
 \mM\cup S}^{-1}\\
&& \left\{n^{-1}\sumi
\{\psi'_i\{\ba_0, \bb(s)\} - Y_i\}
\left[\Z_i\trans, N^{1/2}\left\{\int_{0}^1\X_i (s) \otimes
 \B(s)  ds\right\}\trans\right]\trans\right\}_{\mM
\cup S}. 
\ese
It is easy to see that
\bse
&& \sumi \cov \left(\bPsi^{-1/2}(\bSig, \Q, \bb) \A \Q^{-1}_{\mM\cup S,
 \mM\cup S}\{\ba_0, \bb(\cdot)\}_{\mM\cup S,
 \mM\cup S}^{-1} \right.\\
&&\left.\times n^{-1/2}\{\psi'_i\{\ba_0, \bb(s)\} - Y_i\}
\left[\Z_i\trans, N^{1/2}\left\{\int_{0}^1\X_i (s) \otimes
 \B(s)  ds\right\}\trans\right]\trans_{\mM
\cup S}\right)= \I_{rN \times rN}. 
\ese
Also,
\bse
&& (rN)^{1/4} \sumi E\left[\|\bPsi^{-1/2}(\bSig, \Q, \bb) \A \Q^{-1}_{\mM\cup S,
 \mM\cup S}\{\ba_0, \bb(\cdot)\}_{\mM\cup S,
 \mM\cup S}^{-1} \right.\\
&&\left.\times n^{-1/2}\{\psi'_i\{\ba_0, \bb(s)\} - Y_i\}
\left[\Z_i\trans, N^{1/2}\left\{\int_{0}^1\X_i (s) \otimes
 \B(s)  ds\right\}\trans\right]\trans_{\mM
\cup S}\|_2^3\right]\\
&=& O\{(rN)^{1/4}n^{-1/2}\} = o(1), 
\ese
where the second to the last equality holds by the assumption that
\bse
&& E\left[\|\bPsi^{-1/2}(\bSig, \Q, \bb) \A \Q^{-1}_{\mM\cup S,
 \mM\cup S}\{\ba_0, \bb(\cdot)\}_{\mM\cup S,
 \mM\cup S}^{-1} \right.\\
&&\left.\times \{\psi'_i\{\ba_0, \bb(s)\} - Y_i\}
\left[\Z_i\trans, N^{1/2}\left\{\int_{0}^1\X_i (s) \otimes
 \B(s)  ds\right\}\trans\right]\trans_{\mM
\cup S}\|_2^3\right] = O(1), 
\ese
and the last equality holds by the assumption that $
(rN)^{1/4}n^{-1/2}\to 0$. Using the Lemma \ref{lem:frombentkus} from
the supplementary material, let $\mathcal{Z}$ be a standard Gaussian random
variable, we have
\bse
\lim_{n\to \infty}\sup_{\mC} |\Pr\left[\bPsi^{-1/2}(\bSig, \Q, \bb)
 \bomega_n \in \mC\right] - \Pr(\mathcal{Z} \in \mC)| = 0. 
\ese
Now we choose $\mC_{x} = \left\{\z: \|\z -\c_n \|_2\leq x\right\}$, we have
\bse
\lim_{n\to \infty} |\Pr\left[\bPsi^{-1/2}(\bSig, \Q, \bb)
 \bomega_n \in \mC\right] - \Pr(\mathcal{Z} \in \mC_x)| = 0, 
\ese
which implies
\bse
\lim_{n\to \infty } | \Pr(T_0\leq x) - \Pr\left\{\chi^2(rN, \c_n\trans \bPsi^{-1}(\bSig, \Q, \bb) \c_n) \leq
 x\right\}| = 0, 
\ese
where $\chi^2(r, \gamma)$ is a $r$ degree of freedom non-central chi-square distribution,
with non-centrality parameter $\gamma$.
\qed

\noindent {\bf Proof of Theorem~\ref{th:4}:}
From Theorem \ref{th:3}, we have
\bse
&& \wh{\btheta}_{\mM \cup S} -  \btheta_{ 0\mM \cup S}\nonumber\\
&=& - [{\Q}_{\mM \cup S,
 \mM\cup S}\{\ba_0, \bb(\cdot)\}]^{-1}\left\{\sumi
\{\psi'_i\{\ba_0, \bb(s)\} - Y_i\}
\left[\Z_i\trans, N^{1/2}\left\{\int_{0}^1\X_i (s) \otimes
 \B(s)  ds\right\}\trans\right]\trans\right\}_{\mM
\cup S}\\
&&\times \{1+ o_p(1)\} 
                          \ese
                          and $\wh{\btheta}_{(\mM \cup S)^c}  = \0$
                          with probability greater than $1 -
                          \exp\{-O(n)\} - O\{\max(p, n)^{-1}\}$.
                      Let $\wh{\btheta}_{\mM} \equiv
                      (\wh{\btheta}_{1}\trans, \wh{\btheta}_{2j}\trans, j
                      \in \mM)\trans$, and  ${\btheta}_{0 \mM} \equiv
                      ({\btheta}_{01}\trans, {\btheta}_{02j}\trans, j
                      \in \mM)\trans$.   Then
                          \bse
                      &&    \sqrt{n} \A (\wh{\btheta}_{\mM} -
                      {\btheta}_{0\mM}) \\
                      &=&- \sqrt{n}\A [{\Q}_{\mM \cup S,
                        \mM\cup S}\{\ba_0, \bb(\cdot)\}]^{-1}\\
                      &&\times \left\{\sumi
\{\psi'_i\{\ba_0, \bb(s)\} - Y_i\}
\left[\Z_i\trans, N^{1/2}\left\{\int_{0}^1\X_i (s) \otimes
 \B(s)  ds\right\}\trans\right]\trans\right\}_{\mM
\cup S}\\
&=& \bomega_n\{1 + o_p(1)\}. 
\ese
Furthermore, because when  $\C\bb_{\mM}(s) = \t(s) + \h_n(s)$, we have
\bse
&& \vec\left[\left\{\int_0^1 \B(s) \B(s)\trans\right\}^{-1} ds \int_0^1 \B(s)
 \bb_{\mM}(s)\trans ds \C\trans \right]\\
&=&\vec\left[\left\{\int_0^1 \B(s) \B(s)\trans ds \right\}^{-1} 
\int_0^1 \B(s)\{\t(s)\trans + \h_n\trans(s)\} ds \right]. 
\ese
Hence we have 
\bse
&& \sqrt{n} \bPsi^{-1/2}(\wh{\bSig}, \wh{\Q}, \wh{\btheta}) \left(\A
\wh{\btheta}_{\mM} - N^{-1/2}\vec\left[\left\{\int_0^1 \B(s) \B(s)\trans ds \right\}^{-1} 
 \int_0^1 \B(s)\t(s)\trans ds\right]\right)\\
&=& \sqrt{n} \bPsi^{-1/2}(\wh{\bSig}, \wh{\Q}, \wh{\btheta}) \left(\A
 \wh{\btheta}_{\mM} - \A \wh{\btheta}_{\mM} \right)\\
&& + \sqrt{n} \bPsi^{-1/2}(\wh{\bSig}, \wh{\Q}, \wh{\btheta})\left(\A \wh{\btheta}_{\mM} - N^{-1/2}\vec\left[\left\{\int_0^1 \B(s) \B(s)\trans ds \right\}^{-1} 
   \int_0^1 \B(s)\{\t(s)\trans  + \h_n(s)\trans\}ds\right]\right) \\
&& + \sqrt{n} \bPsi^{-1/2}(\wh{\bSig}, \wh{\Q}, \wh{\btheta})\left(\vec\left[N^{-1/2}\left\{\int_0^1 \B(s) \B(s)\trans ds \right\}^{-1} 
   \int_0^1 \B(s) \h_n(s)\trans ds\right]\right) \\
&=& \sqrt{n}\bPsi^{-1/2}(\wh{\bSig}, \wh{\Q}, \wh{\btheta}) (\bomega_n + \c_n)
\{1 + o_p(1)\} + O_p\{n^{1/2} (rN)^{1/2} N^{-1/2 - \omega}\}, 
\ese
where the last equality holds by the Condition that $\sup_{s \in [0, 1]}|\B(s)\trans
N^{-1/2}\bg_{0j} -N^{-1/2} \beta_j(s)| = N^{-\omega -1/2}$. 
Therefore,
\bse
&& n  \left(\A
\wh{\btheta}_{\mM} - N^{-1/2}\vec\left[\left\{\int_0^1 \B(s) \B(s)\trans ds \right\}^{-1} 
 \int_0^1 \B(s)\t(s)\trans ds\right]\right) \trans \\
&&\times \bPsi^{-1}(\wh{\bSig}, \wh{\Q}, \wh{\btheta}) \left(\A
\wh{\btheta}_{\mM} - N^{-1/2}\vec\left[\left\{\int_0^1 \B(s) \B(s)\trans ds \right\}^{-1} 
 \int_0^1 \B(s)\t(s)\trans ds\right]\right) \\
&=&  n (\bomega_n + \c_n) \trans \bPsi^{-1/2}(\wh{\bSig}, \wh{\Q}, \wh{\btheta}) (\bomega_n + \c_n)
\{1 + o_p(1)\} + o_p\{rN\}\\
&=& n (\bomega_n + \c_n) \trans \bPsi^{-1/2}(\bSig, \Q, \bb) (\bomega_n + \c_n)
\{1 + o_p(1)\} + o_p\{rN\}. 
\ese
The second to the last equality holds by the assumption that $n
N^{-2\omega -1} \to 0$ and the last equality holds by the consistency
of $\wh{\bSig}$, $\wh{\Q}$, and $\wh{\btheta}$.
Therefore, we have $T - T_0 = o_p(rN)$. This proves the result.  \qed

\bibliographystyle{plainnat}
\bibliography{hdfuninfer}

\makeatletter\@input{xx2.tex}\makeatother